\documentclass[fleqn,10pt]{wlscirep}
\usepackage[utf8]{inputenc}
\usepackage[T1]{fontenc}

\usepackage{color}
\usepackage{float}
\usepackage{mathrsfs}
\usepackage{mathtools}
\usepackage{multicol, blindtext}
\usepackage{amsmath}
\usepackage{amsthm}
\usepackage{amssymb}
\usepackage{graphicx}
\usepackage{subfig}
\usepackage{babel}

\usepackage{algorithm}

\usepackage{bm}
\usepackage{booktabs}
\usepackage{enumerate}
\usepackage{epsfig}
\usepackage{epsf,epstopdf}
\usepackage{multirow}
\usepackage{url} 
\usepackage{array}
\usepackage{graphicx}

\PassOptionsToPackage{normalem}{ulem}
\usepackage{ulem}

\providecommand{\tabularnewline}{\\}
\floatstyle{ruled}
\newfloat{algorithm}{tbp}{loa}
\providecommand{\algorithmname}{Algorithm}
\floatname{algorithm}{\protect\algorithmname}
\numberwithin{equation}{section}
\numberwithin{figure}{section}
\theoremstyle{plain}
\newtheorem{thm}{\protect\theoremname}
\usepackage{mathrsfs}
\makeatother

\usepackage{babel}
\usepackage{listings}
\providecommand{\theoremname}{Theorem}

\renewcommand{\thefigure}{\arabic{figure}}

\title{Loss of structural balance in stock markets}

\author[a]{Eva Ferreira}
\author[a]{Susan Orbe}
\author[b]{Jone Ascorbebeitia}
\author[c]{Brais \'Alvarez Pereira}
\author[d,*]{Ernesto Estrada}
\affil[a]{Department of Quantitative Methods, University of the Basque Country UPV/EHU, Avda. Lehendakari Aguirre 81, Bilbao, 48015 Spain}
\affil[b]{Department of Economic Analysis, University of the Basque Country UPV/EHU, Avda. Lehendakari Aguirre 81, Bilbao, 48015 Spain}
\affil[c]{Nova School of Business and Economics (Nova SBE), NOVAFRICA, and BELAB}
\affil[d]{Institute of Mathematics and Applications, University of Zaragoza,
Pedro Cerbuna 12, Zaragoza 50009, Spain; ARAID Foundation, Government
of Aragon, Spain. Institute for Cross-Disciplinary Physics and Complex
Systems (IFISC, UIB-CSIC), Campus Universitat de les Illes Balears
E-07122, Palma de Mallorca, Spain.}

\affil[*]{estrada66@unizar.es}

\begin{abstract}
We use rank correlations as distance functions to establish the interconnectivity between stock returns, building weighted signed networks for the stocks of seven European countries, the US and Japan. We establish the theoretical relationship between the level of balance in a network and stock predictability, studying its evolution from 2005 to the third quarter of 2020. We find a clear balance-unbalance transition for six of the nine countries, following the August 2011 Black Monday in the US, when the Economic
Policy Uncertainty index for this country reached its highest monthly level before the COVID-19 crisis. This sudden loss of balance is mainly caused by a reorganization of the market networks triggered by a group of low capitalization stocks belonging to the non-financial sector. After the transition, the stocks of companies in these groups become all negatively correlated between them and with most of the rest of the stocks in the market. The implied change in the network
topology is directly related to a decrease in stocks predictability, a finding with novel important implications for asset allocation and portfolio hedging strategies.
\end{abstract}
\begin{document}
\flushbottom
\maketitle

\thispagestyle{empty}

\section*{Introduction}

The complex nature of the stock market is condensed in the phrase:
``Economists are as perplexed as anyone by the behavior of the stock
market'' \cite{hall2001struggling}. Understanding this complex behavior--characterized
by a combination of saw-tooth movements and low-frequency upward and
downward switches \cite{preis2011switching}--is a major goal in
finance. However, the importance of stock markets goes beyond finance
to impact macroeconomic modeling and policy discussion \cite{fischer1984macroeconomics}.
In fact, it has been recognized that the stock market is a good predictor
of the business cycle and of the components of gross national product
of a given country \cite{fischer1984macroeconomics}. Special emphasis
has been placed on understanding the role of risk and uncertainty
in stock markets from different perspectives \cite{asgharian2019economic,de1985does,fama1993common,loretan2000evaluating,solnik1996international,spelta2020behavioral,su2019understanding,wen2012measuring,PNAS_2,PNAS_4,PNAS_7,jackson2020systemic,buldyrev2021market},
with a focus on security pricing and corporate investment decisions.
At the end of the day, as said by Fischer and Merton \cite{fischer1984macroeconomics}:
in the absence of uncertainty, much of what is interesting in finance
disappears.

As the majority of complex systems, stock markets have an exoskeleton
formed by entities and complex interactions, which give rise to the
observed patters of the market behavior \cite{haldane2013rethinking,bougheas2015complex,grilli2017networked,iori2018empirical}(see Ref. \cite{bardoscia2021physics} for a review of the recent research in financial networks and their practical applicability).
For the same set of entities, there are different forms to define the
connectivity between them. A typical way of connecting financial institutions
is by means of borrowing/lending relations \cite{battiston2012default,battiston2012debtrank,battiston2016complexity,bardoscia2021physics}.
However, in the case of stock markets, where stocks represent the
nodes of the network, the inter-stock connectivity is intended to
capture the mutual trends of stocks over given periods of times. Mantegna
\cite{mantegna1999hierarchical} proposed to quantify the degree of
similarity between the synchronous time evolution of a pair of stock
prices by the correlation coefficient $\rho_{ij}$ between the two
stocks $i$ and $j$. Then, this correlation coefficient is transformed
into a distance using: $d_{ij}=\sqrt{2\left(1-\rho_{ij}\right)}$.
Mantegna's approach has been widely extended in econophysics \cite{kutner2019econophysics}
where it is ubiquitous nowadays as a way to define the connectivity
between stocks. This approach has been used for instance for the construction
of networks to analyze national or international stock markets \cite{birch2016analysis,brida2016network,kaue2012structure,wang2018correlation,zhao2018stock,guo2018development}.
A characteristic feature of this approach is that although $\rho_{ij}\in\left[-1,1\right]$,
the distance $d_{ij}$ used as a weight of the links between stocks
is nonnegative. This allows the use of classical network techniques
for their analysis (see Ref. \cite{tumminello2010correlation} for a review).
To avoid certain loss of information inherent to the previous approach
\cite{heiberger2014stock}, Tse et al. \cite{chi2010network} decide
that an edge exists between a pair of stocks only if $\rho_{ij}>\left|z\right|$
for a given threshold $z$, which however retains the nonnegativity of the edge weights.

The importance of allowing explicit distinction between positive and
negative interdependencies of stocks, which is lost in the previous
approaches, was recently remarked by Stavroglou et al. \cite{stavroglou2019hidden}.
They built separated networks based on positive and negative interdependencies.
This separation may still hide some important structural characteristics
of the systems under analysis. Fortunately, there is an area of graph
theory which allows to study networks with simultaneous presence of
positive and negative links. These graphs, known as signed graphs,
are known since the end of the 1950's \cite{harary1953notion} (see
Ref. \cite{zaslavsky2012mathematical} for a review), but only recently
have emerged as an important tool in network theory \cite{leskovec2010signed,facchetti2011computing,estrada2014walk,kirkley2019balance,shi2019dynamics}.
A focus on signed networks has been the theory of social balance \cite{heider1946attitudes,harary1953notion}.
It basically states that social triads where the three edges are positive
(all-friends) or where two are negative and one is positive--the
enemy of my enemy should be my friend--are more abundant in social
systems than those having all-negative, or two positive and one negative
ones. The first kind of networks are known as balanced, and the second
ones as unbalanced (see Refs. \cite{leskovec2010signed,facchetti2011computing,estrada2014walk,kirkley2019balance,shi2019dynamics} for applications).

Here we introduce the use of signed networks to represent stock markets
of nine developed countries and to study the degree of balance. First
note that in finance, the time varying behavior of networks is an
important characteristic, since the time varying comovements become a risk factor \cite{ascorbebeitia2021effect}. Hence,
we analyze the time variation of the degree of balance on stock networks
between January 2005 and September 2020, thus addressing a gap pointed out in a recent study \cite{bardoscia2021physics}. To build the graphs we use
the rank correlations between stocks, which provide a measure of
similarity better suited for financial distributions. We find a balance-unbalance
transition (BUT) in six of the countries studied. The BUTs occur around
September/October 2011 for the US, Greece, Portugal, Ireland, and Spain,
and later in France, which take place just after the Black Monday
in August 2011. Neither Germany, nor Italy, nor Japan showed clear
signs of this BUT. We discover that the observed BUTs are mainly triggered
by a reorganization of the topology of the stock market networks.
They consist on the movement of a few low capitalization stocks from
the periphery to the center of the networks by forming cliques of
fully-negative interdependencies among them and with most of the rest
of the stocks in the market. These transitions impact directly on
the naive predictability of stock prices from pairwise rank correlations.
This is a novel finding with a direct impact on optimal asset allocation
and hedging, which opens promising avenues for future research.

\section*{Theoretical Approaches}

\subsection*{Correlations and predictability of interrelated stock prices}\label{subsec:Correlations-and-predictability}

To study the similarity between stock price changes we consider the
time series of the log returns of stocks, $Y_{i}\left(t\right)=\log\left[P_{i}\left(t\right)/\\ P_{i}\left(t-1\right)\right],$ where
$P_{i}\left(t\right)$ is the daily adjusted closing price of the
stock $i$ at time $t$. Let us consider the log return of three stocks
forming the vectors $Y_{1}$, $Y_{2}$ and $Y_{3},$ such that $Y_{i}=\left[Y_{i}\left(1\right),\cdots,Y_{i}\left(S\right)\right]^{T}.$
The similarity between the orderings of the log returns $Y_A$ and $Y_B$ of two stocks $A$ and $B$
can be captured by the Kendall's tau

\begin{equation}
\tau=2\mathcal{P}\left(\left(Y_{A}-Y'_{A}\right)\left(Y_{B}-Y'_{B}\right)>0\right)-1,
\end{equation}
where $Y_{\ell}'$ is an independent copy of the vector $Y_{\ell}$,
$\ell=A,B,$ and $\mathcal{P}(X)$ is the probability of the event $X$. If
two rankings are concordant, $\tau>0$, if two rankings are independent,
$\tau=0$, and if the two rankings are discordant, $\tau<0$.

When considering the correlations between three stocks forming a triad
there are four cases that can emerge: (i) the three pairs of stocks
are correlated: $\tau\left(Y_{1},Y_{2}\right)>0$, $\tau\left(Y_{1},Y_{3}\right)>0$
and $\tau\left(Y_{2},Y_{3}\right)>0$; (ii) one pairs of stocks is
correlated and two pairs are anticorrelated: $\tau\left(Y_{1},Y_{2}\right)>0$,
$\tau\left(Y_{1},Y_{3}\right)<0$ and $\tau\left(Y_{2},Y_{3}\right)<0$;
(iii) two pairs of stocks are correlated and one pair is anticorrelated:
$\tau\left(Y_{1},Y_{2}\right)>0$, $\tau\left(Y_{1},Y_{3}\right)>0$
and $\tau\left(Y_{2},Y_{3}\right)<0$; (iv) the three pairs of stocks
are anticorrelated: $\tau\left(Y_{1},Y_{2}\right)<0$, $\tau\left(Y_{1},Y_{3}\right)<0$
and $\tau\left(Y_{2},Y_{3}\right)<0$.

Let us focus on estimating the trend of one of the stocks from another
in the correlation triad considering a time varying regression model
(see Appendix B in the Supplementary Information (SI)). Let us write $\hat{Y}_{i}(t)\left(Y_{j\neq i}(t)\right)$ for the
nonparametric smoothed estimate of $Y_{i}$ from $Y_{j}$: $\hat{Y}_{i}(t)=\hat{m}_{i}(Y_{j}(t))$,
where $m_{i}(\cdot)$ is a non-specified unknown function. Then, for
any stock, e.g., $Y_{1}$, we can make two predictions from another
stock, e.g., $Y_{2}.$ One of them is simply $\hat{Y}_{1}(t)\left(Y_{2}(t)\right)$,
and the other is using first $Y_{2}$ to estimate $Y_{3}$, and then
estimate $Y_{1}$ from the last, i.e., $\hat{Y}_{1}(t)\left(\hat{Y}_{3}\left(Y_{2}(t)\right)\right)$,
that is $\hat{Y}_{1}(t)=\hat{m}_{1}\big(\hat{Y}_{3}(t)\big)$, where
$\hat{Y}_{3}(t)=\hat{m}_{3}\big(Y_{2}(t)\big)$. In the cases (i)
and (iii) the trends of $Y_{1}$ estimated by $\hat{Y}_{1}(t)\Big(Y_{2}(t)\Big)$
and by $\hat{Y}_{1}(t)\Big(\hat{m}_{3}\big(Y_{2}(t)\big)\Big)$ are
the same. In the case (iii) $\hat{Y}_{1}(t)\Big(Y_{2}(t)\Big)$ estimates
an increase of $Y_{1}$ with the increase of $Y_{2}$, while $\hat{Y}_{1}(t)\Big(\hat{m}_{3}\big(Y_{2}(t)\big)\Big)$
estimates a decrease of $Y_{1}$ with the increase of $Y_{2}$. Therefore,
observing the trend evolution of $Y_{2}$ does not clearly estimate
the trend of $Y_{1}$ independently of the quality of the pairwise
rank correlations. In other words, the trend of $Y_{1}$ is unpredictable
from $Y_{2}$. A similar situation occurs for the case (iv).

\begin{figure}[htpb]
\begin{centering}
{\begin{centering}
\includegraphics[width=0.8\linewidth]{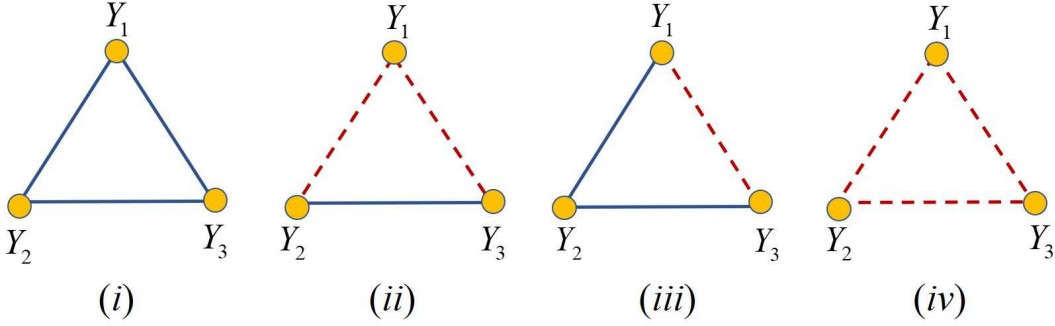}
\par
\end{centering}}
\par\end{centering}
\caption{Graphical representation interdependencies between stocks $Y_{1}$, $Y_{2}$ and $Y_{3}$ in a triad. }\label{correlation_balance}
\end{figure}

In closing, the ``predictability'' of the trend of a stock $Y_{i}$
from that of $Y_{j}$ increases when the signs of the two estimations
coincide. Otherwise, such predictability drops as a consequence of
the different trends predicted by $\hat{Y}_{i}\left(t\right)\left(Y_{j}\left(t\right)\right)$
and $\hat{Y}_{i}(t)\Big(\hat{m}_{k}\big(Y_{j}(t)\big)\Big)$. If we
represent the three stocks at the vertices of a triangle and the signs
of the estimates as the corresponding edges, we have that the four
cases analyzed before can be represented as in Fig. \ref{correlation_balance}.
Because the magnitude of $\tau_{ij}(t)$ quantifies the quality of the
estimation for each $t$, we suggest to replace the edges of the triangles in Fig.
\ref{correlation_balance} by the corresponding values of the Kendall's
tau instead of their estimates signs (for more details see Appendix B in the SI). In this case, a measure of the predictability
of the stocks in a given triad is given by $\tilde{K}(t)=\tau_{ij}(t)\tau_{ik}(t)\tau_{jk}(t)$,
with $\tilde{K}(t)\rightarrow+1$ corresponding to larger predictability
and with $\tilde{K}(t)\rightarrow-1$ to poorer predictability at time $t$. Obviously,
we should extend this measure beyond triangles, which is what we do
in the next Section.

A triangle like those illustrated in Fig. \ref{correlation_balance}
is known as a signed triangle. A signed triangle for which the product
of its edge signs is positive is known as a balanced triangle (cases (i) and (ii) in Fig. \ref{correlation_balance}). Otherwise,
it is said to be unbalanced (cases (iii) and (iv) in Fig. \ref{correlation_balance}). These concepts are naturally extended
to networks of any size by means of the concept of signed graphs (see
next Section). A signed graph where all the triangles (more generally
all its cycles) are balanced is said to be balanced. Therefore, we
move our scenario of stock predictability to that of the analysis
of balance in signed stock networks.

\subsection*{Balance in Weighted Signed Stock Networks}

Let the stocks in a given market be represented as a set of nodes
$V$ in a graph $G=\left(V,E,W\right)$, where $E\subseteq V\times V$
is a set of edges which represent the correlations between pairs of
stocks, and $W:E\rightarrow[-1,+1]$ is a mapping that assigns a value between $-1$ and $+1$ to each edge. We call these graphs weighted
signed stock networks (WSSNs). The edges of the WSSN are defined on
the basis of time varying Kendall's tau between the log returns of
the stocks. The details about the edges definition and the adjacency
matrix construction of the WSSN are given in Appendix A of the SI.

For any cycle in a WSSN we will say that it is positive if the product
of all Kendall's taus forming the edges of the cycle is positive.
A WSSN is balanced if all its cycles are positive. However, the question
is not to reduce the problem to a yes or not classification but to
quantify how close or far a WSSN is from balance. To do so, we consider
a hypothetical scenario in which a given WSSN $G=\left(V,E,W\right)$
has evolved from a balanced network $G'=\left(V,E,\left|W\right|\right)$.
That is, we are interested in quantifying the departure of a given
WSSN from a balanced version of itself. We define the equilibrium
constant $K$ for the equilibrium $G'\rightleftarrows G$ as: $K=\exp\left(-\beta\triangle F\right)$,
where $\beta=\left(k_{B}T\right)^{-1}$ with $k_{B}$ a constant and
$T$ the temperature, and $\triangle F$ is the change of the Gibbs
free energy of the system \cite{estrada2014walk}. The equilibrium
constant is bounded as $0<K\leq1$, with the lower bound indicating
a large departure of the WSSN from its balanced analogous and reaching
$K=1$ when it is balanced. We prove in Appendix C of the SI that this $K$ is related
to $\tilde{K}$, which was defined in the previous section to account
for the product of the signs of all edges in the cycles of the WSSN.

To calculate balance in a signed network, Facchetti et al. \cite{facchetti2011computing}
used an intensive computational technique that assigns a $+1$ or
a $-1$ to all the nodes so as to minimize the energy functional

\begin{equation}
\hat{\mathcal{\mathcal{\mathscr{H}}}}\left(\Gamma\right)\coloneqq-\sum_{\left(v,w\right)\in E}J_{vw}\sigma_{v}\sigma_{w},\label{eq:Hamiltonian}
\end{equation}
where $\sigma_{k}\in\left\{ +1,-1\right\} $, $k=1,\ldots,n$ with
$n$ equal to the number of nodes and $J_{vw}$ accounts for the sign
of the edge in the signed graph. Here we use a different approach,
which avoids the minimization of that energy functional, and which
is based on the free energies $F\left(G\right)$ and $F\left(G'\right)$
appearing in $\triangle F=F\left(G\right)-F\left(G'\right)$. In this
case, we notice that the energy functional (\ref{eq:Hamiltonian})
is minimized if $J_{vw}\sigma_{v}\sigma_{w}>0$ and maximized if $J_{vw}\sigma_{v}\sigma_{w}<0$.
Therefore, we identify this term with the Kendall's tau: $J_{vw}\sigma_{v}\sigma_{w}\coloneqq\tau_{vw}$,
which means that $\hat{\mathcal{\mathcal{\mathscr{H}}}}\left(\Gamma\right)=-A\left(\Gamma\right)$,
such that correlated pairs of stocks contribute to the minimization
of the energy functional and anticorrelated ones contribute in the
opposite direction. Then, we have $\triangle F=-\beta^{-1}\ln\left(Z\left(G\right)/Z\left(G'\right)\right)$,
where $Z\left(\Gamma\right)=tr\left(\exp\left(-\beta\hat{\mathcal{\mathcal{\mathscr{H}}}}\left(\Gamma\right)\right)\right)$.

For the current work we consider the Economic Policy Uncertainty (EPU)
index \cite{baker2016measuring} as a proxy for the inverse temperature
$\beta$. EPU is compiled for every country on a monthly basis and
represents a level of ``agitation'' of the market at a given date.
We then define a relative inverse temperature as: $\beta_{rel}=EPU/\max\left(EPU\right)$,
where $\max\left(EPU\right)$ is the maximum EPU reported for that
country in the period of analysis. In this case the risk is high when
the ``temperature'' of the system at a given time is approaching
the maximum temperature reached by that country in the whole period,
$\beta_{rel}\rightarrow1$. On the other hand, the risk is minimum
when the temperature at a given time is much smaller that the maxim
temperature of the period, $\beta_{rel}\rightarrow0$.

This finally gives our measure of structural balance for an WSSN \cite{estrada2014walk,estrada2019rethinking}:

\begin{equation}
K=\dfrac{Z\left(G\right)}{Z\left(G'\right)}=\dfrac{tr\left(\exp\left(-\beta_{rel}\hat{\mathcal{\mathcal{\mathscr{H}}}}\left(G\right)\right)\right)}{tr\left(\exp\left(-\beta_{rel}\hat{\mathcal{\mathcal{\mathscr{H}}}}\left(G'\right)\right)\right)}=\dfrac{tr\left(\exp\left(\beta_{rel}A\left(G\right)\right)\right)}{tr\left(\exp\left(\beta_{rel}A\left(G'\right)\right)\right)}.
\end{equation}

In the Supplementary Information we prove that $K=1$ if and only
if the WSSN is balanced. The departure of $K$ from unity characterizes
the degree of unbalace that the WSSN has.

\section*{Results}

\begin{figure}[htpb]
\centering %
\begin{tabular}{@{}cccccc@{}}
\multicolumn{3}{c}{\includegraphics[width=0.4\linewidth]{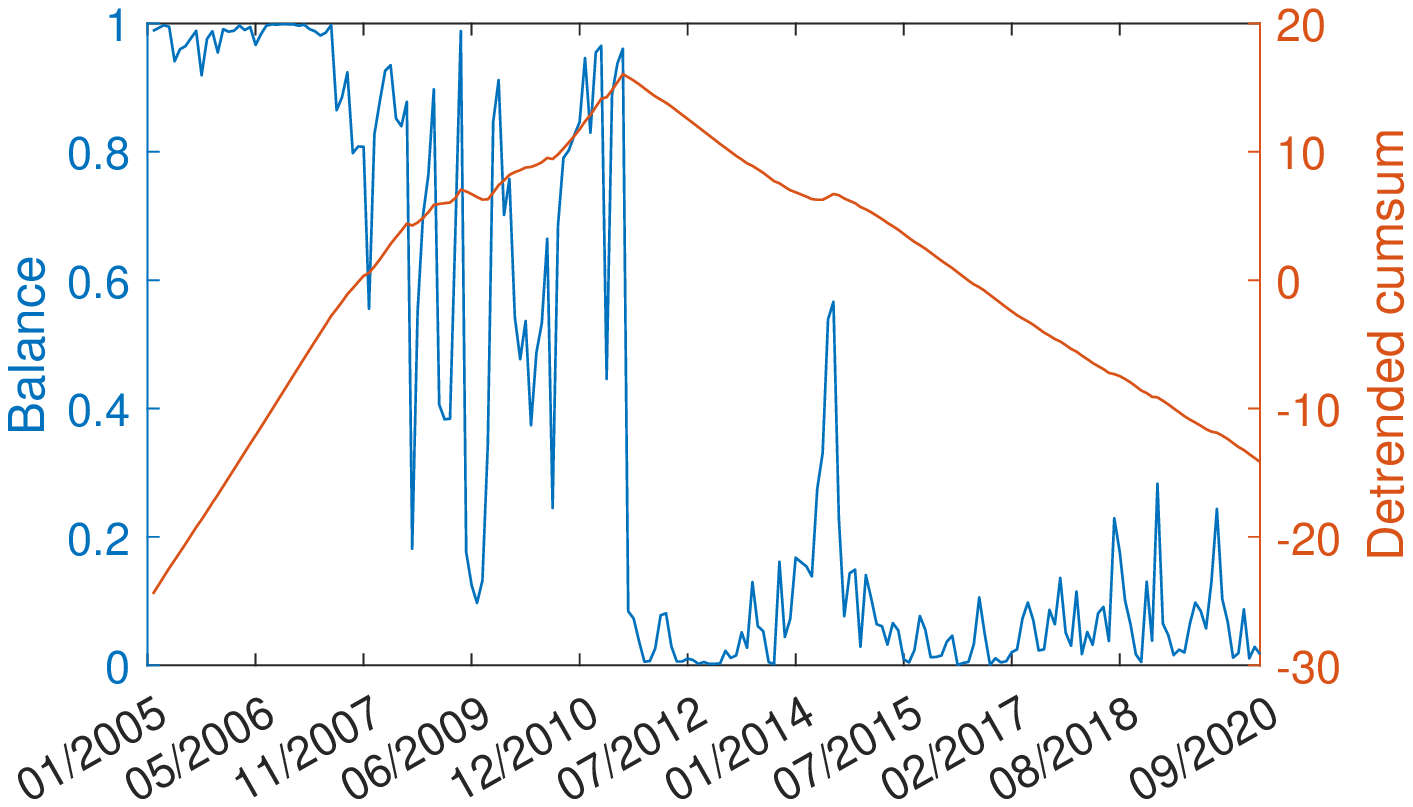}} & \multicolumn{3}{c}{\includegraphics[width=0.4\linewidth]{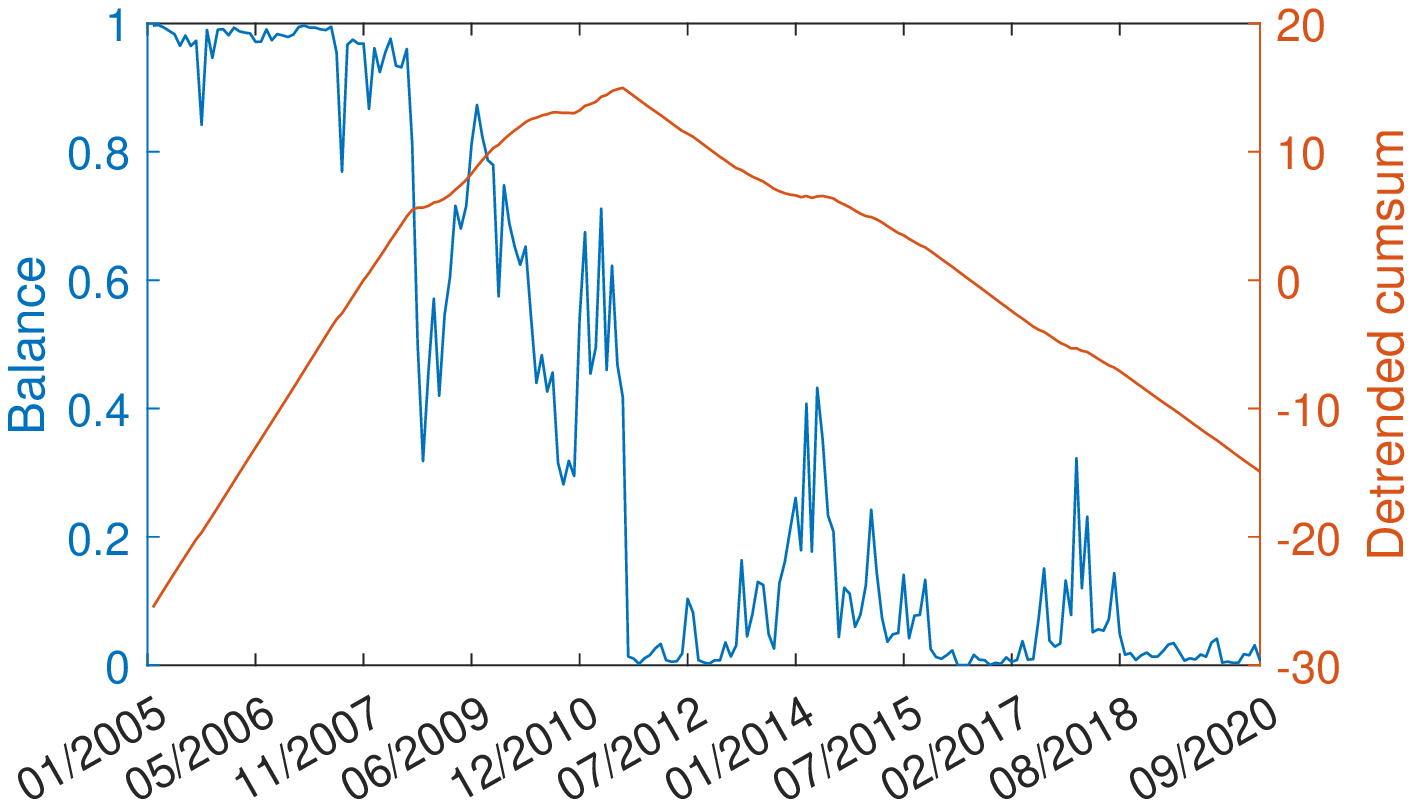}}\tabularnewline
 &  &  &  &  & \tabularnewline
\includegraphics[trim=0.5cm 0.25cm 0cm 0.25cm,clip,width=0.12\linewidth]{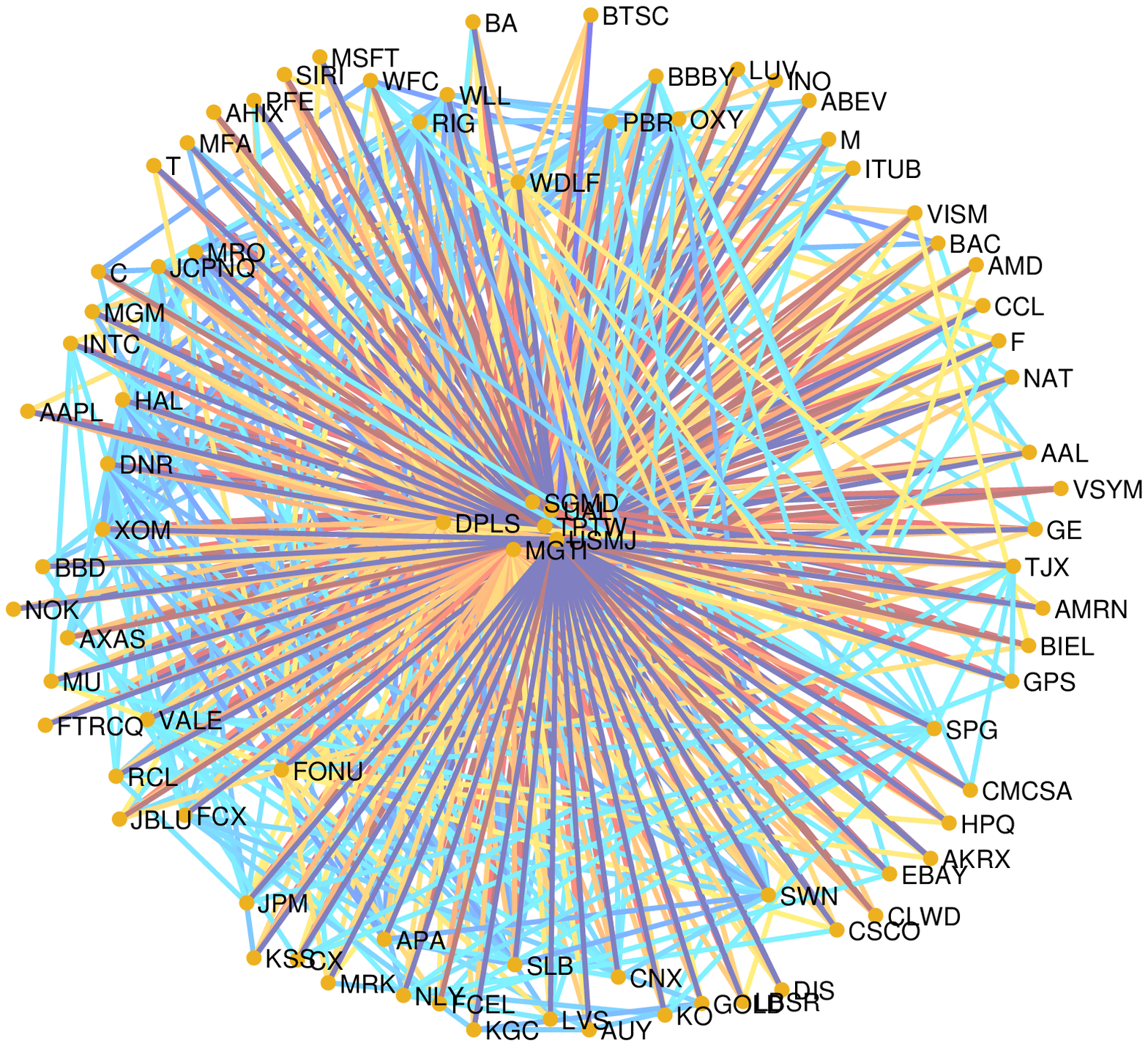} & \includegraphics[trim=0.5cm 0.25cm 0cm 0.25cm,clip,width=0.12\linewidth]{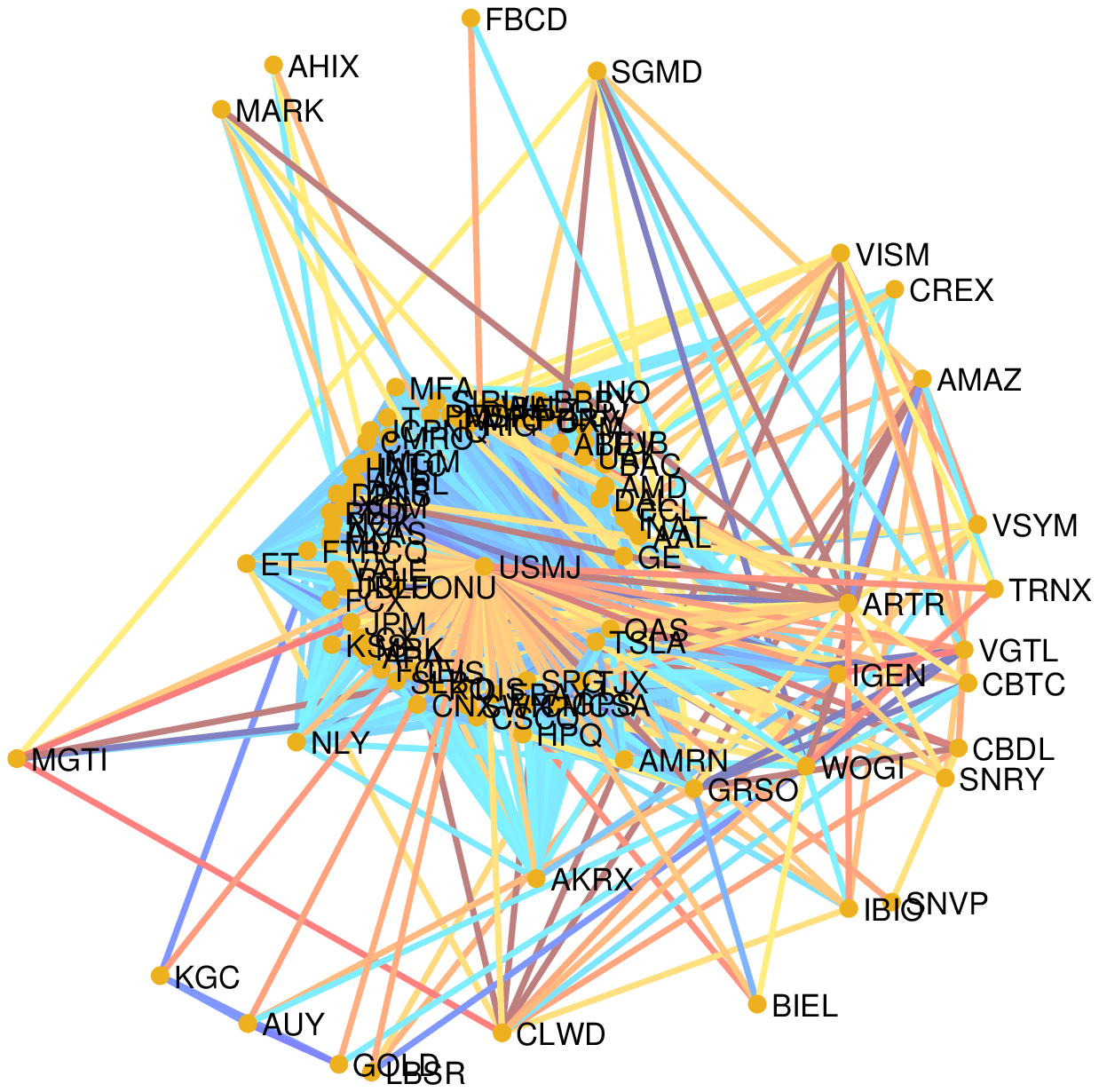} & \includegraphics[trim=0.5cm 0.25cm 0cm 0.25cm,clip,width=0.12\linewidth]{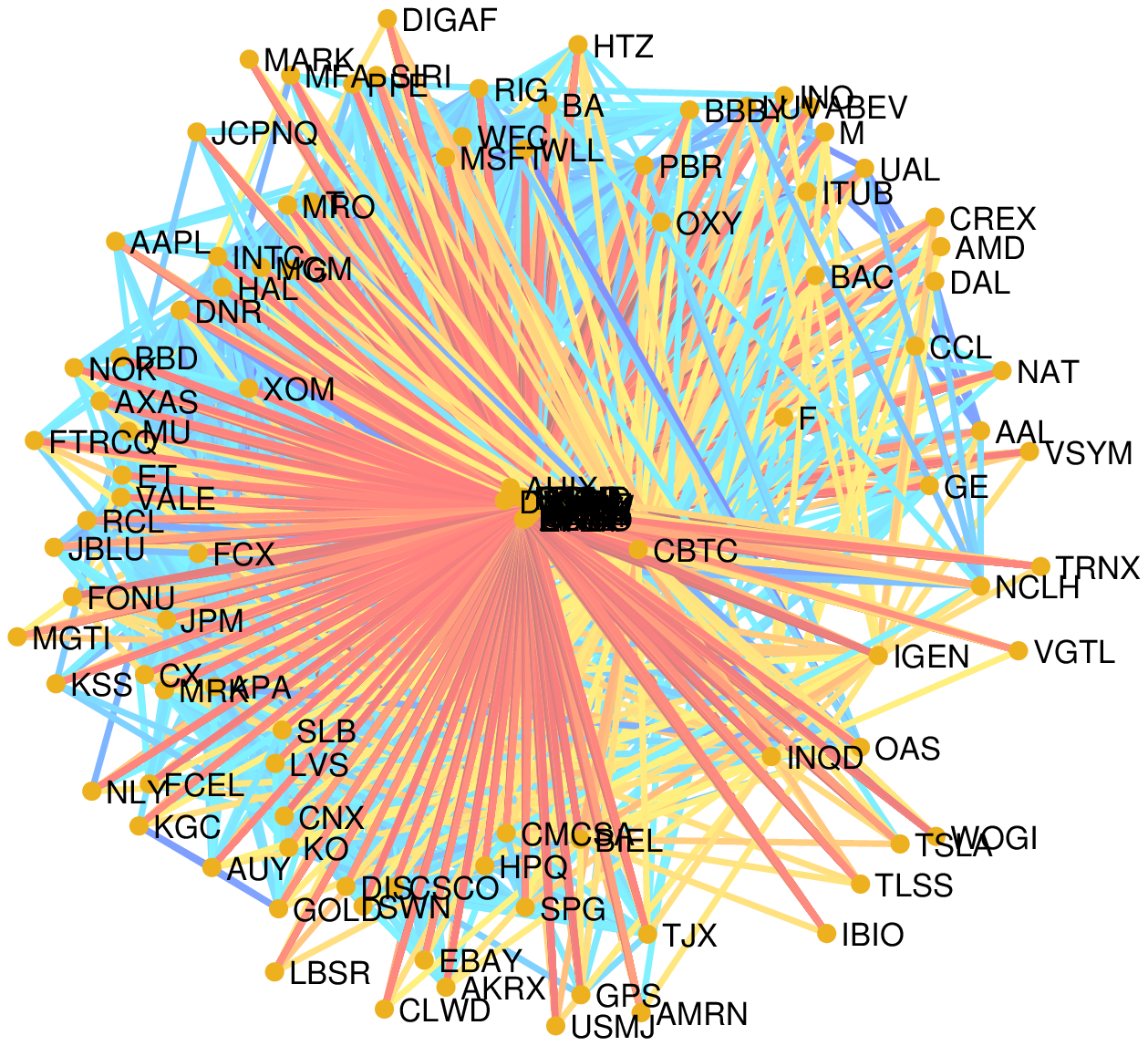} & \includegraphics[trim=0.5cm 0.25cm 0cm 0.25cm,clip,width=0.12\linewidth]{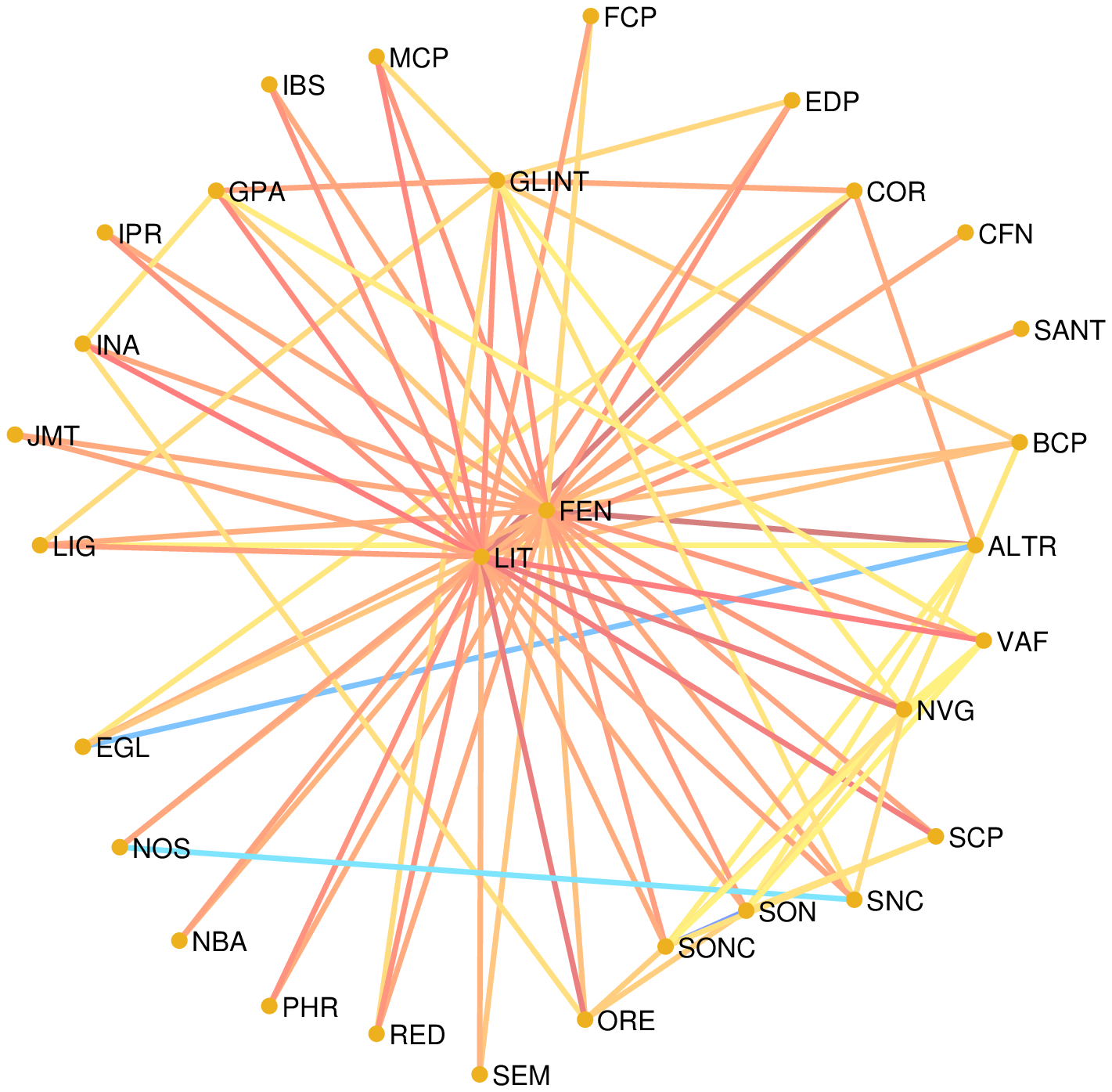} & \includegraphics[trim=0.5cm 0.25cm 0cm 0.25cm,clip,width=0.12\linewidth]{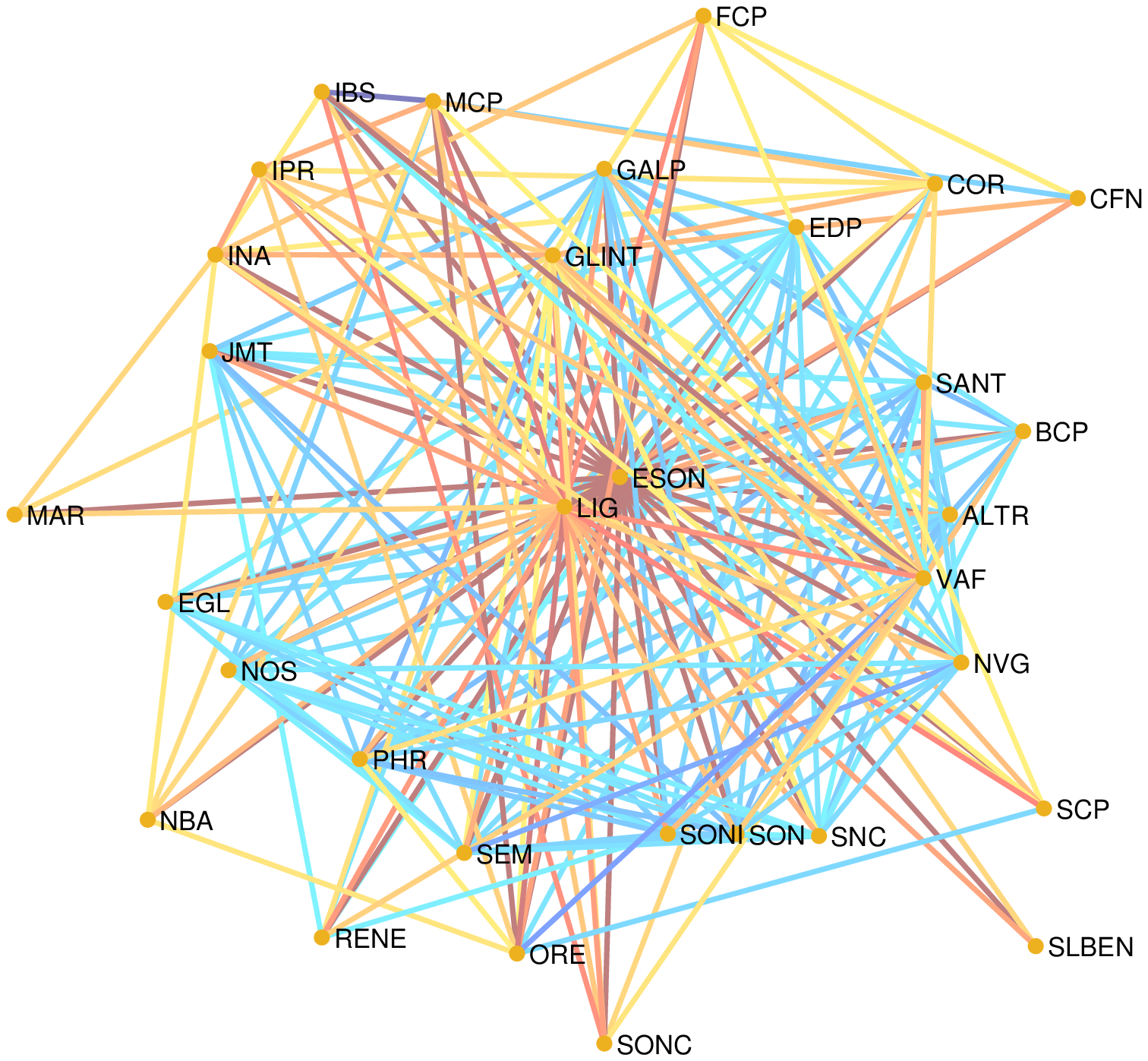} & \includegraphics[trim=0.5cm 0.25cm 0cm 0.25cm,clip,width=0.12\linewidth]{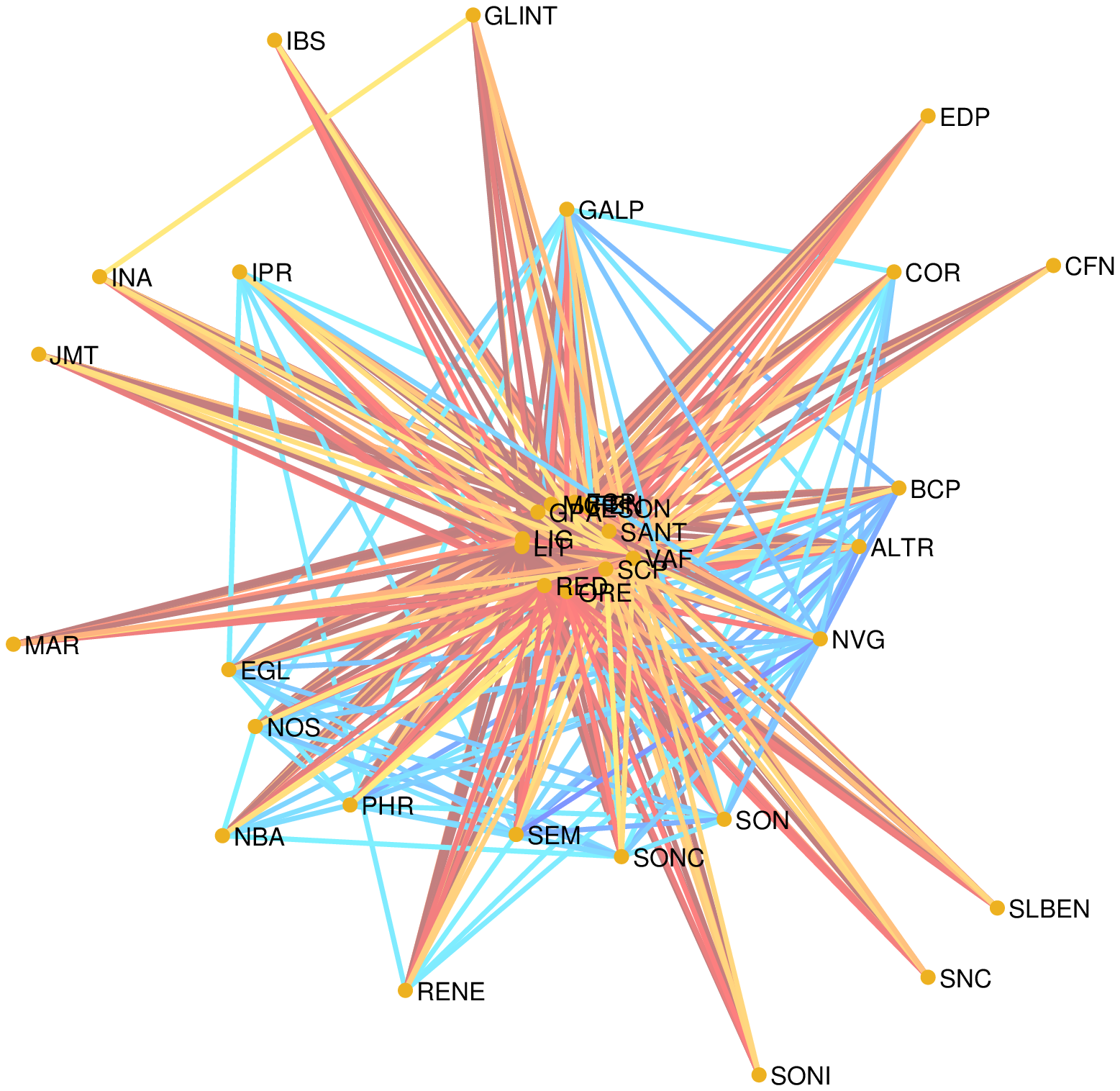}\tabularnewline\tabularnewline
\multicolumn{3}{c}{\textbf{(a)} the US} & \multicolumn{3}{c}{\textbf{(b)} Portugal}\tabularnewline\tabularnewline
\multicolumn{3}{c}{\includegraphics[trim=0cm 0cm 0cm 0cm,clip,width=0.4\linewidth]{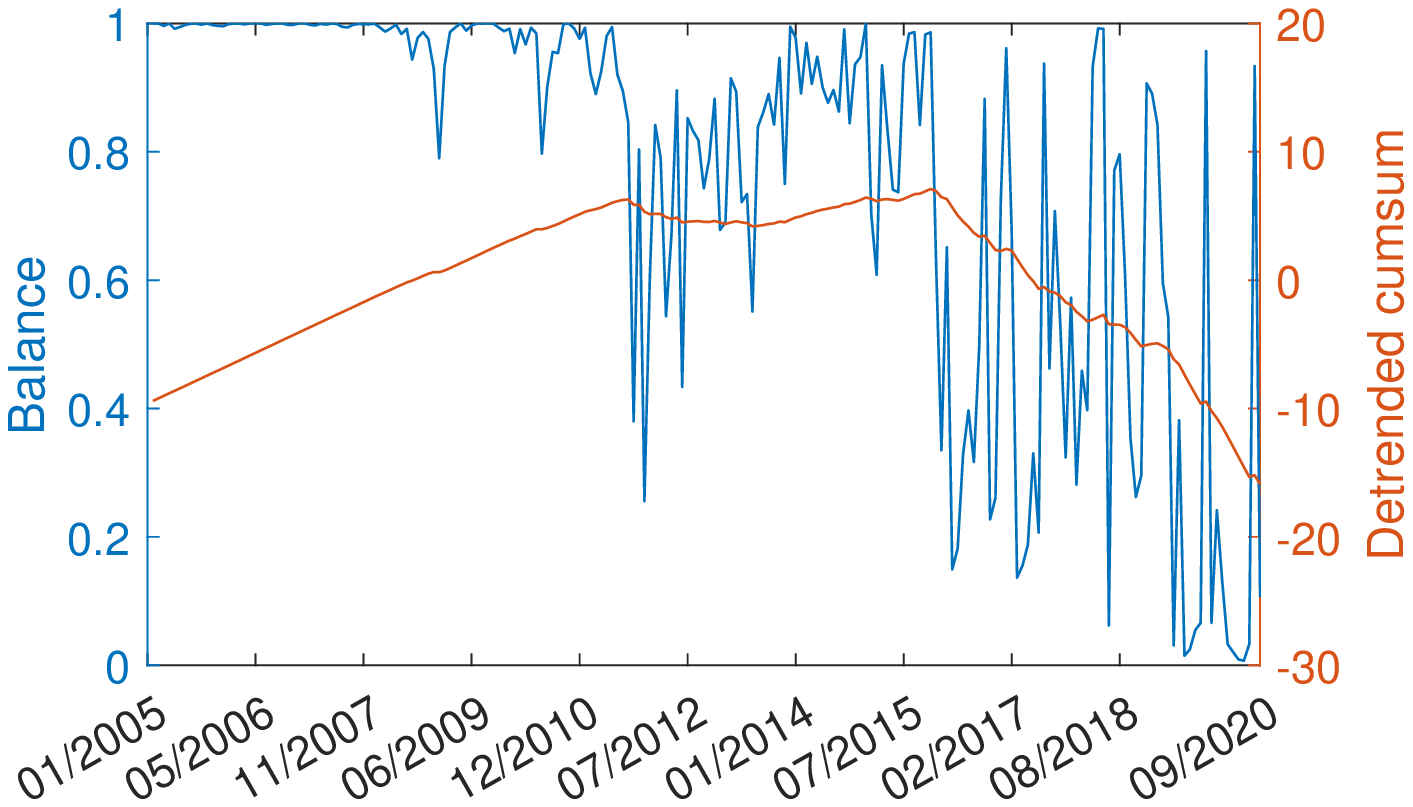}} & \multicolumn{3}{c}{\includegraphics[trim=0cm 0cm 0cm 0cm,clip,width=0.4\linewidth]{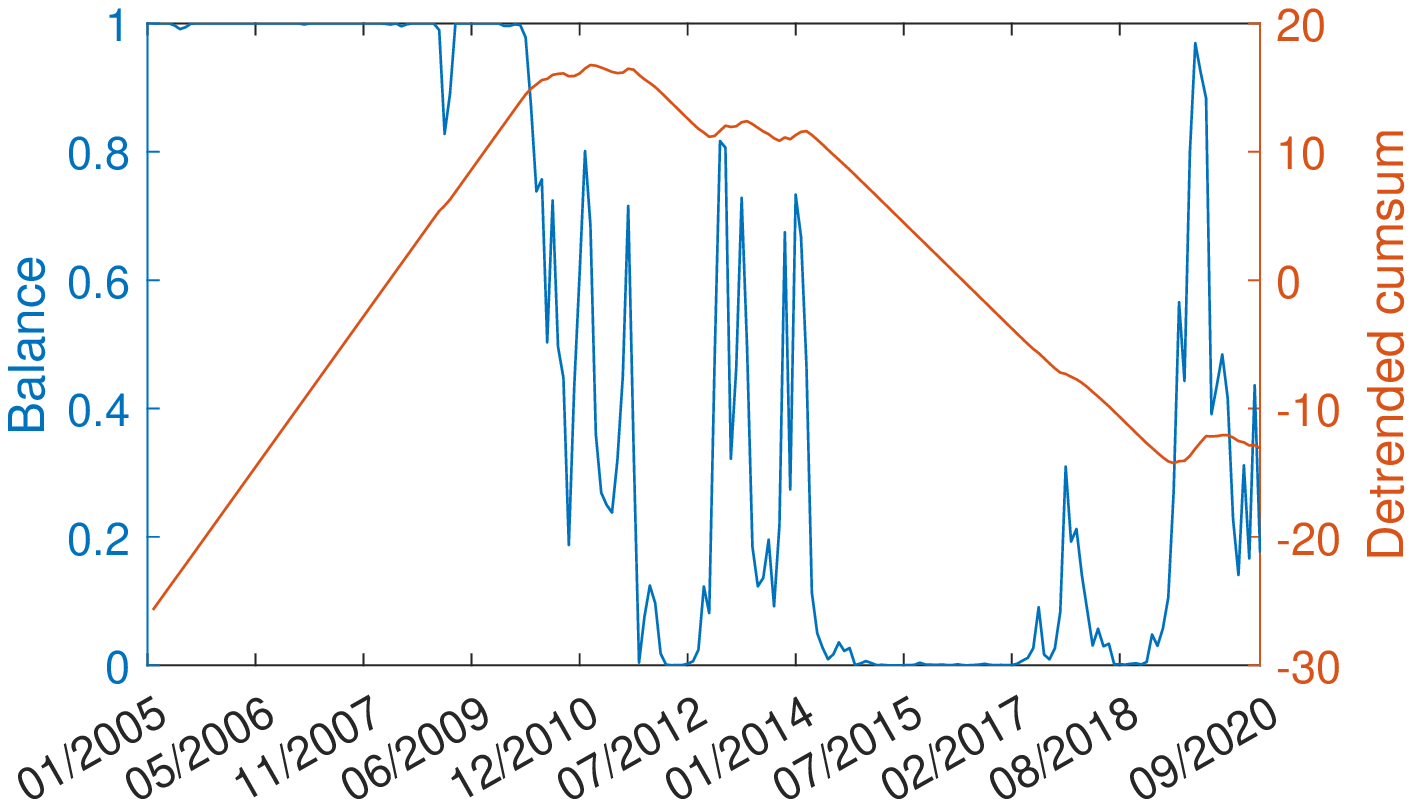}}\tabularnewline
\includegraphics[trim=0.5cm 0.25cm 0cm 0.25cm,clip,width=0.12\linewidth]{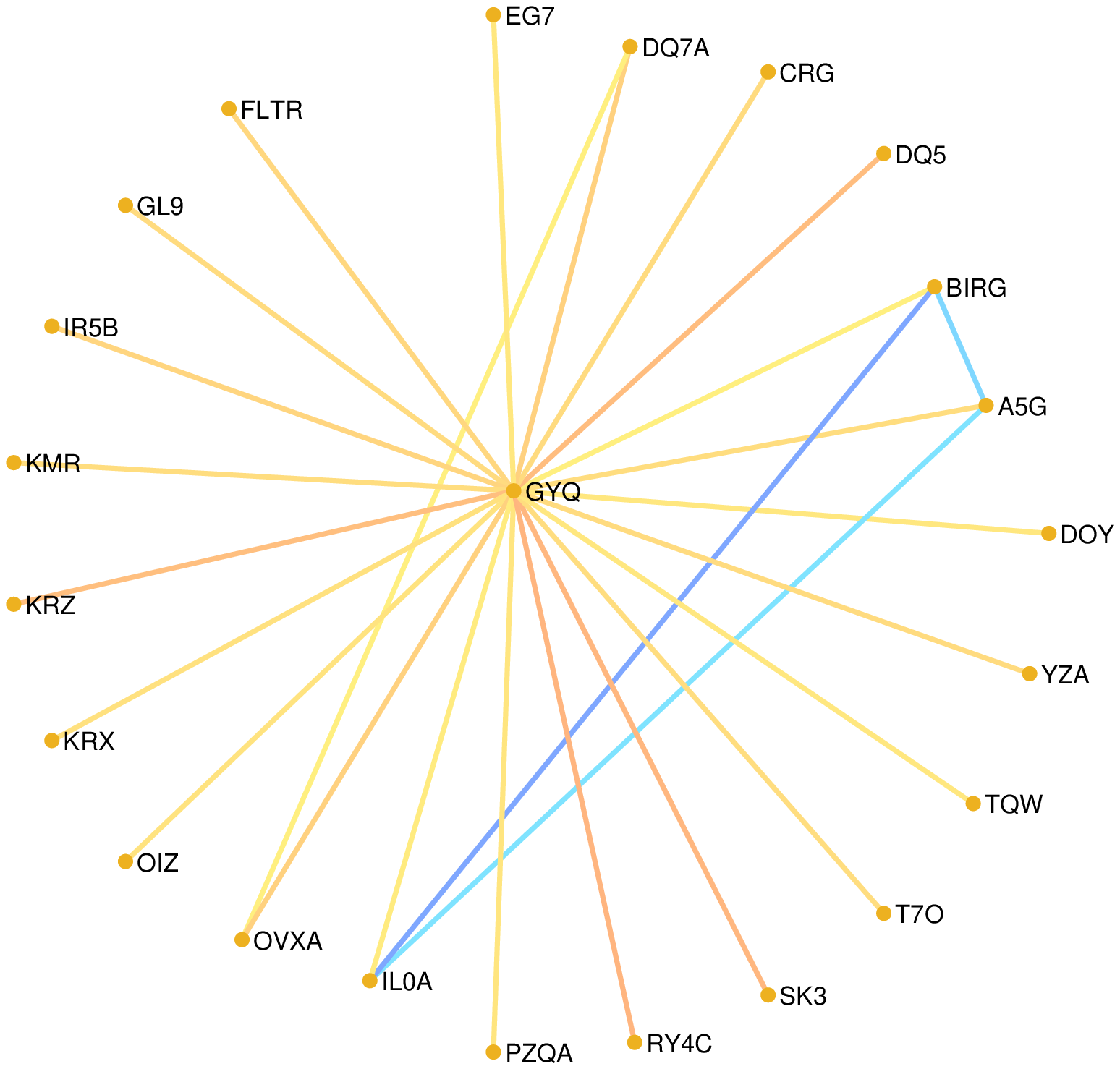} & \includegraphics[trim=0.5cm 0.25cm 0cm 0.25cm,clip,width=0.12\linewidth]{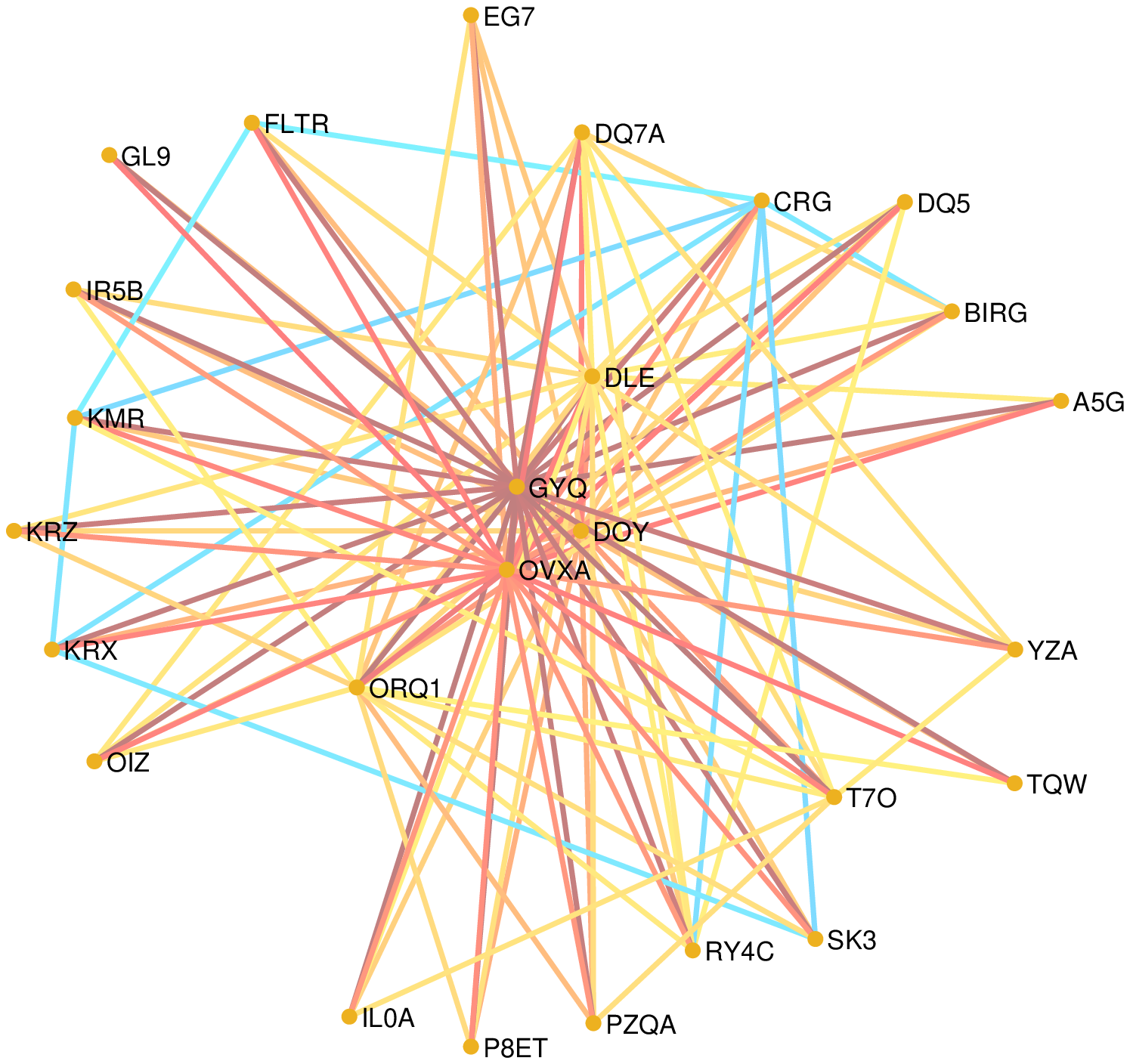} & \includegraphics[trim=0.5cm 0.25cm 0cm 0.25cm,clip,width=0.12\linewidth]{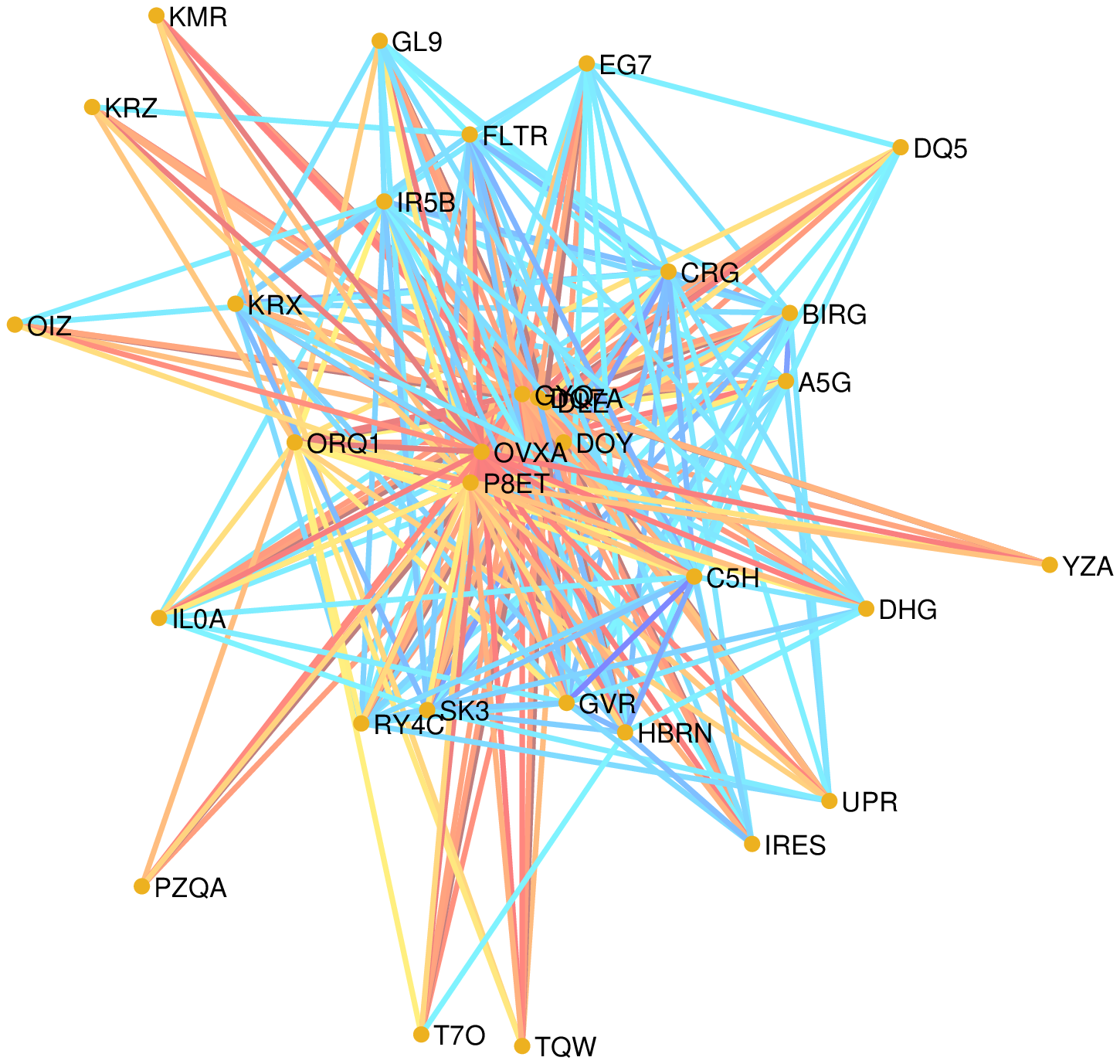} & \includegraphics[trim=0.5cm 0.25cm 0cm 0.25cm,clip,width=0.12\linewidth]{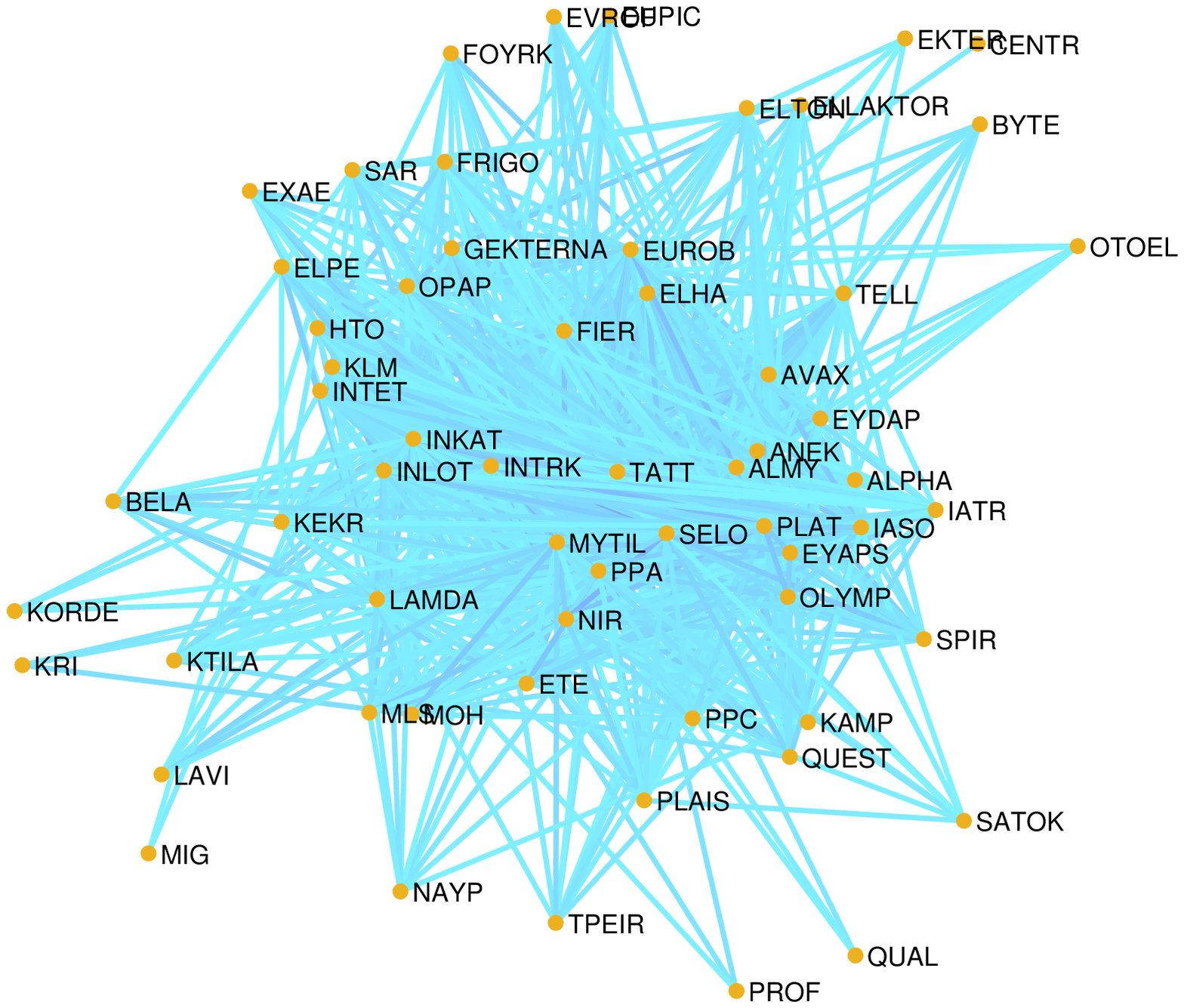} & \includegraphics[trim=0.5cm 0.25cm 0cm 0.25cm,clip,width=0.12\linewidth]{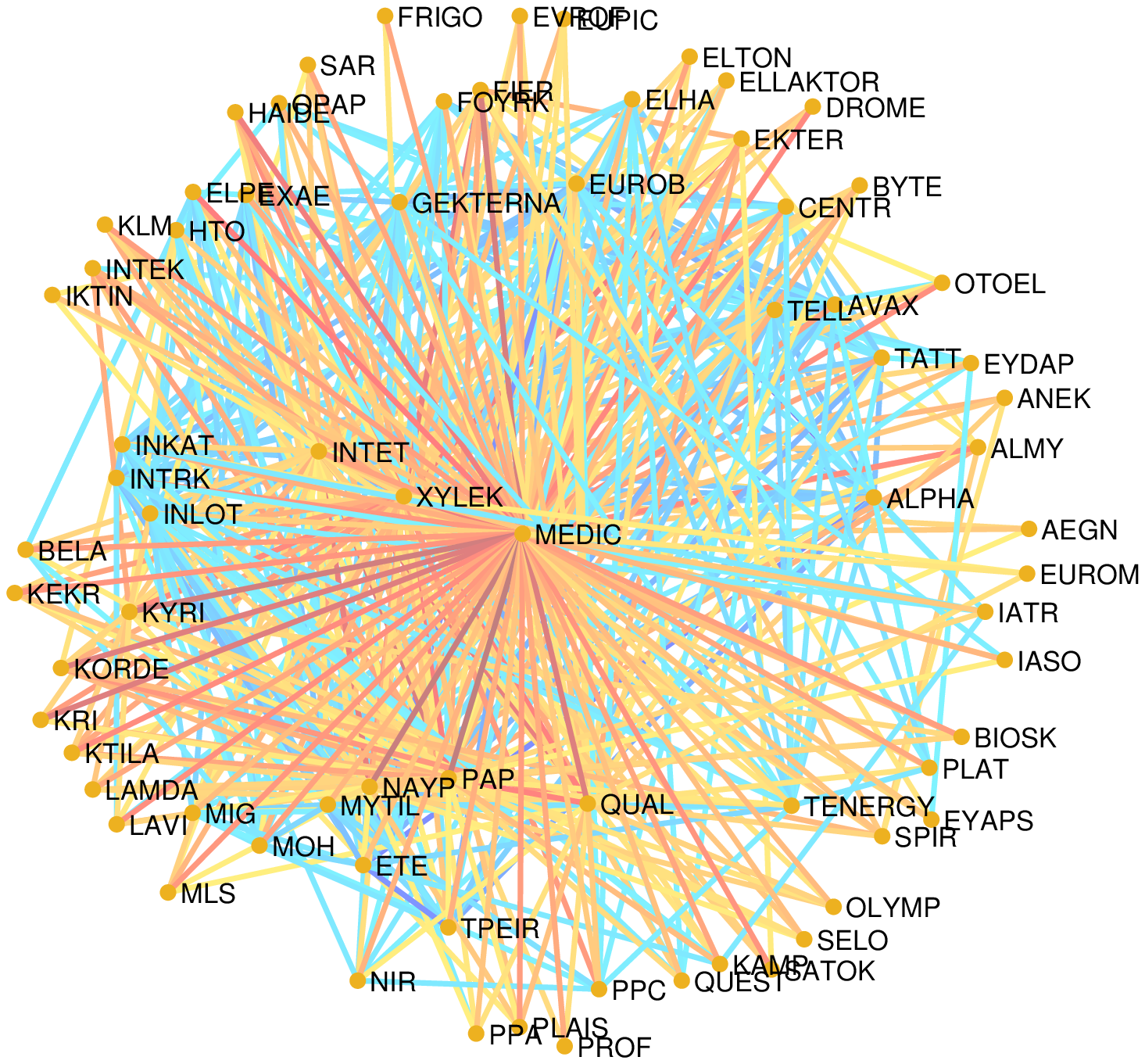} & \includegraphics[trim=0.5cm 0.25cm 0cm 0.25cm,clip,width=0.12\linewidth]{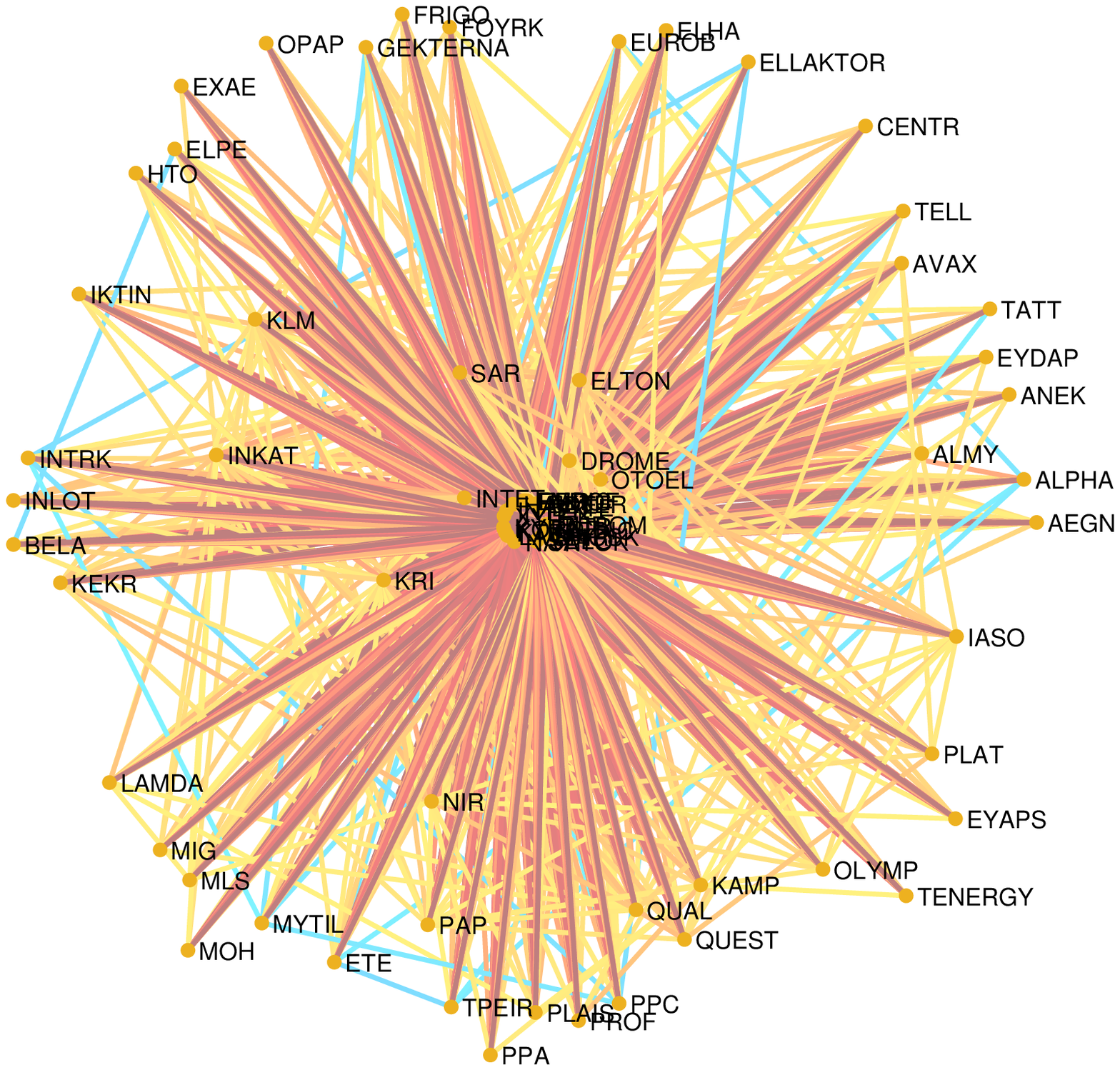}\tabularnewline\tabularnewline
\multicolumn{3}{c}{\textbf{(c)} Ireland} & \multicolumn{3}{c}{\textbf{(d)} Greece}\tabularnewline\tabularnewline
\multicolumn{3}{c}{\includegraphics[trim=0cm 0cm 0cm 0cm,clip,width=0.4\linewidth]{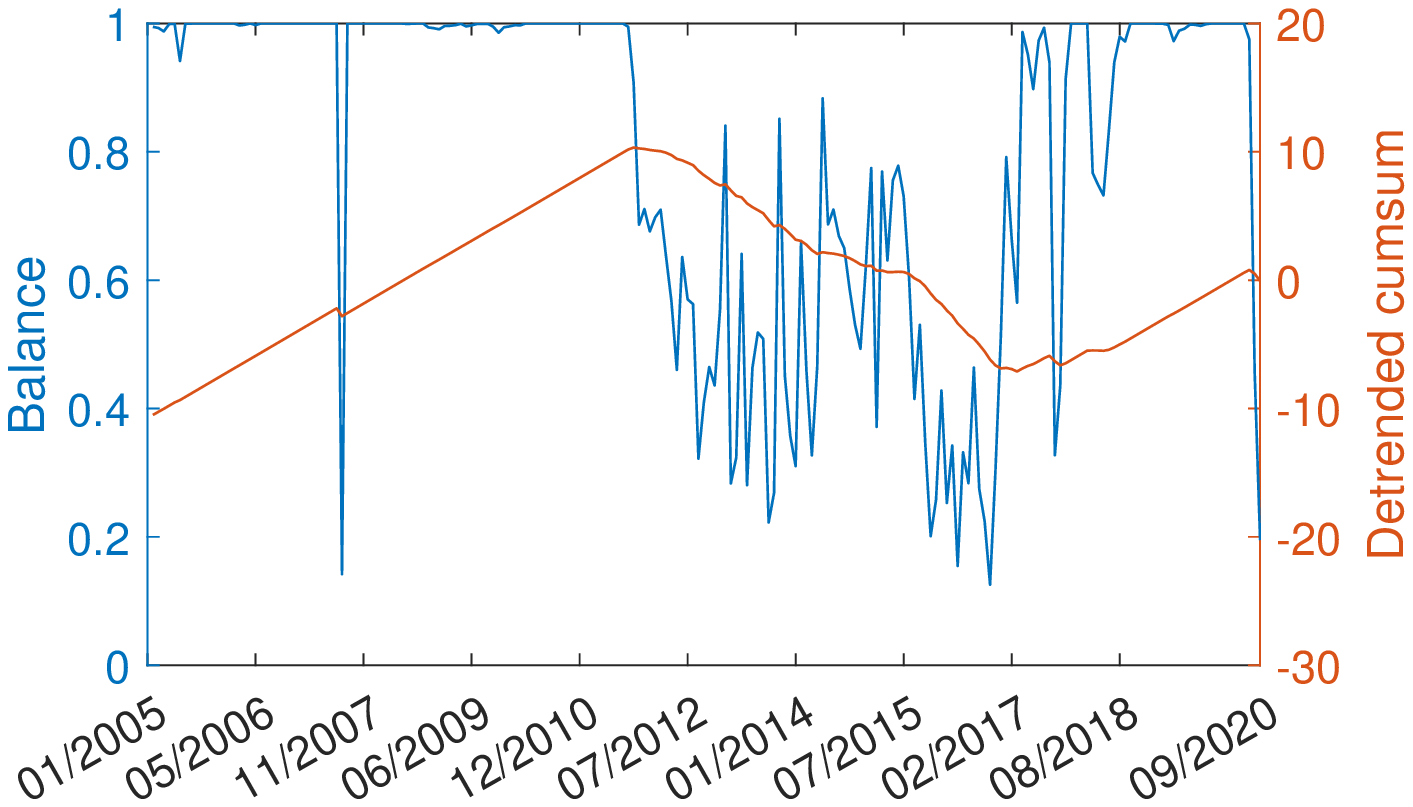}} & \multicolumn{3}{c}{\includegraphics[trim=0cm 0cm 0cm 0cm,clip,width=0.4\linewidth]{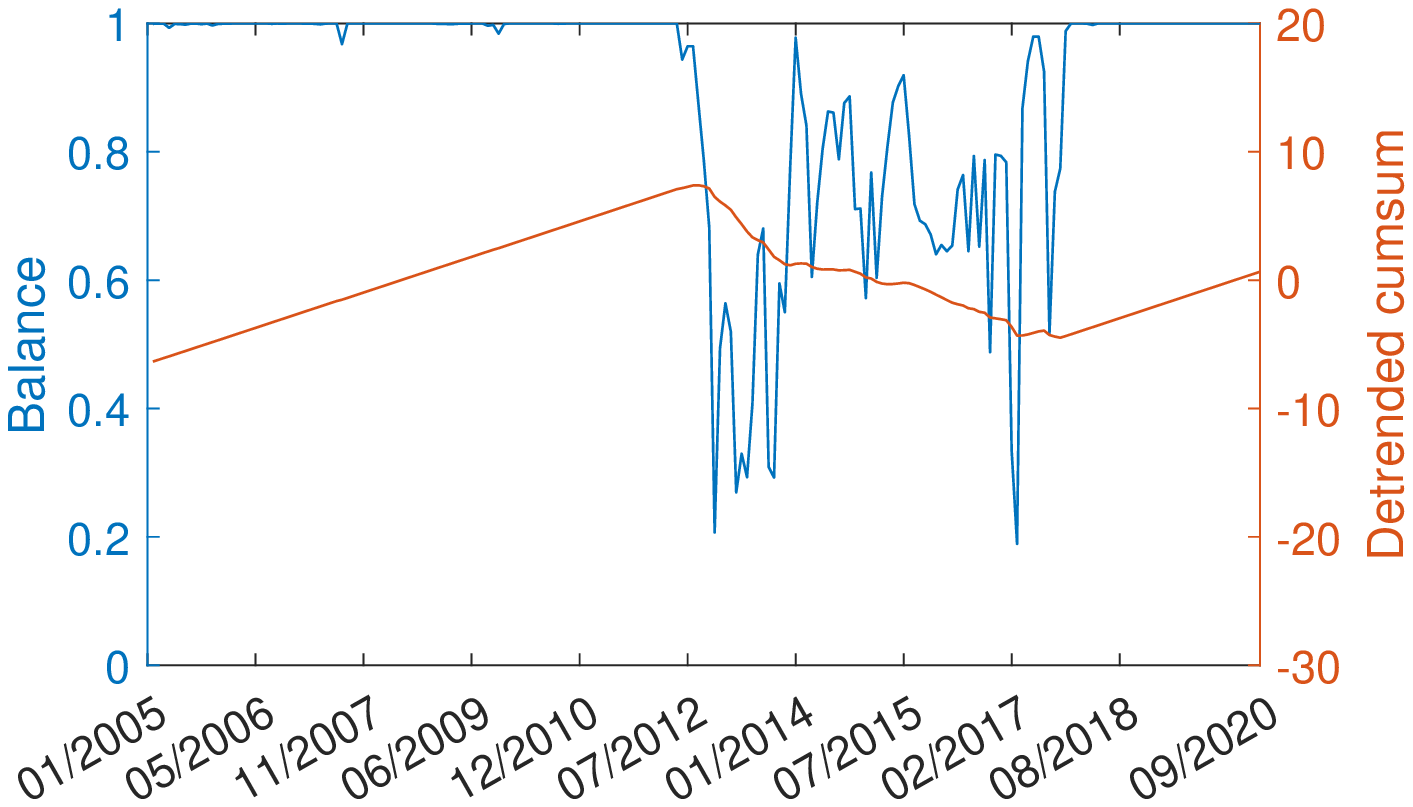}}\tabularnewline
\includegraphics[trim=0.5cm 0.25cm 0cm 0.25cm,clip,width=0.12\linewidth]{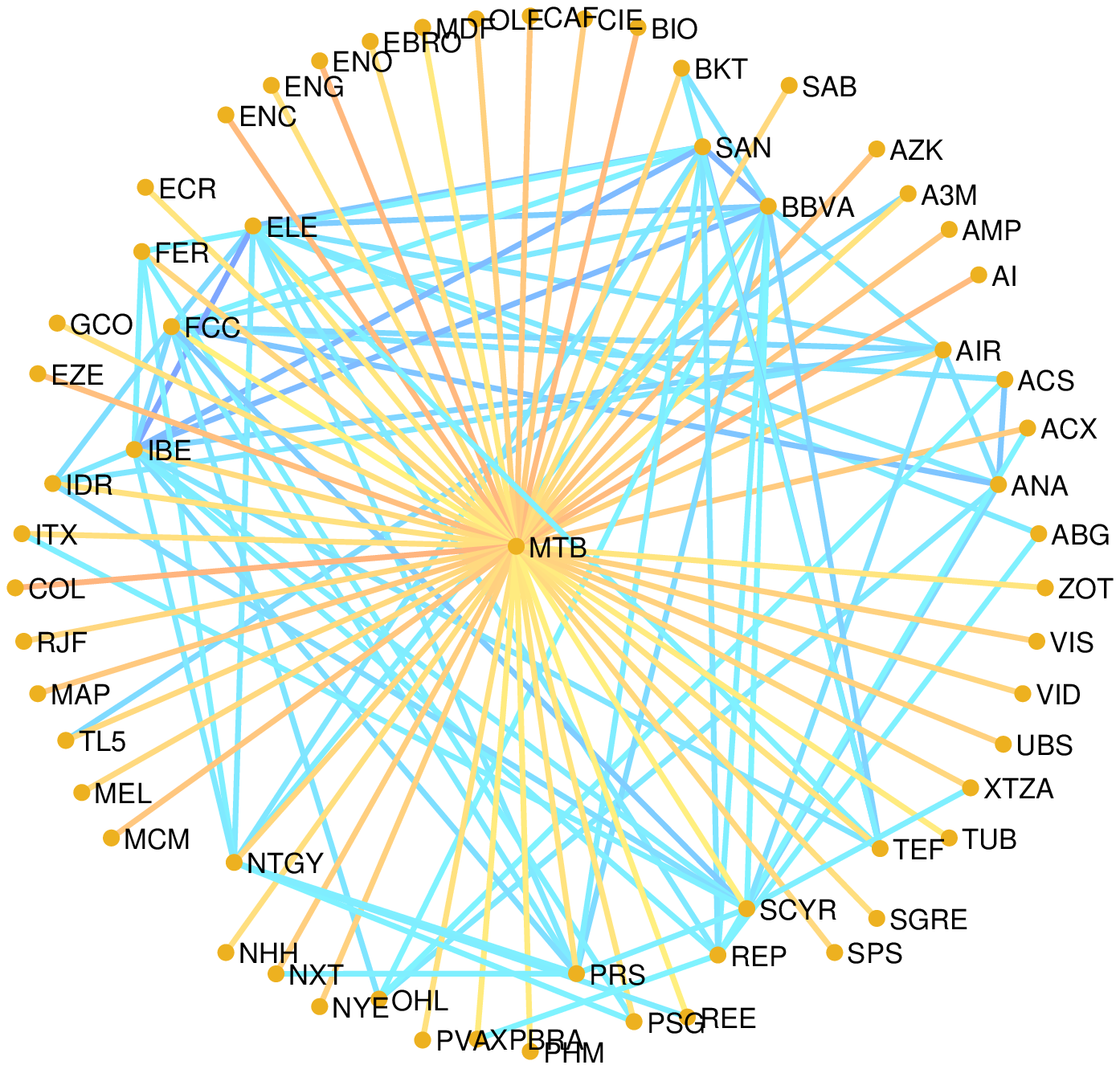} & \includegraphics[trim=0.5cm 0.25cm 0cm 0.25cm,clip,width=0.12\linewidth]{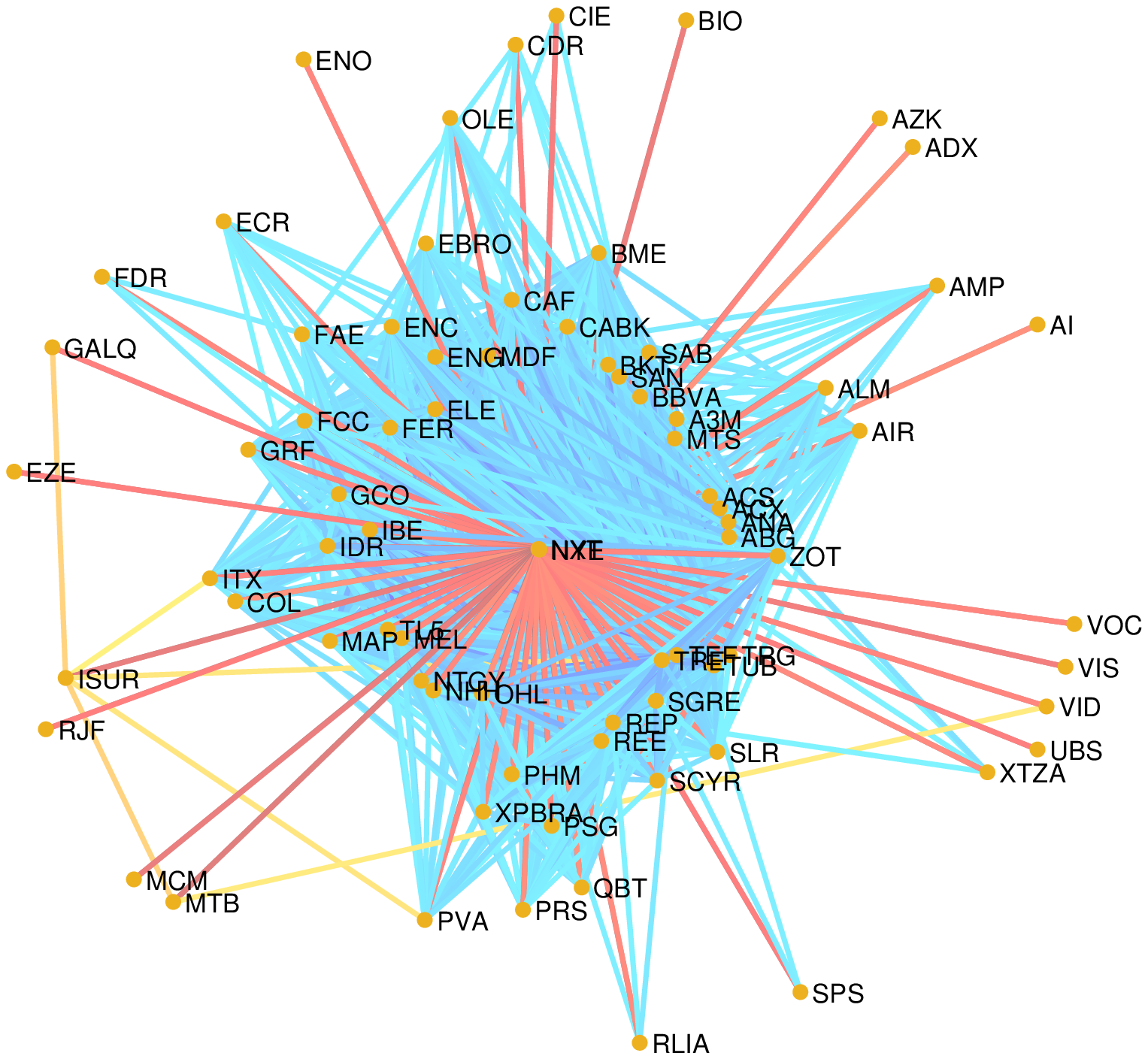} & \includegraphics[trim=0.5cm 0.25cm 0cm 0.25cm,clip,width=0.12\linewidth]{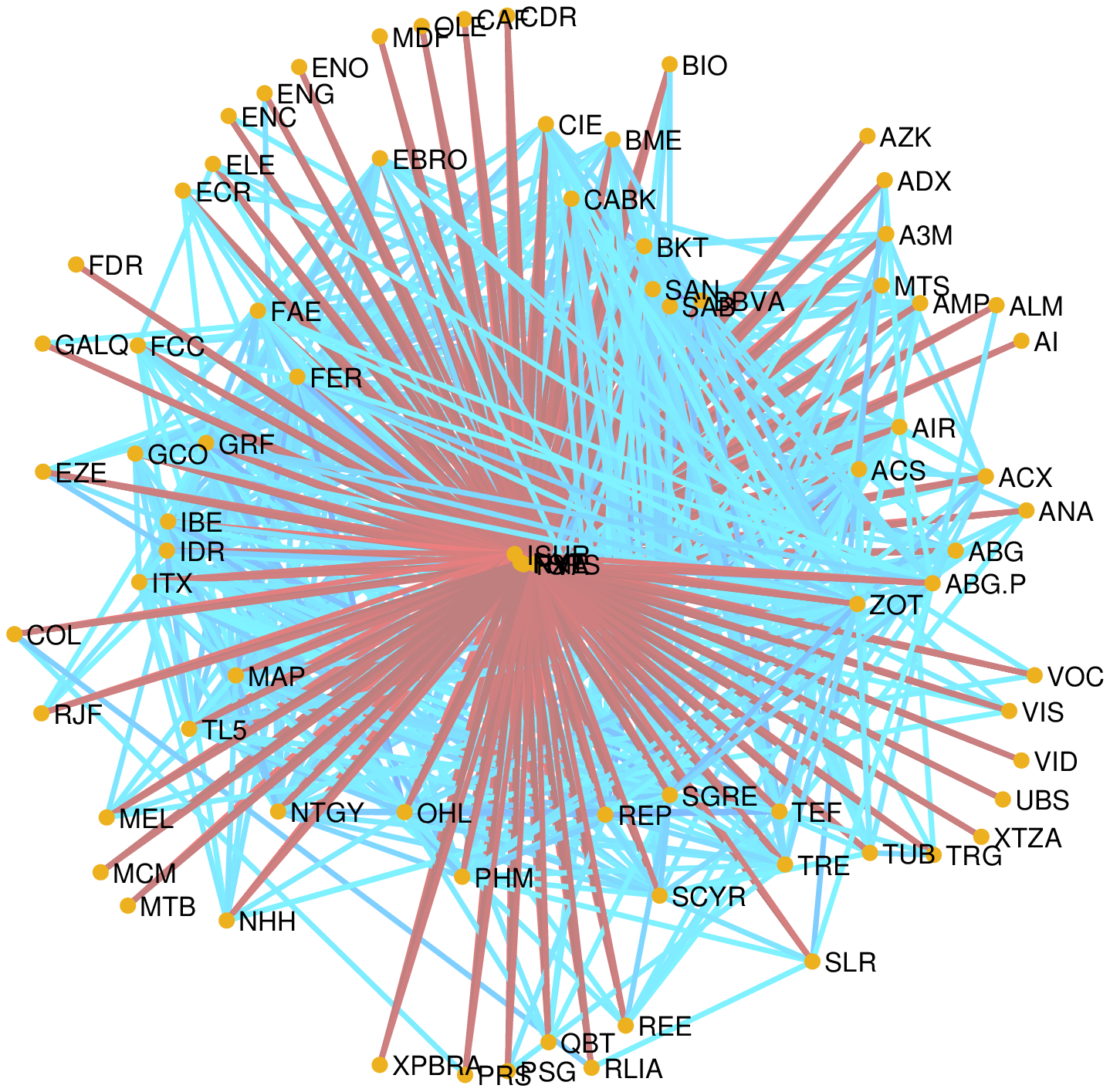} & \includegraphics[trim=0.5cm 0.25cm 0cm 0.25cm,clip,width=0.12\linewidth]{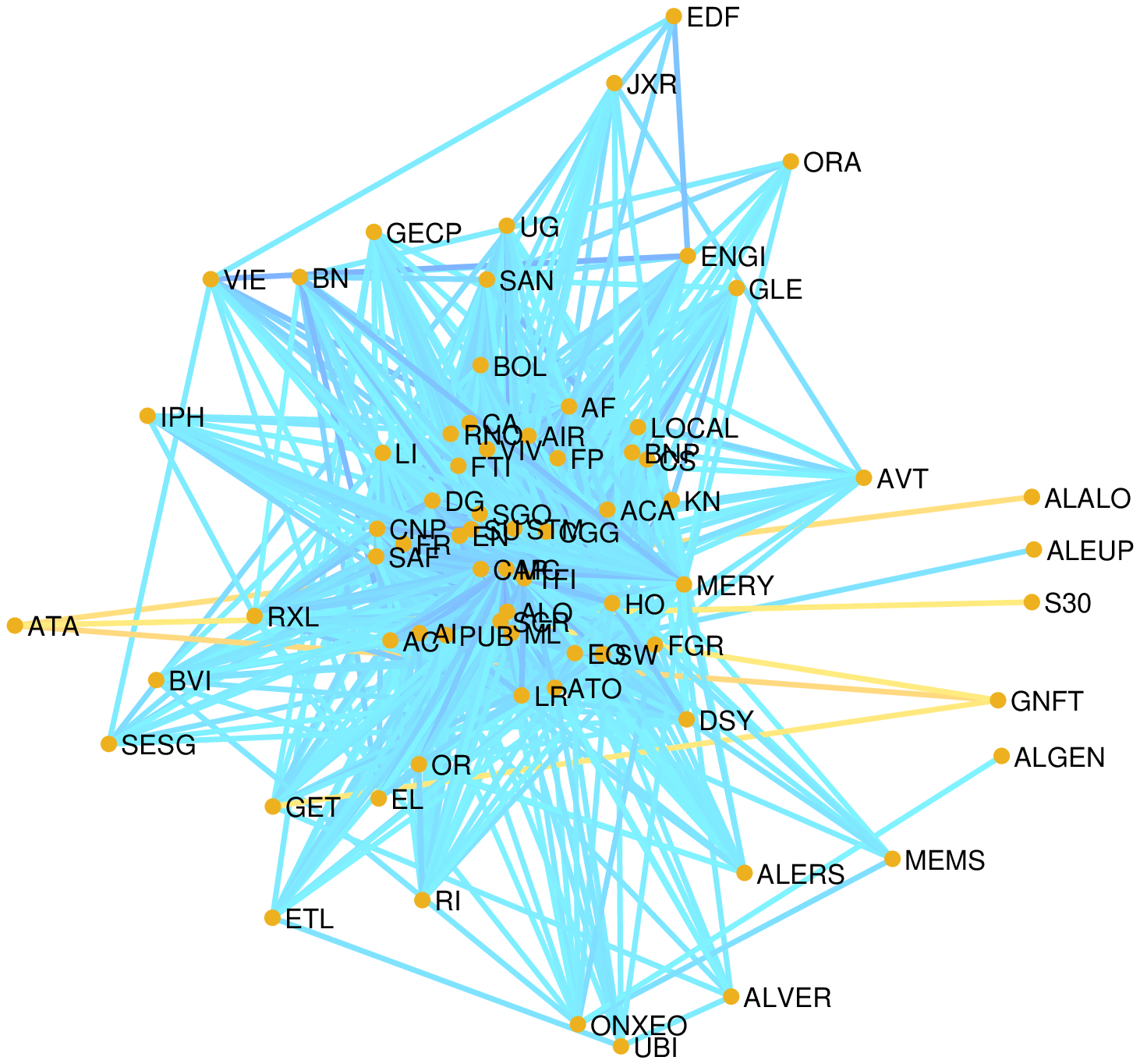} & \includegraphics[trim=0.5cm 0.25cm 0cm 0.25cm,clip,width=0.12\linewidth]{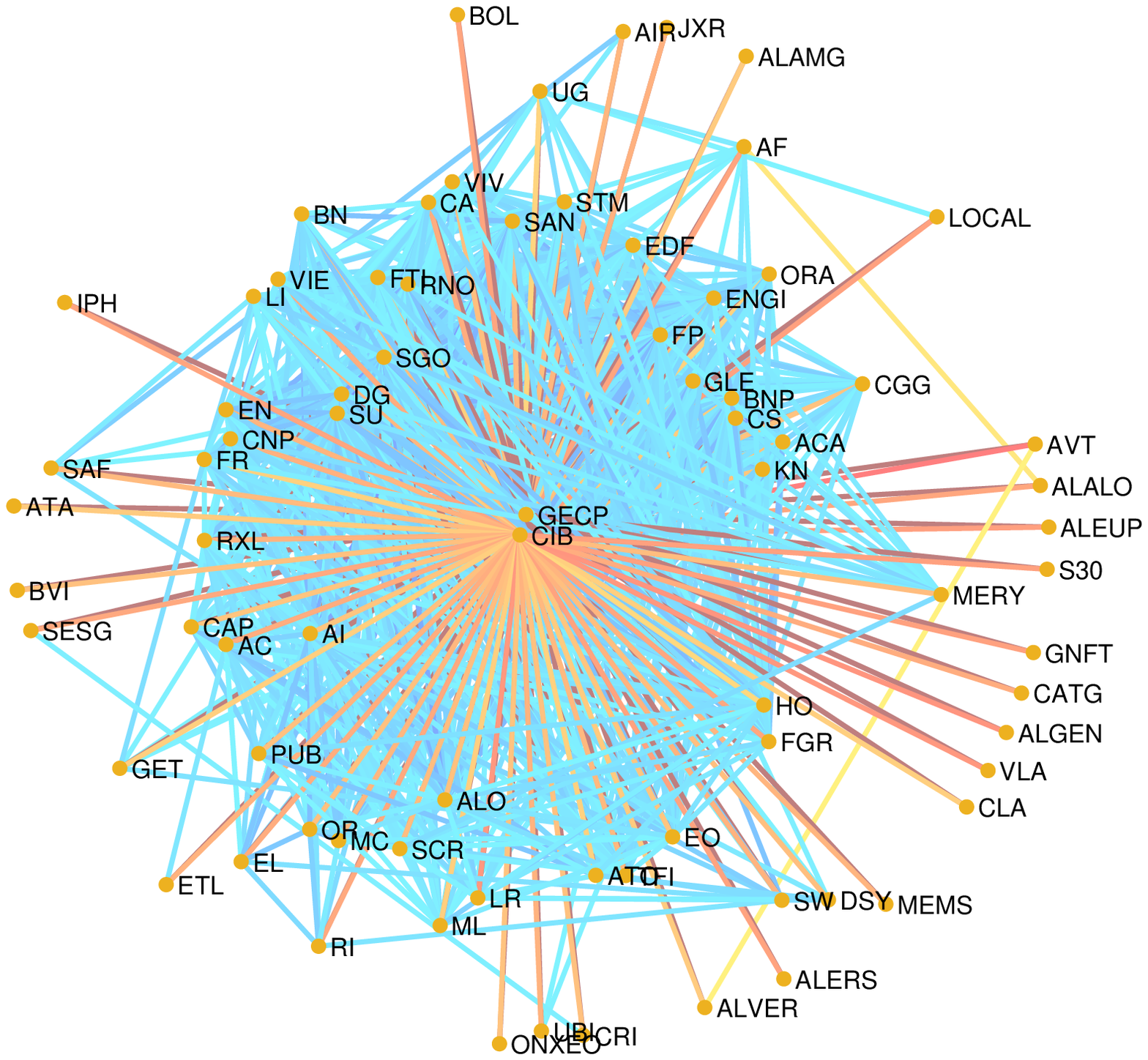} & \includegraphics[trim=0.5cm 0.25cm 0cm 0.25cm,clip,width=0.12\linewidth]{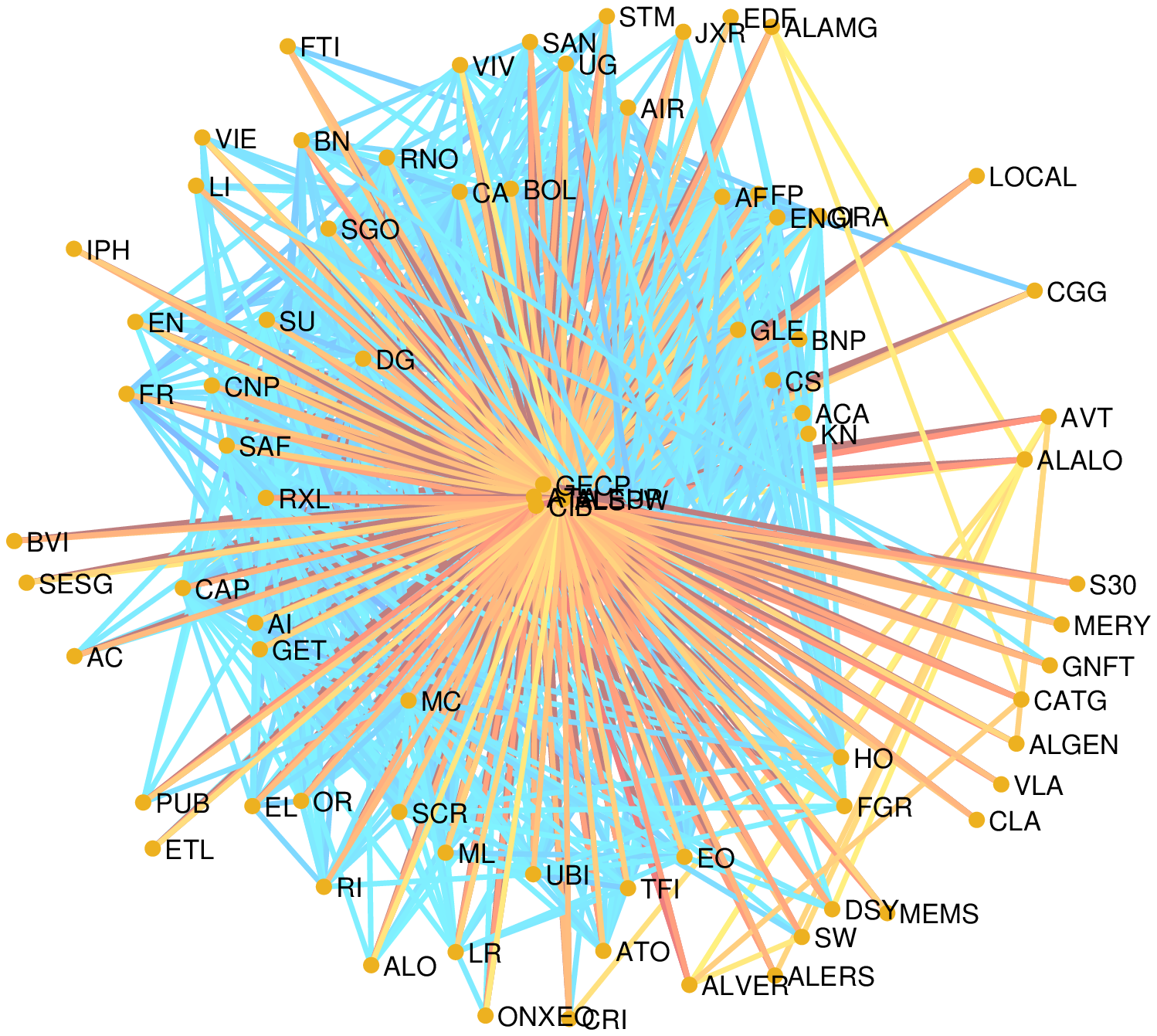}\tabularnewline
\multicolumn{3}{c}{\textbf{(e)} Spain} & \multicolumn{3}{c}{\textbf{(f)} France}\tabularnewline
\end{tabular}\caption{Balance degree evolution.}
\label{Transition_countries}
\end{figure}

We study the stock markets of France, Germany, Greece, Italy, Ireland,
Japan, Portugal, Spain, and the US during the period \textcolor{black}{between
January 2005 and September 2020 (for more details about the data  see SI, Appendix D). The first interesting observation
is the large balance observed for all markets between January 2005
and August 2007, when all of them display $K\geq0.98$. However, in
five countries there are sudden drops of balance around September/October
2011. This transition from highly balanced to poorly balanced markets
occurs }in Ireland in September 2011 and in the US, Portugal, Greece
and Spain in October 2011. The transitions are observed by naked eye
in the plots of the temporal evolution of balance (see top panels
of Fig. \ref{Transition_countries}(a-e)). For a quantitative analysis
of these BUTs as well as the identification of the dates at which
they occurred we use the detrended cumulative sums (DCS) of the balance
(see Fig. \ref{Transition_countries}). In these five markets the
general trend is a decay of the balance from 2005 to 2020. Therefore,
the DCS increases for those periods in which the balance does not
follow the general trend, i.e., when it does not decay with time.
A negative slope of the DCS in some periods indicates that the balance
drops more abruptly during this period than the general decreasing
trend. Then, we can observe that DCS has a positive slope in all five
markets with BUT between January 2005 and September/October 2011.
At this point, the DCS changes its slope indicating an abrupt decay
of balance. We select the point in which this change of slope occurs
as the date marking the BUT. We should notice that there are some
differences in the behavior of the DCS in each specific market. The US,
Portugal and Greece display an initial increasing period followed
by a continuous decay one. Ireland displays a short period in which
DCS has slope slightly positive but close to zero between September
2011 and January 2016 when it definitively starts to decay. The market
in Spain interrupted the decay of its balance on April 2017 when it
starts to recover balance at a rate similar to the one of the period
between January 2005 to October 2011. The value of the balance drops
dramatically again after the crisis produced by COVID-19 as can be
seen in Fig. \ref{Transition_countries} (e). Finally, we have included
the stock market of France in Fig. \ref{Transition_countries} (f)
because it displays a behavior similar to that of Spain, although
the BUT occurs significantly later with respect to the other countries
in this group, i.e., on September 2012.

\begin{figure}[ht]
\centering %
\begin{tabular}{@{}cccccc@{}}
\multicolumn{3}{c}{\includegraphics[trim=0cm 0.1cm 0cm 0cm,clip,width=0.4\linewidth]{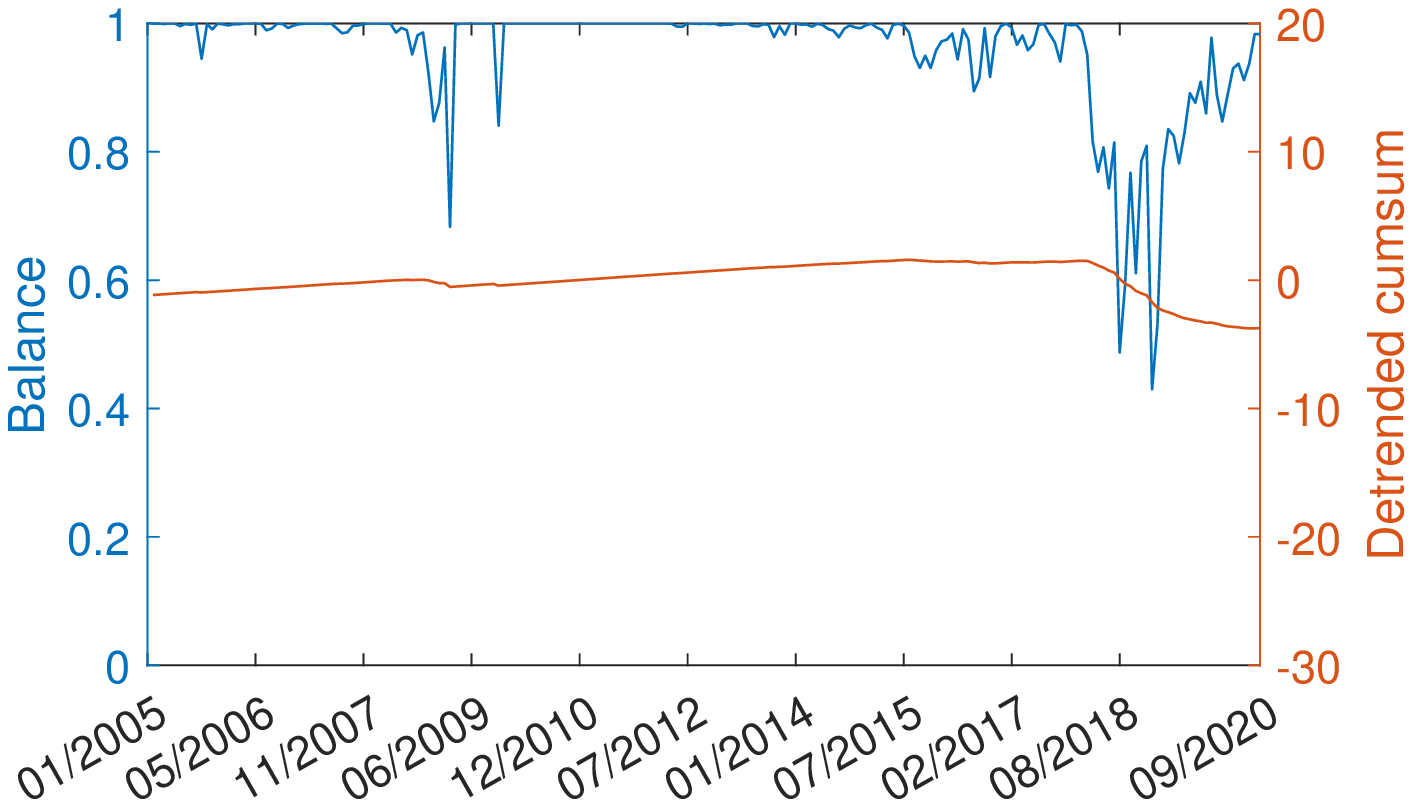}} & \multicolumn{3}{c}{\includegraphics[trim=0cm 0.1cm 0cm 0cm,clip,width=0.4\linewidth]{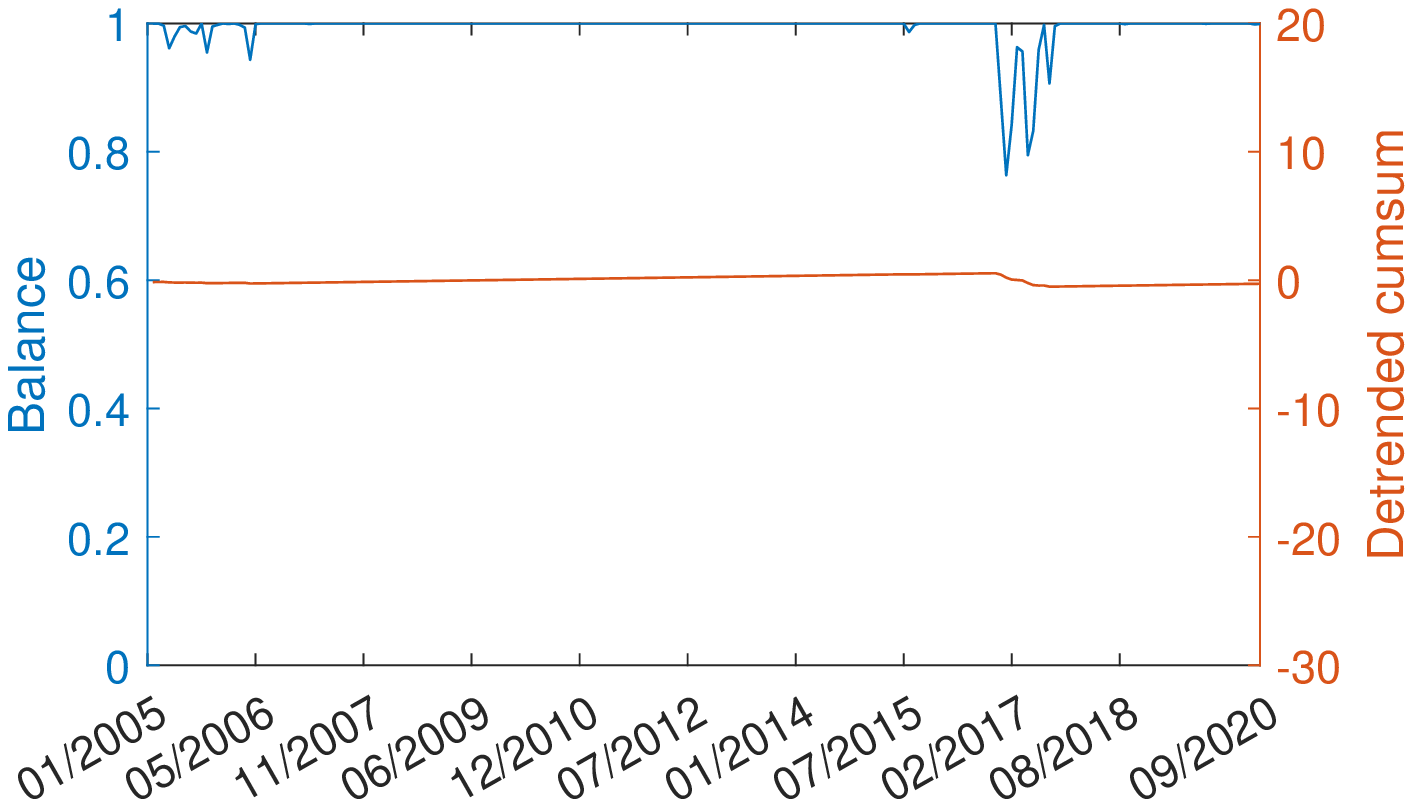}}\tabularnewline
 &  &  &  &  & \tabularnewline
\includegraphics[trim=0.5cm 0cm 0cm 0.25cm,clip,width=0.12\linewidth]{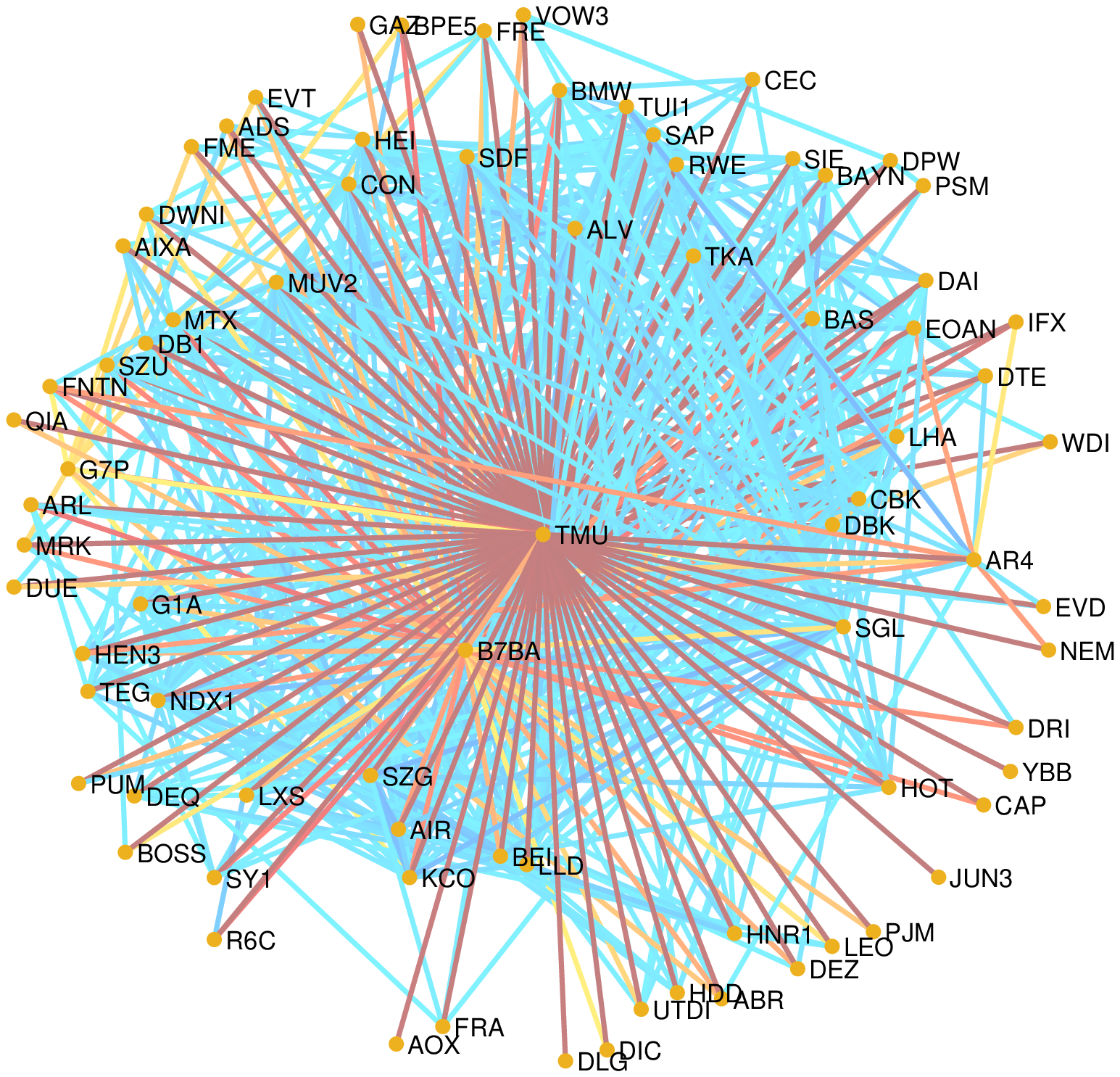} & \includegraphics[trim=0.5cm 0cm 0cm 0.25cm,clip,width=0.12\linewidth]{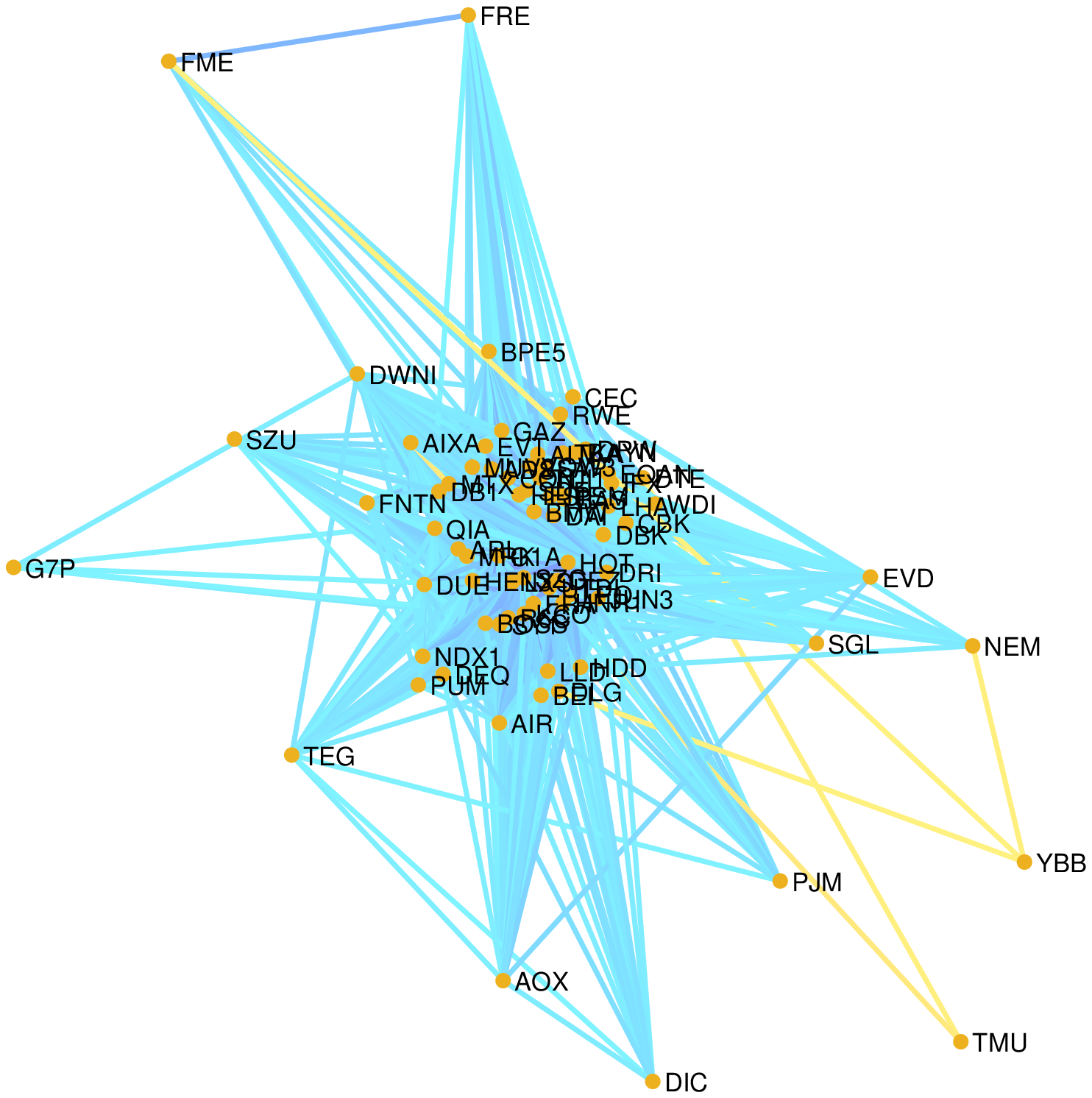} & \includegraphics[trim=0.5cm 0cm 0cm 0.25cm,clip,width=0.12\linewidth]{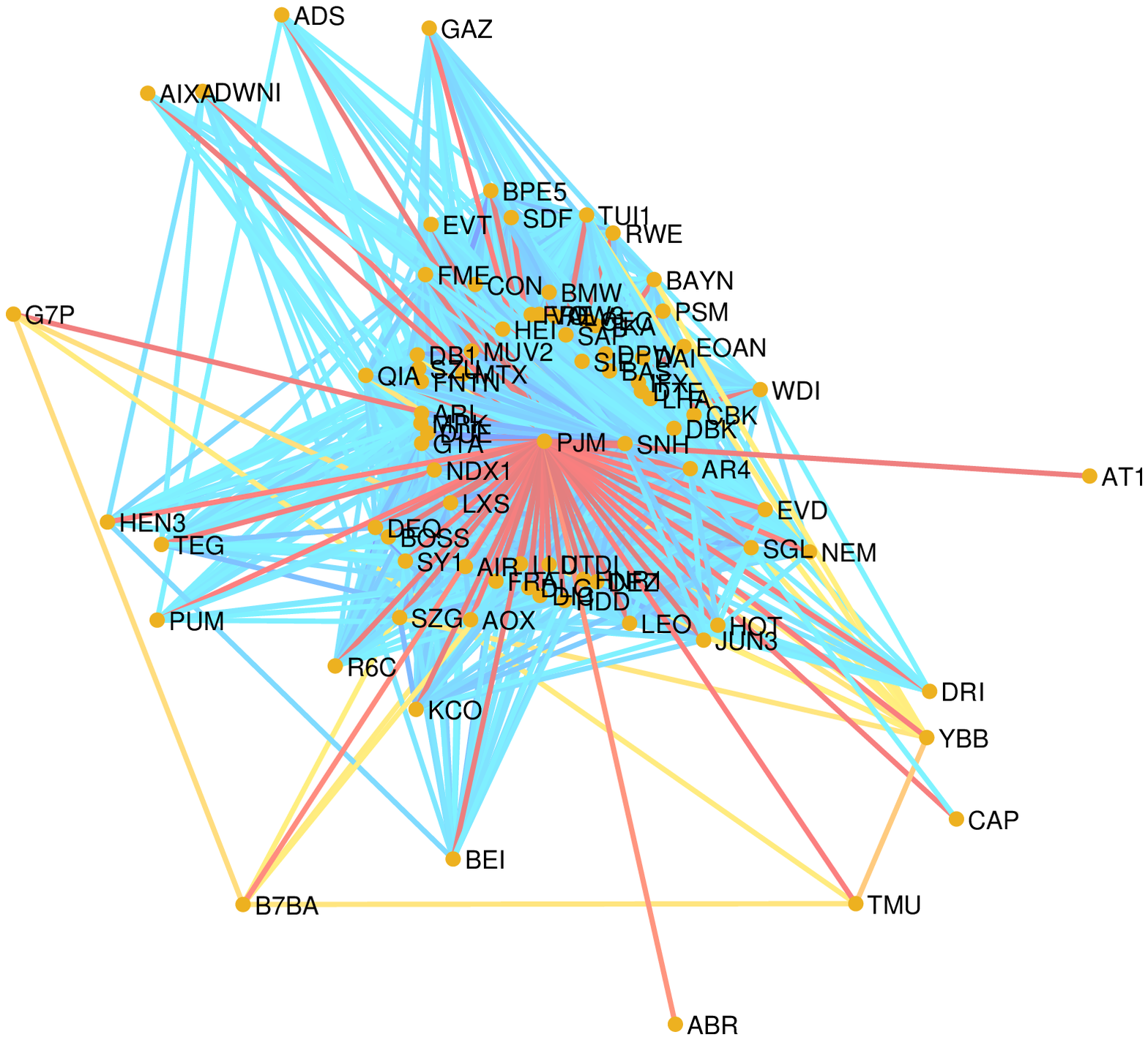} & \includegraphics[trim=0.5cm 0cm 0cm 0.25cm,clip,width=0.12\linewidth]{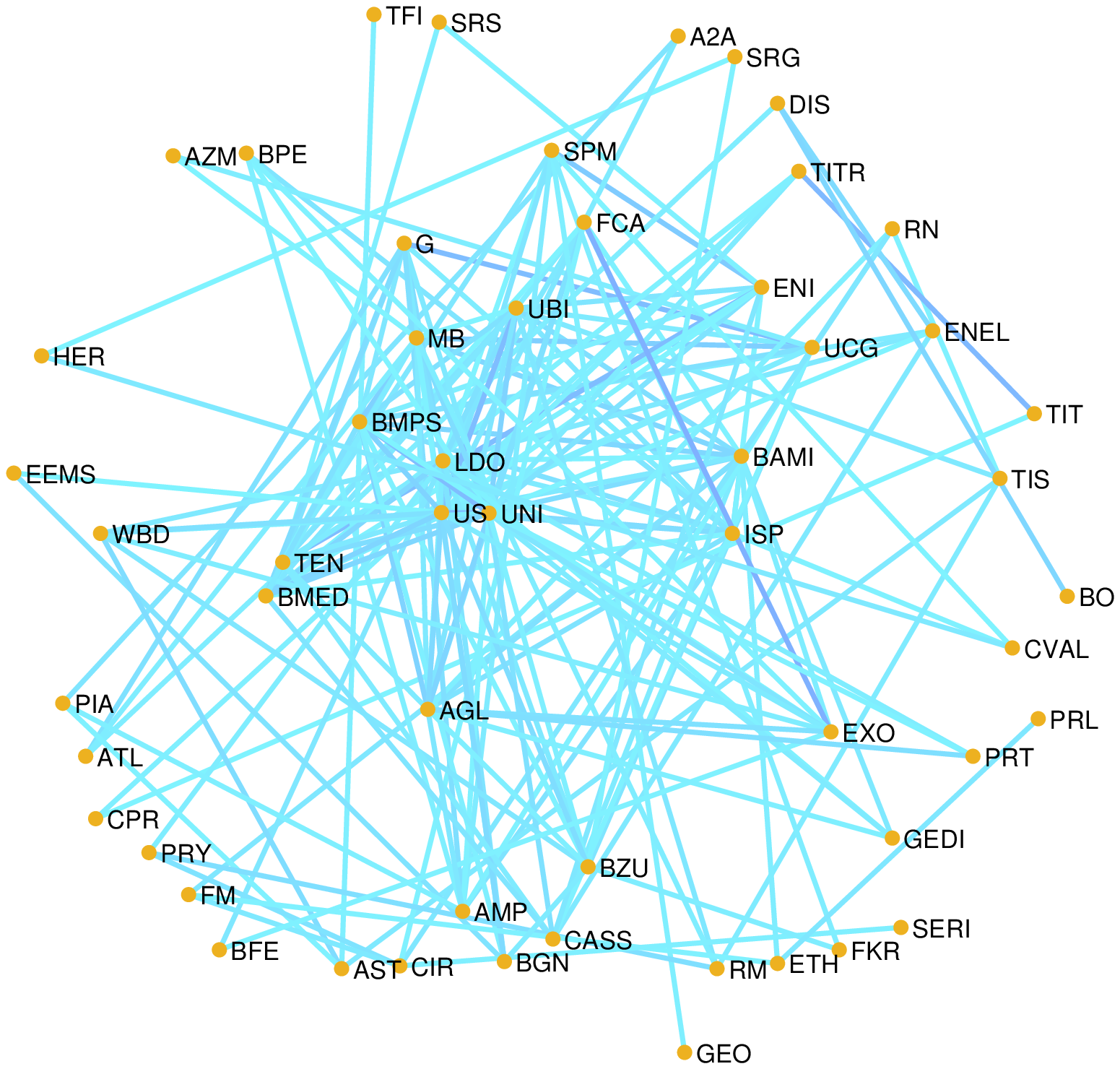} & \includegraphics[trim=0.5cm 0cm 0cm 0.25cm,clip,width=0.12\linewidth]{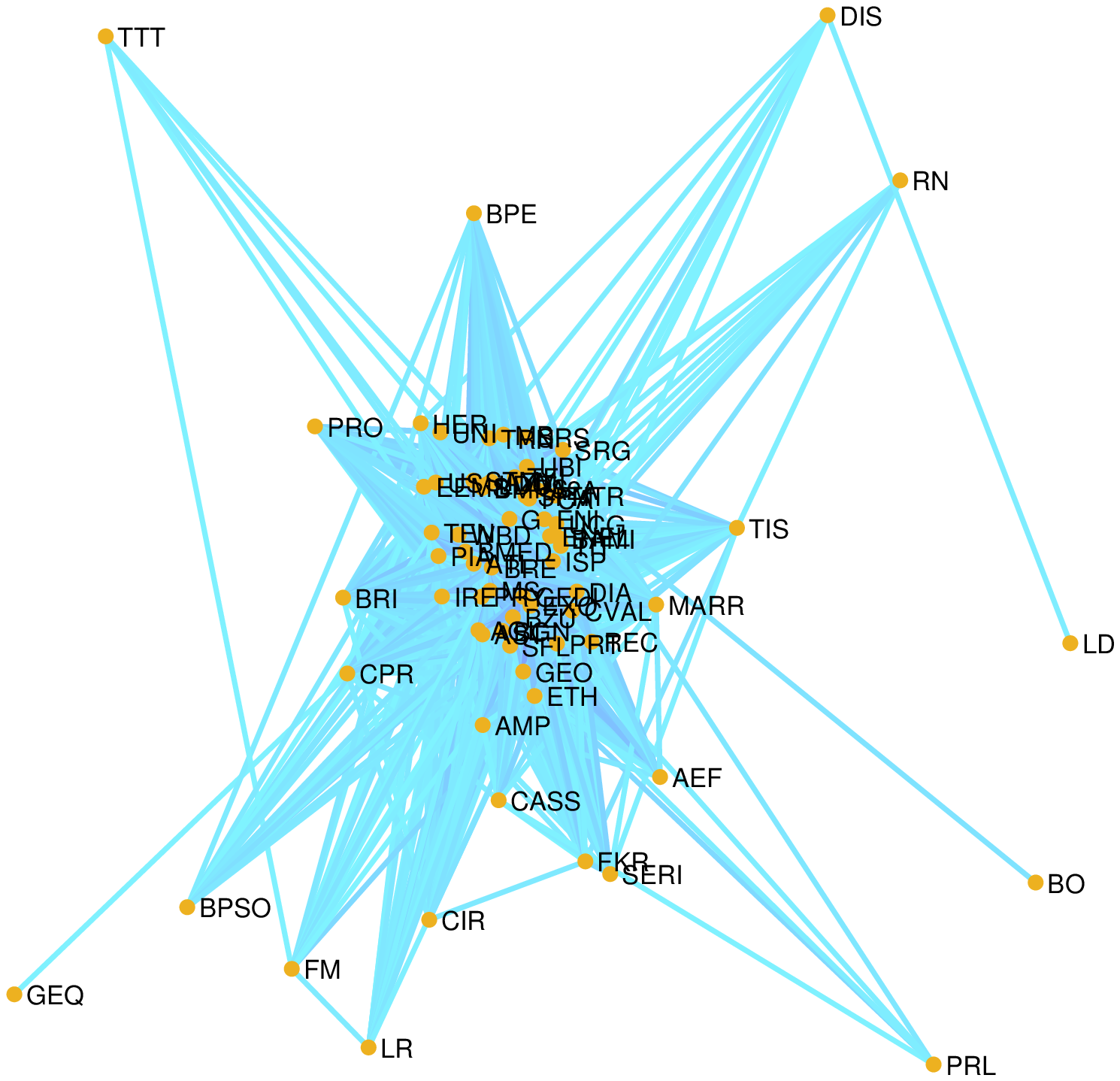} & \includegraphics[trim=0.5cm 0cm 0cm 0.25cm,clip,width=0.12\linewidth]{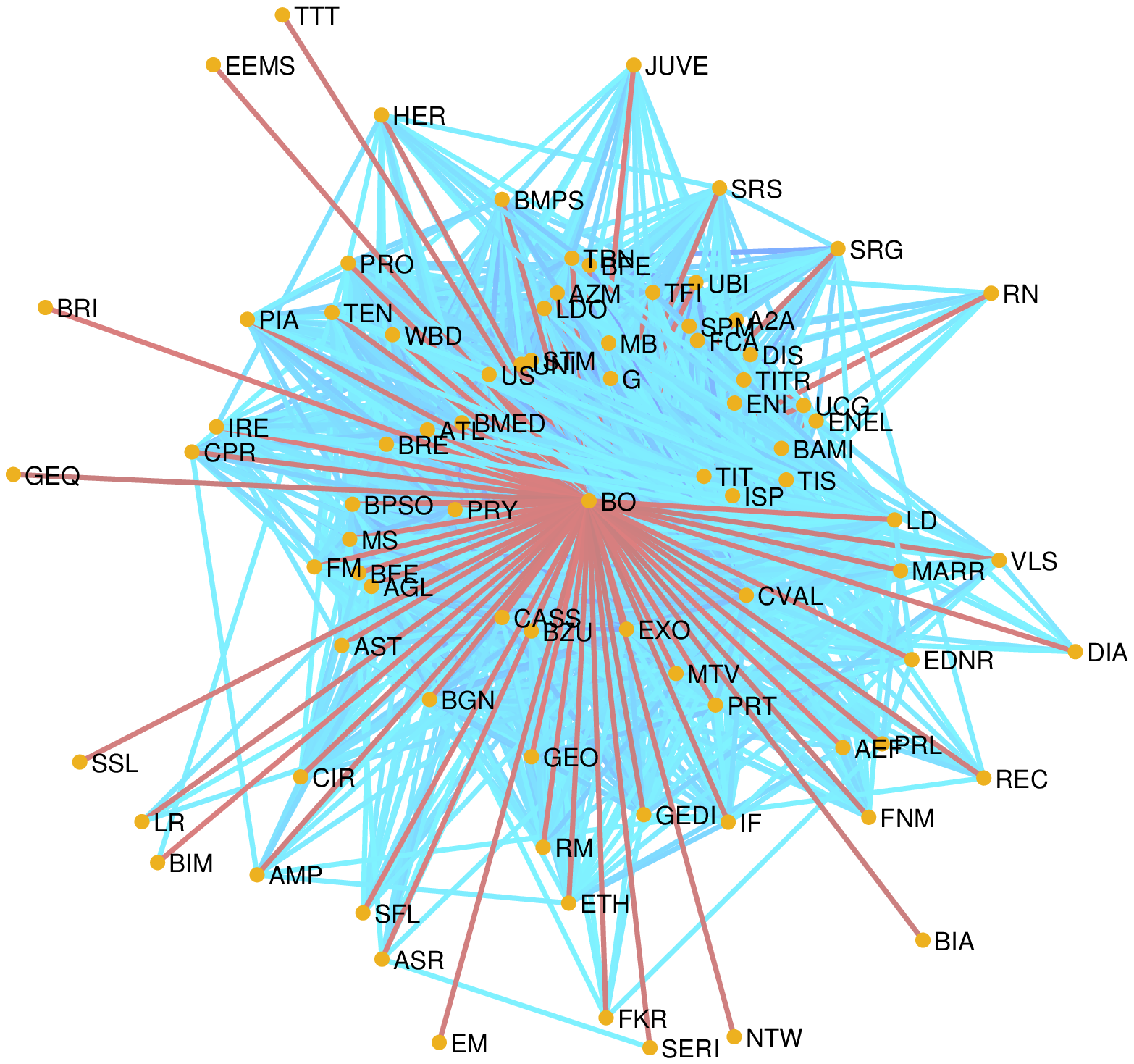}\tabularnewline
\multicolumn{3}{c}{\textbf{(a)} Germany} & \multicolumn{3}{c}{\textbf{(b)} Italy}\tabularnewline\tabularnewline\tabularnewline
\multicolumn{6}{c}{\includegraphics[trim=0cm 0cm 0cm 0cm,clip,width=0.4\linewidth]{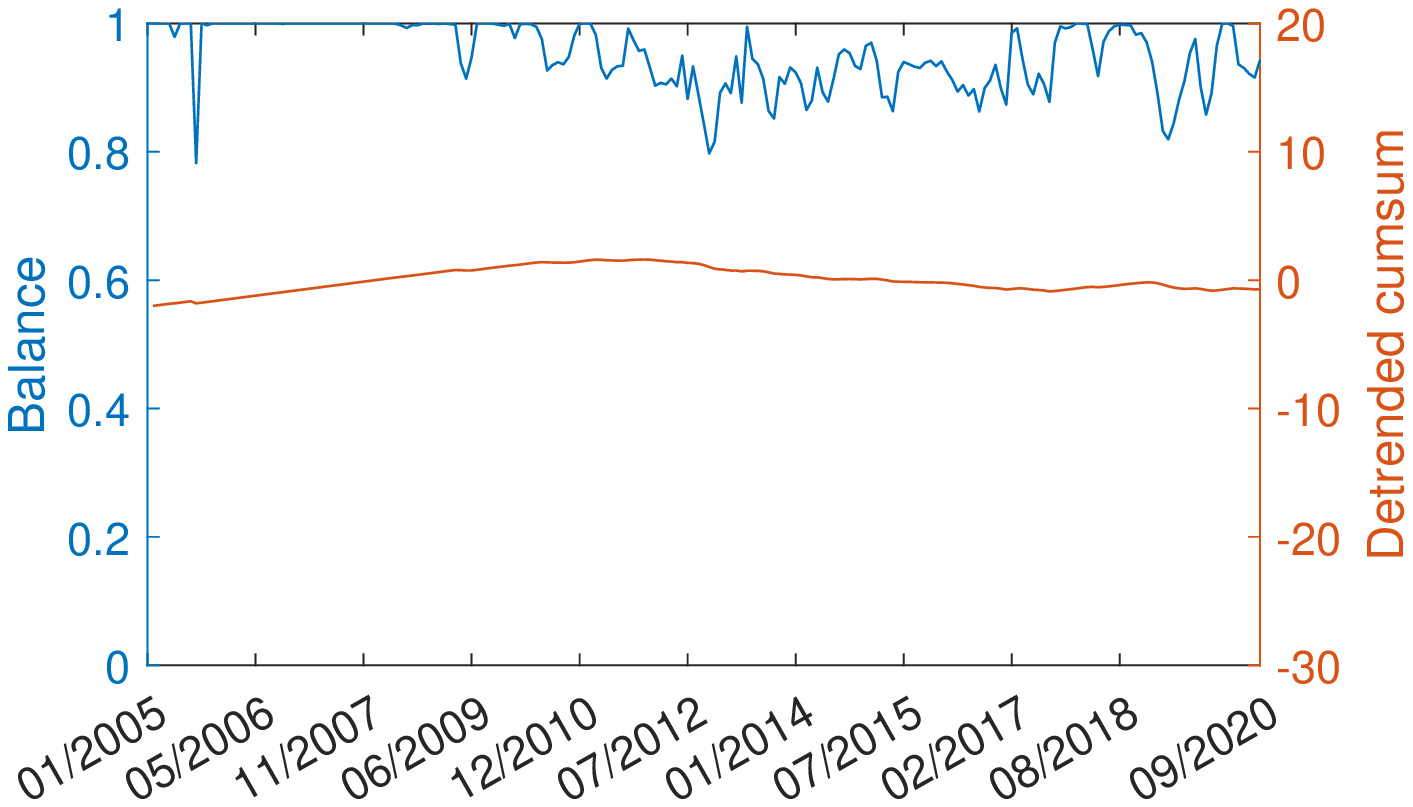}}\tabularnewline\tabularnewline
\multicolumn{6}{c}{\includegraphics[trim=0.5cm 0.25cm 0cm 0.25cm,clip,width=0.12\linewidth]{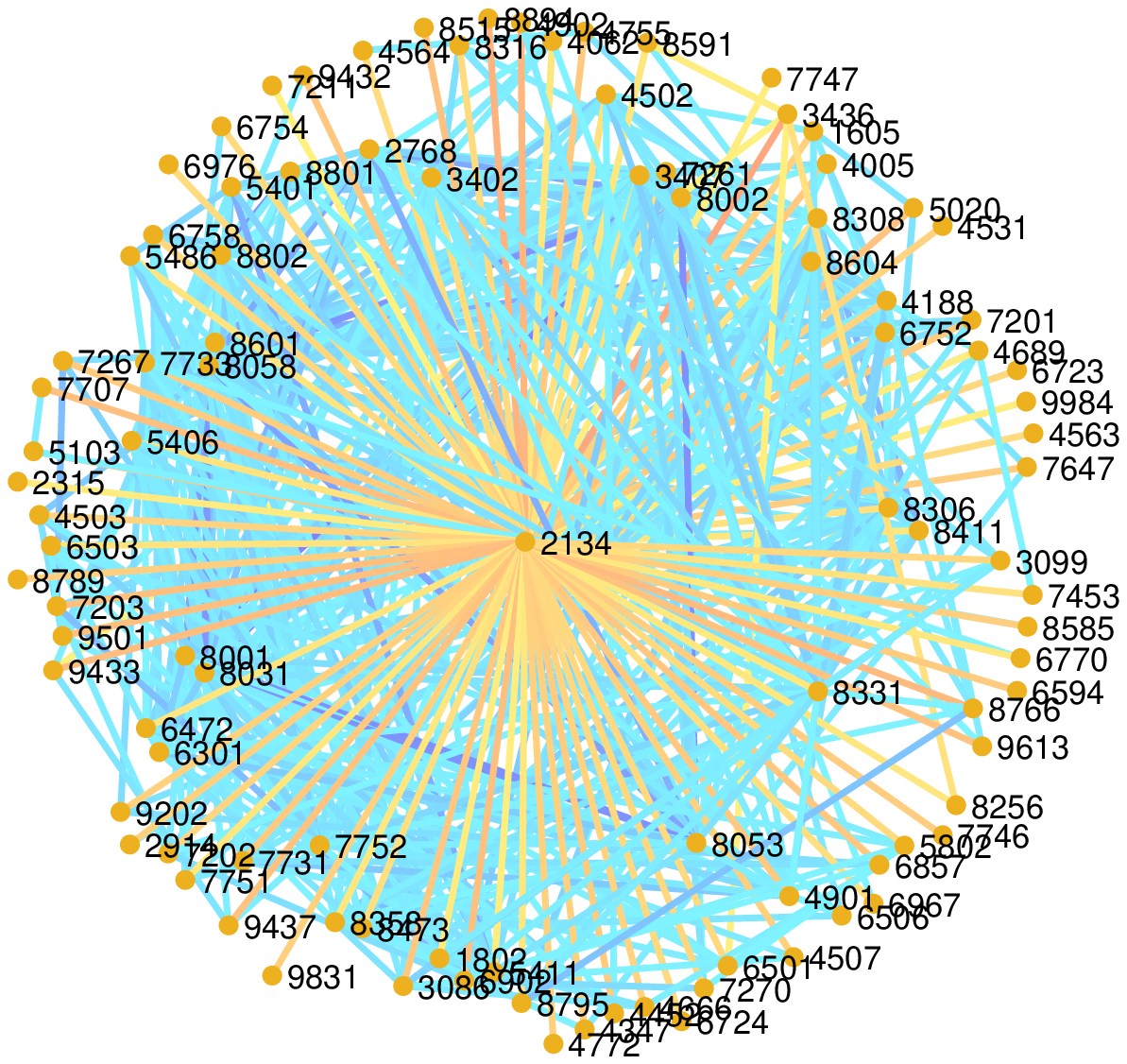}
\includegraphics[trim=0.5cm 0cm 0cm 0.25cm,clip,width=0.12\linewidth]{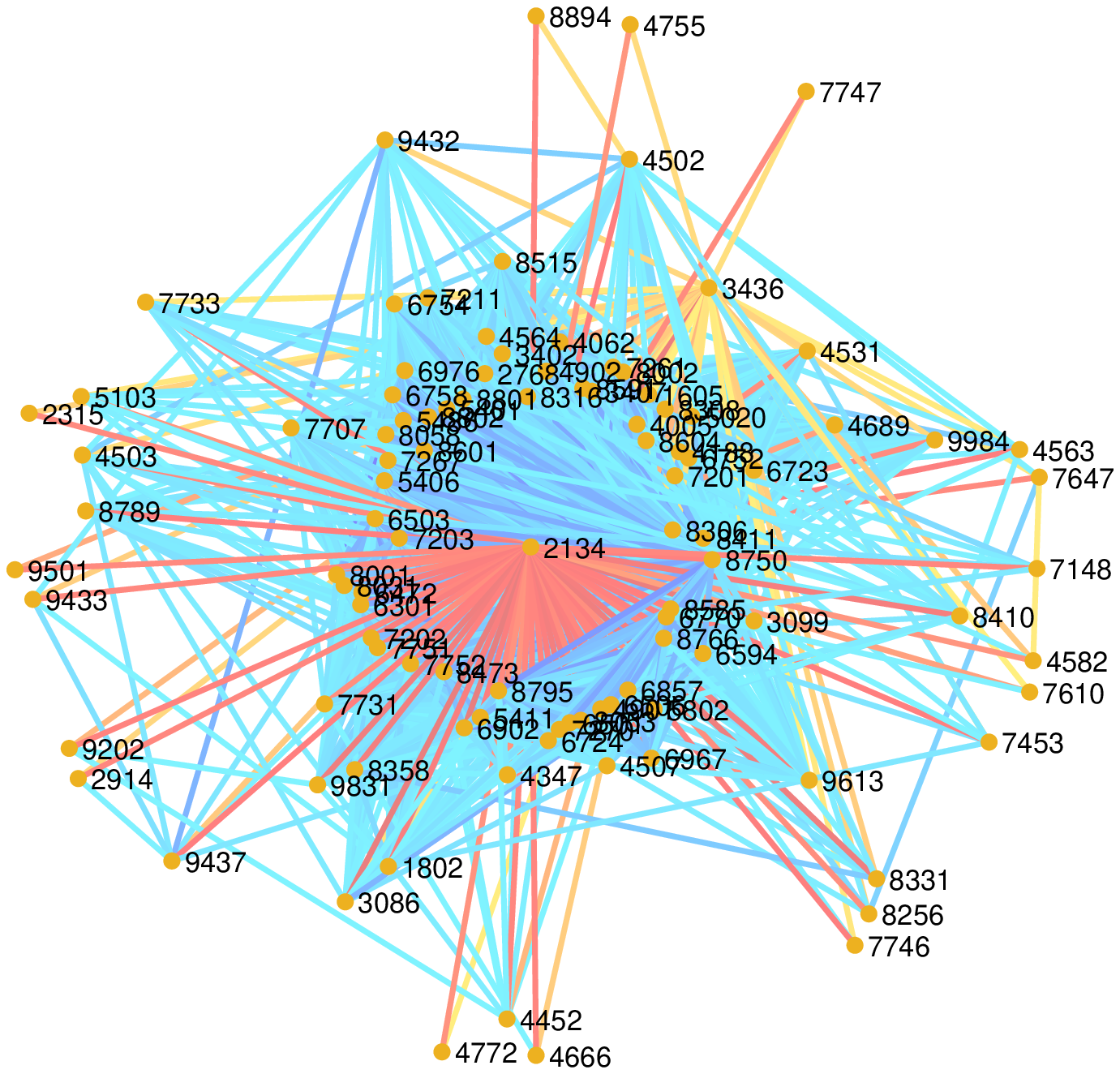}
\includegraphics[trim=0.5cm 0cm 0cm 0.25cm,clip,width=0.12\linewidth]{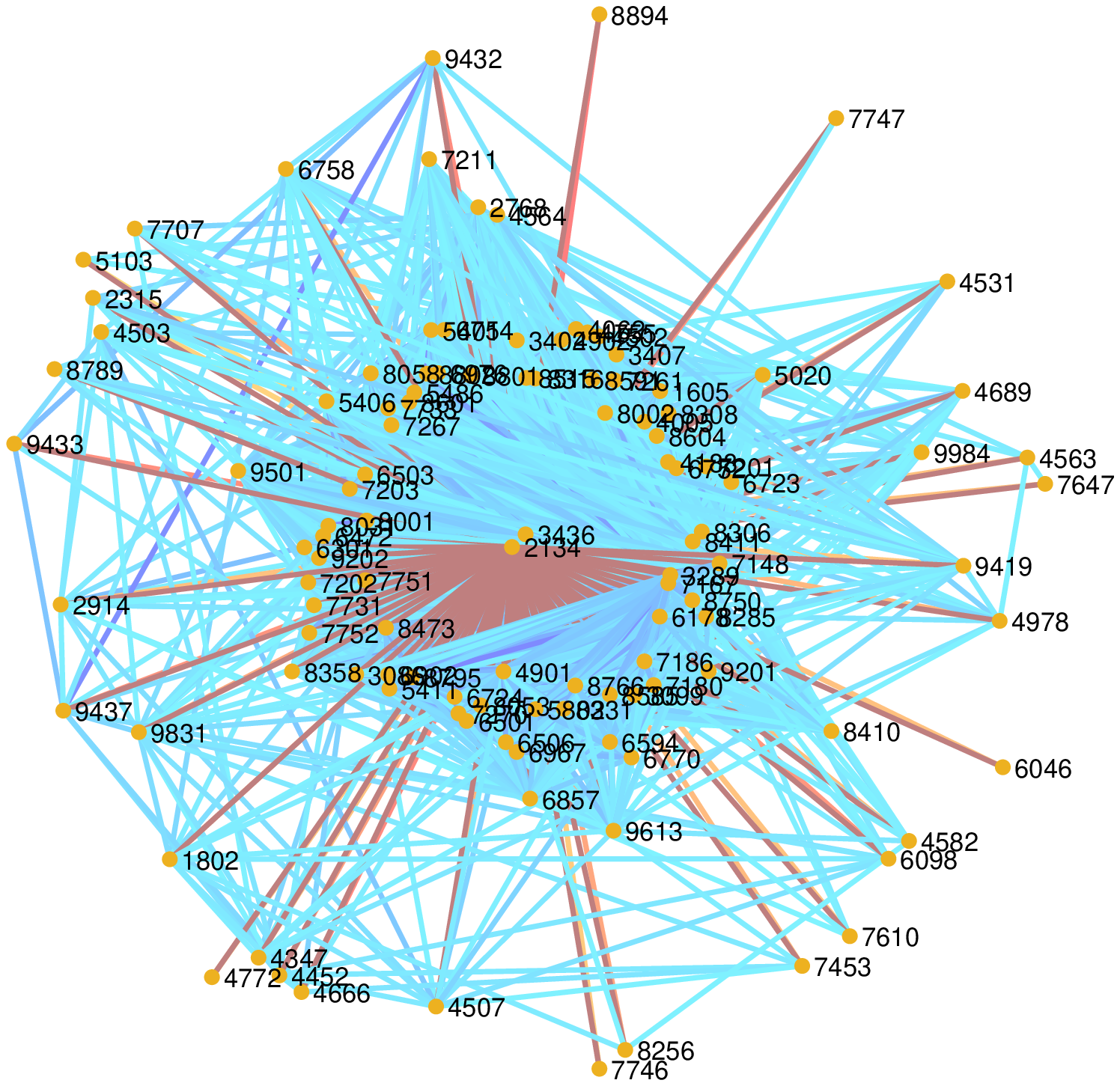}}\tabularnewline
\multicolumn{6}{c}{\textbf{(c)} Japan}\tabularnewline
\end{tabular}

\caption{Balance degree evolution.}

\label{No_Transition}
\end{figure}

The markets of Germany, Italy and Japan display very constant values
of their balance across the whole period of analysis. Although there
are some oscillations at certain specific dates, their DCS display
an almost zero slope confirming the constancy of the balance of these
markets (see Fig. \ref{No_Transition}).


\section*{Discussion}

Among the factors that may influence the BUT in six out of nine countries
studied, we should mention the global level of risk at which a given
market is exposed to at a given time. We account for this factor through
the parameter $0<\beta_{rel}\leq1$, which is based on the relative
EPU index. Following our theoretical model, when the level of global
risk is very low, $\beta_{rel}\rightarrow0$, we have that $K\rightarrow1$
and the corresponding WSSN tends to be balanced independently of its
topology and edge weights. The analysis of the relative EPU values
for all countries indicates that the level of risk accounted for by
this indicator was relatively low from January 2005 to August 2007.
This is exactly the same period for which a high balance was observed
in all markets with $K\geq0.97$. However, the relative EPU indices
alone are not able to explain the BUTs observed in most markets around
September/October 2011. For instance, the relative EPU index peaked
at different values for different markets: Portugal (June 2016), the US
(May 2020), Spain (October 2017), France (April 2017), Greece (November
2011) and Ireland (May 2020). It has been found that the US EPU can achieve
better forecasting performance for foreign stock markets than the
own national EPUs. This is particularly true during very high uncertainty
episodes \cite{bijsterbosch2013characterizing} (see Appendix E in the SI for more details and references). Following this empirical
observation, we can speculate that a plausible cause that triggered
the BUT in five European countries were the events of August 2011,
when the US EPU index rocketed to its highest value between 2005 and the
COVID-19 crisis.

The topological analysis of BUT should necessarily start by considering
the role of negative interdependencies. It is obvious that they are
necessary for unbalance, i.e., a totally positive network cannot be
unbalanced. However, it has been widely demonstrated in the mathematical
literature that this is not a sufficient condition for unbalance,
i.e., there are networks with many negative edges which are balanced
\cite{estrada2019rethinking,kirkley2019balance}. As an illustrative
example let us consider the WSSN of the US of September 3, 2010, which
has 950 negative edges and a balance of $K\approx0.245$ indicating
its lack of balance. However, the US WSSN of February 18, 2011 also
has 950 negative edges, but it is balanced with $K\approx0.946$.
Similarly, the WSSN of Portugal on July 6, 2012 ($K\approx0.700$) and
on May 3, 2019 ($K\approx0.407$), both have 720 negative edges. The
ratio of negative to total edges is neither a sufficient condition
for balance. For instance\textcolor{black}{, in Spain on August 12,
2005 the market was perfectly balanced $(K=1)$, while on September
2007 it was unbalanced ($K\approx0.142)$, although both WSSNs have
the same proportion of negative to total edges, i.e., 1.4\%.}

\begin{figure}[htb]
\centering %
\begin{tabular}{@{}cccccc@{}}
\multicolumn{1}{c}{\includegraphics[width=0.4\linewidth]{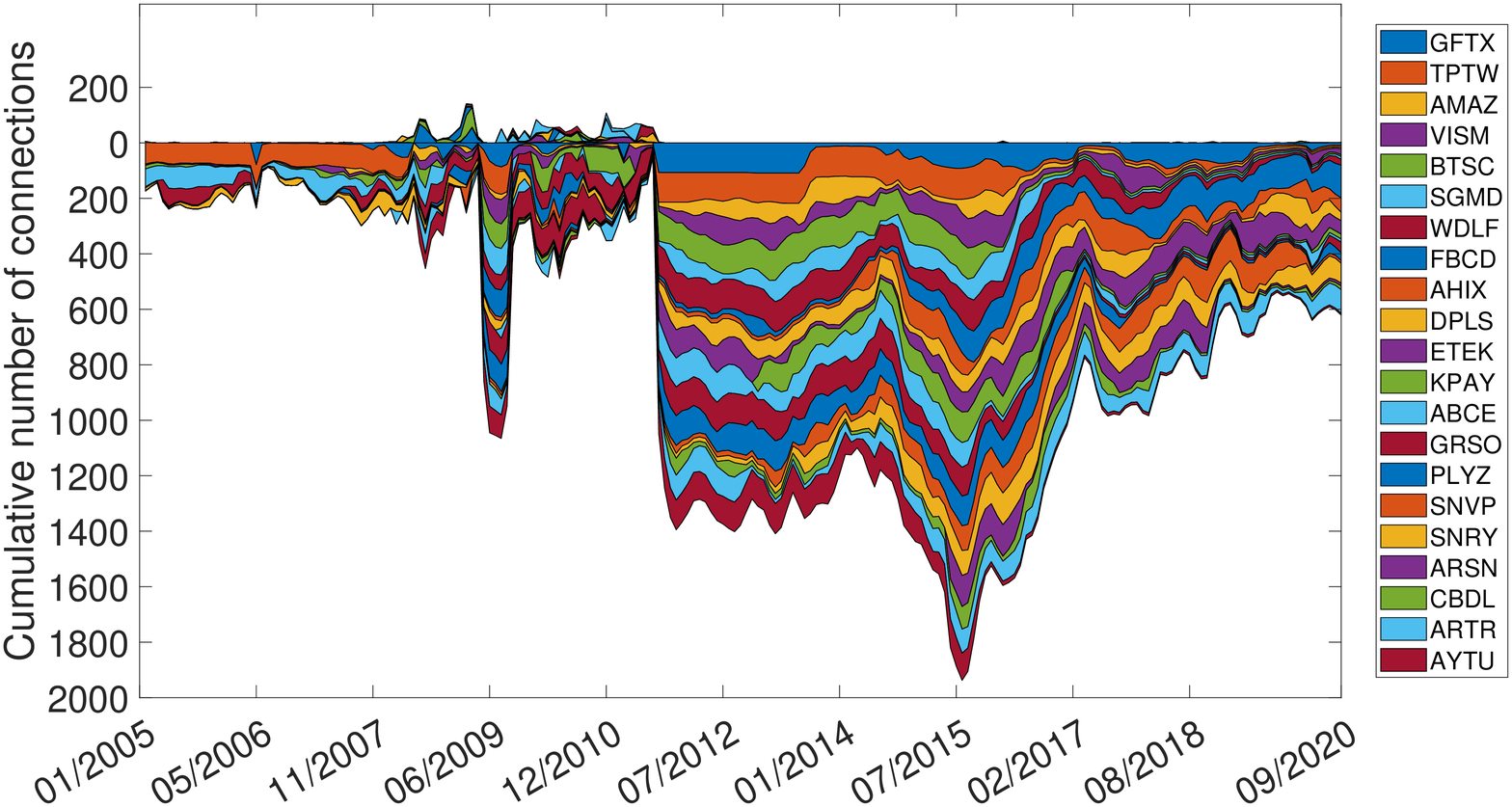}} & \multicolumn{1}{c}{\includegraphics[width=0.4\linewidth]{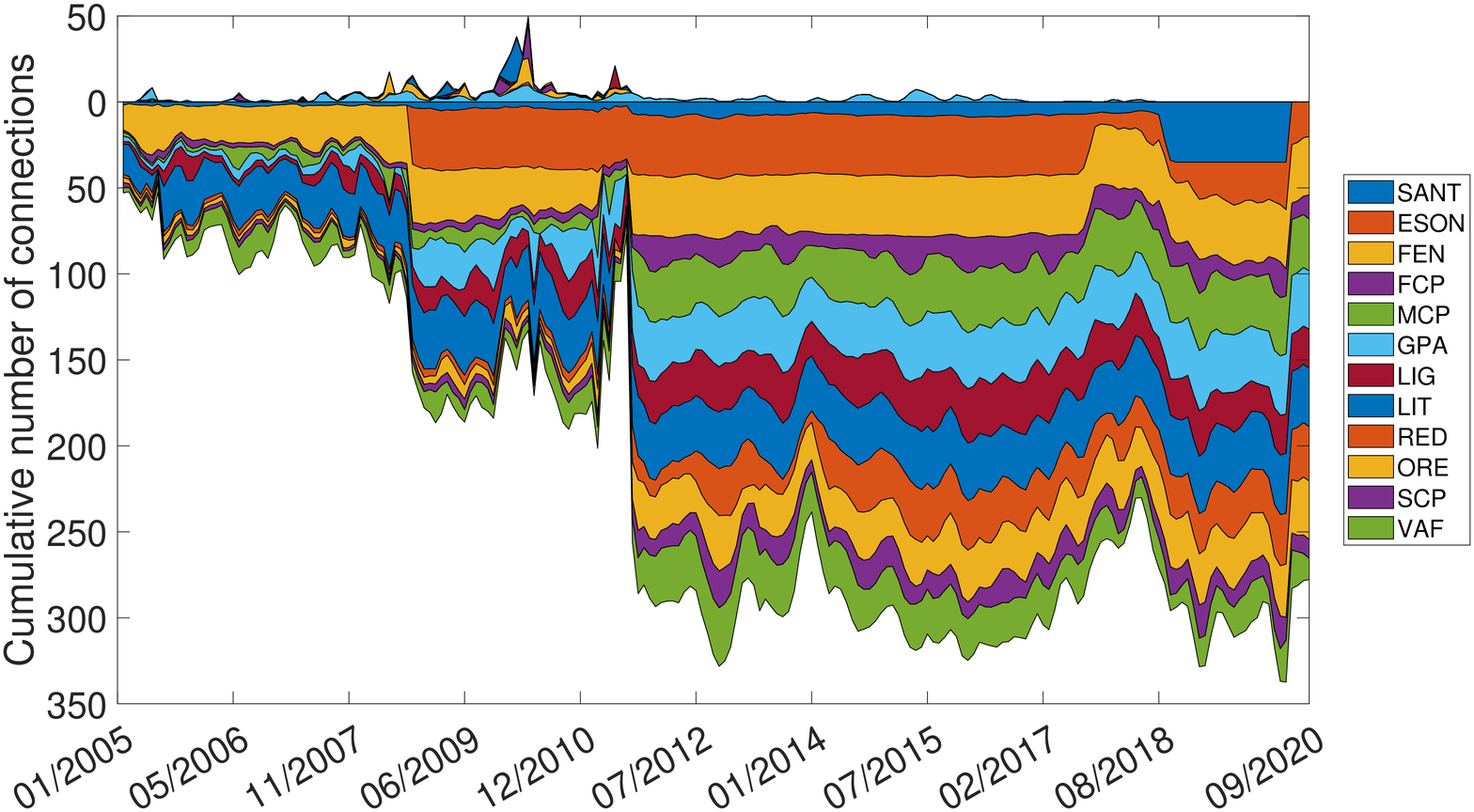}}\tabularnewline
\multicolumn{1}{c}{(a) the US} & \multicolumn{1}{c}{(b) Portugal}\tabularnewline\tabularnewline\tabularnewline
\multicolumn{1}{c}{\includegraphics[width=0.4\linewidth]{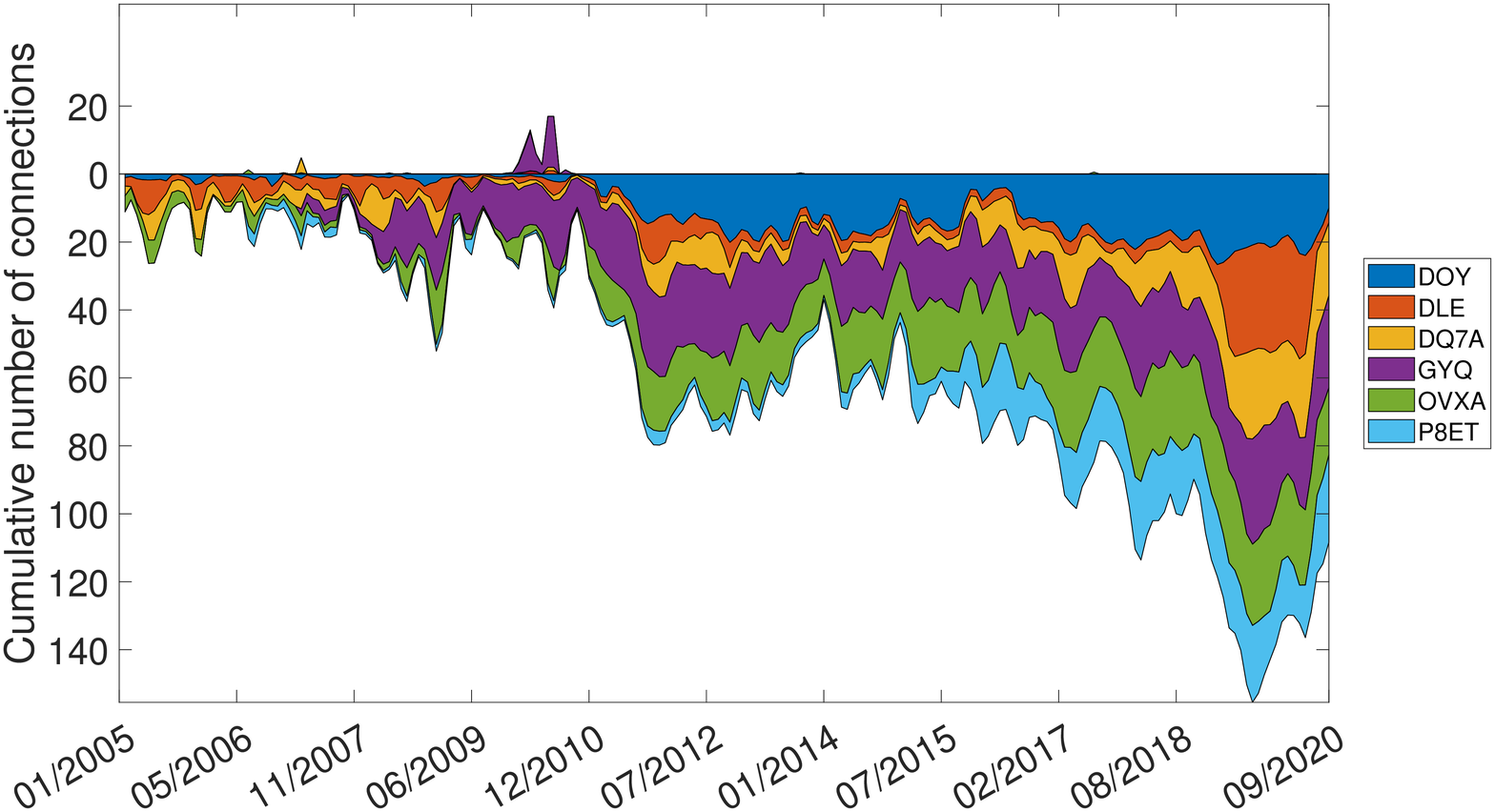}} & \multicolumn{1}{c}{\includegraphics[width=0.4\linewidth]{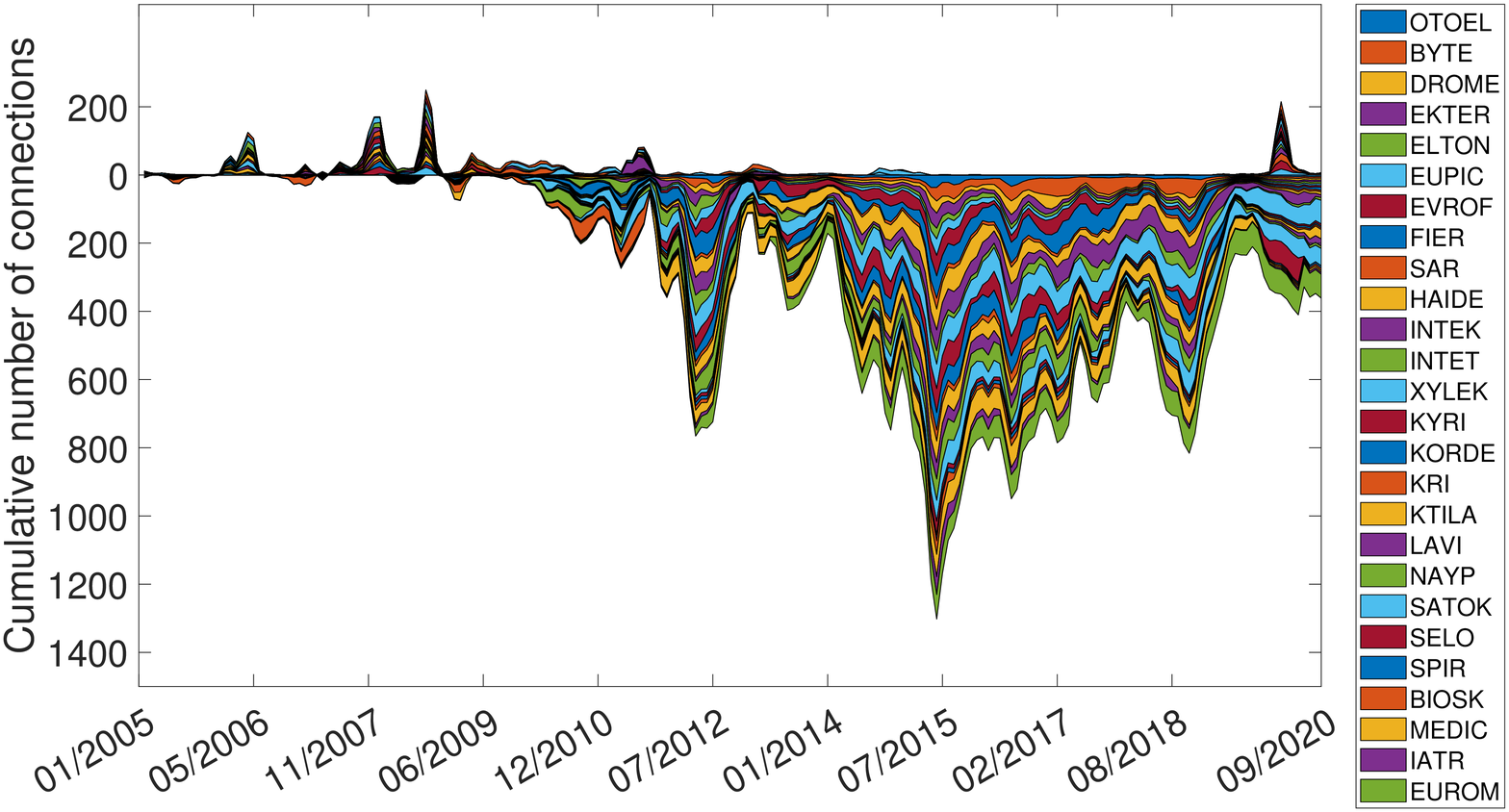}}\tabularnewline
\multicolumn{1}{c}{(c) Ireland} & \multicolumn{1}{c}{(d) Greece}\tabularnewline\tabularnewline\tabularnewline
\multicolumn{1}{c}{\includegraphics[width=0.4\linewidth]{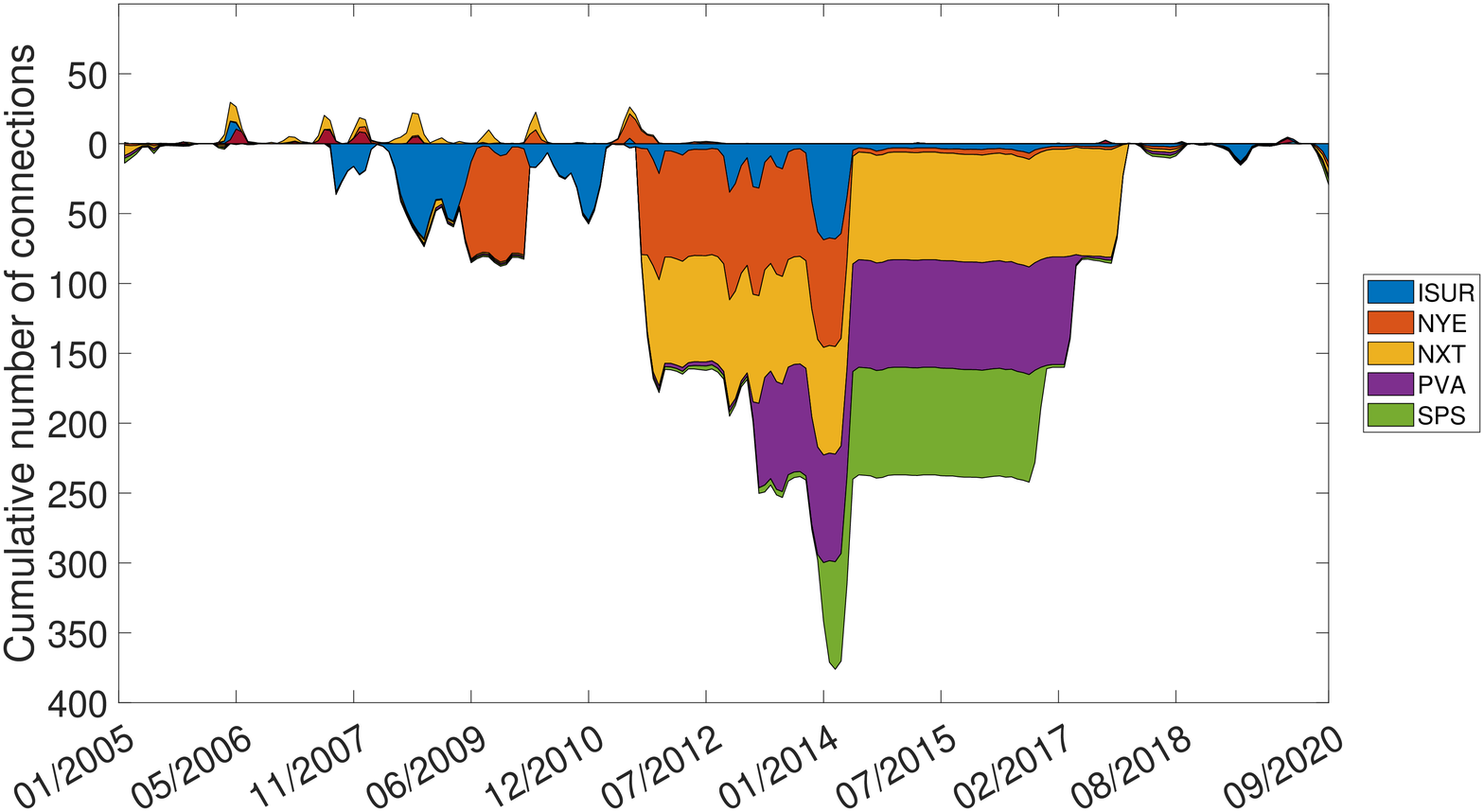}} & \multicolumn{1}{c}{\includegraphics[width=0.4\linewidth]{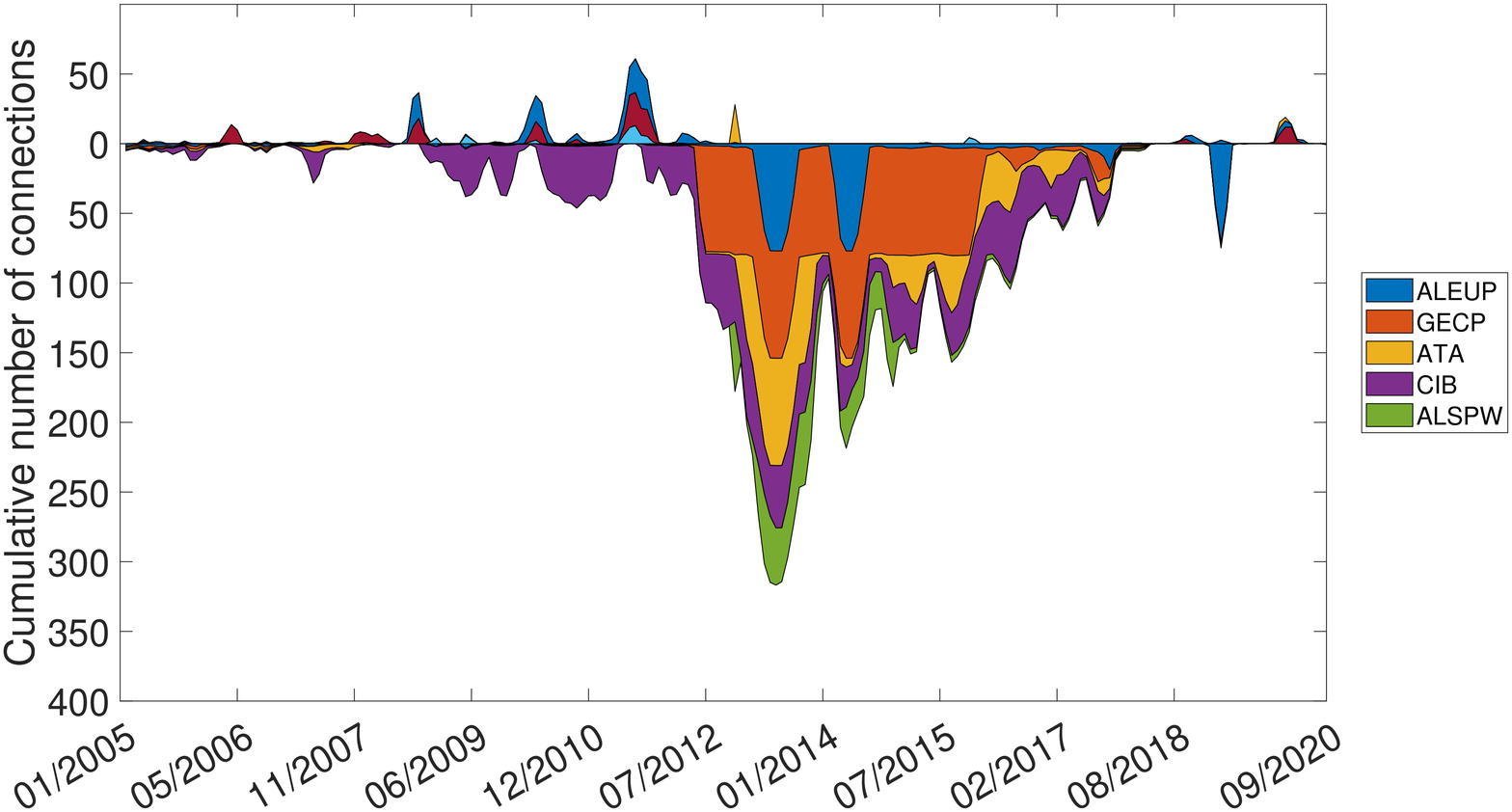}}\tabularnewline
\multicolumn{1}{c}{(e) Spain} & \multicolumn{1}{c}{(f) France}\tabularnewline
\end{tabular}

\caption{Positive and negative cumulative centrality degrees. }

\label{plots_degrees}
\end{figure}

A detailed exploration of the WSSNs with low balance occurring after
BUT (see Fig. \ref{Transition_countries}) revealed the existence
of small fully-negative cliques (FNC) formed by a group of $s$ stocks. A FNC
is highly unbalanced. That is, the balance of an FNC of $s$ stocks
is $K\!=\!\left(\exp\left(-\beta_{rel}\left(s\!-\!1\right)\right)\!+\!\left(s\!-\!1\right)e^{\beta_{rel}}\!\right)\!/\!\big(\!\exp\left(\beta_{rel}\left(s\!-\!1\right)\right)\!+\!\left(s\!-\!1\right)$ $e^{-\beta_{rel}}\big)$,
which for a fixed value of $\beta_{rel}$ clearly tends to zero as
$s\rightarrow\infty$. Therefore, the unbalanced nature of the post-transition
structure of stock markets is mainly due to the presence of these
cliques of mutually anti-correlated stocks. Additionally, each of
the stocks in the FNC increases significantly
its number of negative connections with the rest of stocks after BUT
(see Fig. \ref{plots_degrees}), being also negatively connected to
almost every other stock in the WSSN. This subgraph resembles a kind
of graph known as \textit{complete split graph} (CSG) \cite{estrada2017core}.

To explore the determinants of the lack of balance in these networks
we simulate the structure of the networks displaying BUT by means
of a quasi-CSG-WSSN (see Appendix F in the SI for details) and compare it with a simulated
random WSSN. The quasi-CSG-WSSNs have the same number of nodes and
edges as the WSSNs but the size of the central clique is variable.
We determine the ``optimal'' value of the sizes in the simulated
networks, $s_{opt}$, by minimizing the root mean square error (RMSE)
between the spectrum of the real WSSN and that of the quasi-CSG-WSSN.
In SI Appendix F, Table S1, we report that the RSME of the random model
is about twice bigger than that of the quasi-CSG for the six countries
where BUT occurs. This clearly indicates that the topological organization,
more than the number of positive/negative connections, is what determines
the lack of balance in these networks.

The size of the FNC correlates very well with the intensity of the
drop in the mean values of $K$ before and after the BUT ($r\approx0.857)$.
Two other important characteristics of these FNC are that: (i) they
are mainly formed by non-financial entities (see SI Appendix F, Table
S1), and (ii) they are composed by small and micro-caps with the exception
of SAR in Greece which is a mid-cap. We have shown (see Appendix G in the SI) that
if we split the WSSN by financial (F) and non-financial (NF) sectors,
then the BUT is observed mainly for the networks containing NF-NF
interactions.

All in all, these BUTs change the predictability of the stocks in the way we have explained in Section \ref{subsec:Correlations-and-predictability}
from more predictable markets to more unpredictable ones. 

\section*{Conclusions}

Considering stock-stock correlations in stock markets as signed graphs
allowed us to introduce the concept of balance into stock market networks
for the first time. We related the level of balance in these networks
with stock predictability, and identified a previously unknown transition
between balanced markets to unbalanced ones in six out of nine countries
studied. These balance-unbalance transitions occur in the US, Greece,
Portugal, Spain, and Ireland around September 2011, following the
Black Monday, and later on in France. No transition is observed in
the stock markets of Germany, Italy and Japan for the same period
of time. The balance-unbalance transition is driven by a reorganization
of the stock-stock correlations of a group of low capitalization stocks,
mostly of non-financial entities which collapse into a fully-negative
clique of anticorrelated pairs of stocks. Further studies are needed
to understand the reasons and implications of this reorganization
of non-financial entities in a given number of markets, and the associated
loss of network balance and stock predictability.


Figure 1. Stocks are represented as the vertices
of a triangle and edges represent the correlation between them as
accounted for by Kendall's tau. Positive $\tau_{ij}$ are represented
as solid (blue) lines between the corresponding stocks, and negative
$\tau_{ij}$ are represented as dashed (red) lines.\\
Figures 2-3.  (Top panels) Illustration of the evolution of the balance for the period between January 2005 and September 2020 in the WSSN (blue line) and of its detrended cumulative sum (red line). (Bottom
panels) Illustration of three snapshots of representative networks
at different times (before, at, and after the BUT). The WSSNs are
illustrated using the degree of the nodes as a proxy for the location
of the nodes. The most central nodes are at the center and the low
degree nodes are located at the periphery of the graph. The colors
of the edges correspond to the Kendall's tau estimates, with links
going from dark red for the most negative to blue for the positive
ones.\\
Figure 4. Temporal evolution of the cumulative number of connections (degrees)
of every stock in the FNC of the six countries where BUT occurred.
The cumulative number of positive connections with other stocks are 
shown over the horizontal line, and the cumulative number of negative 
connections are shown below that line.

\section*{Acknowledgements}

E.F., S.O., and J.A. were supported by the Spanish Ministry of the Economy and Competitiveness under grant ECO2014-51914-P; the UPV/EHU under
grants BETS-UFI11/46, MACLAB-IT93-13 and PES20/44; and the Basque Government under BiRTE-IT1336-19. J.A. also acknowledges financial support under PIF16/87 from UPV/EHU. E.E. thanks partial financial support from from Ministerio de Ciencia, Innovacion y Universidades, Spain, grant PID2019-107603GB-I00.

\section*{Author contributions statement}

E.F. and E.E. designed and E.E. directed the investigation. E.F., S.O., and J.A. constructed the data sets, and defined and constructed the adjacency matrices. B.A. and E.E. defined, constructed, and analyzed the networks. E.E. wrote the paper with inputs from all the authors. All the authors contributed to preparing the SI.

\section*{Additional Information}
\subsection*{Competing interests}
The authors declare no competing interests.

\newpage{}
\pagebreak

\setcounter{equation}{0}
\setcounter{figure}{0}
\setcounter{table}{0}

\renewcommand{\thefigure}{S\arabic{figure}}
\renewcommand{\thetable}{S\arabic{figure}}
\renewcommand\thesection{\Alph{section}}
\renewcommand\thesubsection{\thesection.\Alph{subsection}}

\begin{center}
\LARGE Supplementary Information\\ 
\Large\textbf{Loss of structural balance in stock markets}\\ 
\end{center}

 \large
Eva Ferreira\textsuperscript{a}, Susan Orbe\textsuperscript{a}, Jone Ascorbebeitia\textsuperscript{b}, Brais \'Alvarez Pereira\textsuperscript{c}, Ernesto Estrada\textsuperscript{d,*}\\ \\ \normalsize
\textsuperscript{a}Department of Quantitative Methods, University of the Basque Country UPV/EHU, Avda. Lehendakari Aguirre 81, Bilbao, 48015 Spain\\
\textsuperscript{b}Department of Economic Analysis, University of the Basque Country UPV/EHU, Avda. Lehendakari Aguirre 81, Bilbao, 48015 Spain\\
\textsuperscript{c}Nova School of Business and Economics (Nova SBE), NOVAFRICA, and BELAB\\
\textsuperscript{d}Institute of Mathematics and Applications, University of Zaragoza,
Pedro Cerbuna 12, Zaragoza 50009, Spain; ARAID Foundation, Government
of Aragon, Spain. Institute for Cross-Disciplinary Physics and Complex
Systems (IFISC, UIB-CSIC), Campus Universitat de les Illes Balears
E-07122, Palma de Mallorca, Spain.\\
\textsuperscript{*}estrada66@unizar.es\\ \\

\setcounter{page}{1}
\pagenumbering{arabic}

\Large{\textbf{Methods}}\\ \\
\large \textbf{A. Weighted Signed Stock Networks}\normalsize

In order to define the weight of the edges in the WSSN we use Kendall's tau rank correlation coefficients. For two stocks $A$ and $B$ in a given market, our interest is to investigate whether the relationship between the stock log returns $(Y_{A},Y_{B})^{T}$ changes with time. 
In this context, copulas are very useful since they give a flexible structure for modeling multivariate dependences \cite{nelsen2007introduction}. Let $F_A$, $F_B$ and, $F$ be the continuous marginals and the joint distribution function of $(Y_{A},Y_{B})^{T}$.
 Then, based on Sklar's theorem \cite{sklar1959fonctions}, there is a unique copula function $C(\cdot,\cdot,t):[0,1]^2\rightarrow[0,1]$ such that $F(y_{At}, y_{Bt},t)=C(F_{A}(y_{At},t), F_{B}(y_{Bt},t),t)$ for any $y_{jt}\in Y_j$, $j=A,B$.
Therefore, time varying Kendall's tau is defined in terms of copulas as $$\tau_{A,B} (t)\!=\! 4 \int_{[0,1]^2} C(u_1,u_2,t)dC(u_1,u_2,t)\!-\!1,$$ where $(u_1,u_2)\in [0,1]^2$. To estimate the time varying dependence, a nonparametric estimation method is used because of its advantage of overcoming the rigidity of parametric estimators. Specifically, it allows to remove the restriction that the joint distribution function belongs to a parametric family. In this line, Ascorbebeitia et al. \cite{ascorbebeitia2021effect} propose to estimate the time varying copula $C(u_1,u_2,t)$ as $$\hat{C}(u_1,u_2,t)=\frac{1}{Sh}\sum_{s=1}^S w_{h}(t-s)I\{Y_{A}(s)\leq\hat{F}_{A}^{-1}(u_1,t), Y_{B}(s)\leq\hat{F}_{B}^{-1}(u_2,t)\},$$where $w_{h}(t-s)=(Sh)^{-1}k\left(\left(t-s\right)/\left(Sh\right)\right)$ is a sequence of kernel weights that smooths over the time space, $h>0$
is the bandwidth that regulates the degree of smoothness, $\hat{F}_{j}(y,t)=(Sh)^{-1}\sum_{s=1}^S w_{h}(t-s)I\{Y_{j}(s)\leq y\}$ denotes the nonparame\-tric time varying estimator of the $j$-marginal distribution, and $\mathbb{I}\left\{ \cdot\right\} $
is the indicator function. They also derive the following consistent nonparametric estimator for the time varying Kendall's tau under $\alpha$- mixing local stationary variables:

\begin{equation}\tag{S1}
\hat{\tau}_{A,B}\left(t\right)=\dfrac{4}{1-\sum_{s=1}^{S}w_{h}(t-s)^{2}}\sum_{s,r=1}^{S}w_{h}(t-s)w_{h}(t-r)\mathbb{I}\left\{ Y_{A}(s)<Y_{A}(r),Y_{B}(s)<Y_{B}(r)\right\} -1\label{eq:estimator}
\end{equation}
 In all our calculations we consider the Epanechnikov kernel $k\left(x\right)=\tfrac{3}{4}\left(1-x^{2}\right) \mathbb{I}\left\{ \left|x\right|<1\right\} $. This kernel assigns a higher weight to those values of the indicator
function that are close in time and lower
weights to observations of the indicator farther away from $t$. The smoothing parameter $h$ is selected minimizing the mean squared error of the Kendall's tau estimator in (\ref{eq:estimator}) (for more details about the estimators see Ref. \cite{ascorbebeitia2021effect}). 

Let us now fix $t$ and calculate $\hat{\tau}_{ij}\left(t\right)$
for every pair of stocks $(i,j)$ in a given market and define the rank correlations
matrix $M\left(t\right)$ whose entries are the values of $\hat{\tau}_{ij}\left(t\right)$.
Let $\varepsilon\in\mathbb{R}^{+}$ be a given threshold. We create
the matrix $A\left(t\right)$ whose entries are $A_{ij}\left(t\right)=\hat{\tau}_{i,j}\left(t\right)\mathbb{I}\{\left|\hat{\tau}_{i,j}\left(t\right)\right|\geq\varepsilon\}$. Different candidates of the  threshold have been considered and after some robustness checks $\varepsilon =0.3$ has been selected. The matrix $A\left(t\right)$ is then the adjacency
matrix of a Weighted Signed Stock/Equities Network (WSSN) at time
$t$. Here we construct weighted adjacency matrices based on daily returns from January 2005 until September 2020. Therefore, we have a set of WSSNs obtained with matrices $\mathscr{\mathcal{A}}=\left\{ A\left(t=1\right),\ldots,A\left(t=S\right)\right\} $.\\ \\

\large \textbf{B. Time Varying Nonparametric Regression}\normalsize

Let us consider the following regression model $Y_i(t)=m_i(Y_j(t))+\epsilon_i(t)$, where $t=1,...,S$, $Y_i(t)$ is the dependent variable,  $m_i(\cdot)$ is a non-specified unknown smooth function, $Y_{j\neq i}(t)$ is the explanatory variable and $\epsilon_i(t)$ is the error term. As pointed out in several studies (see, e.g., Refs. \cite{krishnan2009correlation,ferreira2011conditional,casas2020sure}), the time varying behavior of the variables is an important characteristic in finance to take into account. Since financial variables are dependent and not stationary, a nonparametric estimator for the regression model that accounts for time variation is considered under local stationary and $\alpha$-mixing variables \cite{robinson1989nonparametric}.  The time varying estimator of $m_i(y_{jt})$ for any $y_{jt}\in Y_j$ is defined as 
$$\hat m_i(y_{jt})=\left(\sum_{s=1}^S w_h(t-s) w_h(y_{jt}-Y_{j}(s))\right)^{-1}\sum_{s=1}^S w_h(t-s)w_h(y_{jt}-Y_j(s))Y_i(s),$$
where $w_h(y_{jt}-Y_j(s))=(Sh)^{-1}k\left((y_{jt}-Y_j(s))/(Sh)\right)$ and $k(\cdot)$ denotes the kernel weights. To choose the smoothing parameter $h$, cross-validation methods proposed in the literature for nonparametric regression can be used (for details see, e.g., Ref. \cite{fan2008nonlinear}).

If one is interested in the relation between regression slopes and  correlation coefficients, a time varying relationship between variables can be assumed $m_i(Y_j(t))=\beta_i(t)Y_j(t)$  (see Ref. \cite{robinson1989nonparametric}) leading to a semiparametric regression model. In such  case, the time varying correlation between two variables, $\rho_{ij}(t)$, is related to the time varying slope $\beta_i(t)$ as $\rho_{ij}(t)=\beta_i(t)\left(\sigma_i(t)\right)^{-1}\sigma_j(t),$ where $\sigma^2_{i}(t)$ and $\sigma^2_{j}(t)$ are the time varying variances of variables $Y_i$ and $Y_j$ that can be estimated by smoothing the corresponding squared residuals. Then, $\hat\rho_{ij}(t)$ can be estimated through the time varying slope defined as 
$$\hat \beta_i(t)=\left(\sum_{s=1}^S w_{h,ts}Y_j(s)^2\right)^{-1}\sum_{s=1}^S w_{h,ts}Y_i(s)Y_j(s)$$
In this setting, the estimation of $Y_i(t)$ is given by $\hat Y_i(t)=\hat m_i(Y_j(t))= \hat \beta_i(t) Y_j(t)$.  This relation would allow to replace the product of slopes of the regression models estimating the trend of the different pairs of stocks in a triad by the product of their respective conditional correlation and replace the study of predictability in stock markets by that of balance in WSSNs. Nevertheless, linear correlation is not appropriate for financial variables as we have mentioned before. Hence, we consider the rank correlation as the connectivity measure between nodes in the network. Note that if the variables were gaussian, there is a one-to-one relationship between the linear correlation and Kendall's tau, i.e., $\rho_{ij}(t)=sin((\pi/2)\tau_{ij}(t))$, so using Kendall's tau generalizes the analysis made with linear correlation.\\

\large \textbf{C. Balance}\normalsize

The definition of balance given in the main text is:

\begin{equation*}
K=\dfrac{tr\left(\exp\left(\beta_{rel}A\left(G\right)\right)\right)}{tr\left(\exp\left(\beta_{rel}A\left(G'\right)\right)\right)},
\end{equation*}
which has also been defined in \cite{estrada2014walk}.

Let us first express the exponential of the adjacency matrix of the
WSSN as a Taylor series (see for instance Ref. \cite{estrada2012structure}):

\begin{equation*}
tr\left(\exp\left(\beta_{rel}A\left(G\right)\right)\right)=n+\dfrac{\beta_{rel}^{2}}{2!}tr\left(A^{2}\left(G\right)\right)+\dfrac{\beta_{rel}^{3}}{3!}tr\left(A^{3}\left(G\right)\right)+\cdots.
\end{equation*}

Notice that $tr\left(I\right)=n$ and $tr\left(A\left(G\right)\right)=0.$
The term $tr\left(A^{k}\left(G\right)\right)$ represents the sum
of all closed walks of length $k$ in the WSSN. A \textit{walk of
length} $k$ between the nodes $v$ and $u$ in an WSSN is the product
of the Kendall's tau for all (not necessarily different) edges in
the sequence $e_{v,1},e_{1,2},\ldots e_{k-1,u}$. The walk is closed
if $v=u$. Then, $tr\left(A^{2}\left(G\right)\right)=\sum_{\left(i,j\right)\in E}\tau_{ij}^{2}$
and $tr\left(A^{3}\left(G\right)\right)=\sum_{\left(i,j,k\right)\in\triangle}\tau_{ij}\tau_{ik}\tau_{jk}$,
where $\triangle$ is the triangle with vertices $i,j,k$. We can
continue with higher order powers of $A\left(G\right)$, which are
related to squares, pentagons, and so forth, apart from other cyclic
and non-cyclic subgraphs. Obviously, we have already shown here that
for a signed triangle $tr\left(A^{3}\left(G\right)\right)=6\tilde{K}$,
where $\tilde{K}=\tau_{ij}\tau_{ik}\tau_{jk}$ was defined in the
main text of the paper as an index of predictability of the stock
trends in a triad.

Let us first prove that $K=1$ if and only if the WSSN is balanced.
For that we first state a result proved by Acharya \cite{acharya1980spectral}.
\begin{thm}
For any signed graph, the matrices $A\left(G\right)$ and $A\left(G'\right)$
are isospectral if and only if the signed graph is balanced.
\end{thm}

This means that both matrices $A\left(G\right)$ and $A\left(G'\right)$
have exactly the same eigenvalues $\lambda_{j}\left(A\left(G\right)\right)$
and $\lambda_{j}\left(A\left(G'\right)\right)$, respectively, if
and only if the graph $G$ is balanced. Then, we can write

\begin{equation*}
K=\dfrac{\sum_{j=1}^{n}e^{\beta_{rel}\lambda_{j}\left(A\left(G\right)\right)}}{\sum_{j=1}^{n}e^{\beta_{rel}\lambda_{j}\left(A\left(G'\right)\right)}},
\end{equation*}
which is equal to one if and only if the graph is balanced.

Let us now show that $K$ quantifies the departure of a WSSN from
balance for non-balanced ones. Let us designate $W_{k}=tr\left(A^{k}\left(G\right)\right)$
the total number of CWs of length $k$ in the WSSN. Obviously, $W_{k}=\sum_{i=1}^{n}W_{k}\left(i\right)$,
where $W_{k}\left(i\right)$ is the number of CW of length $k$ that
starts (and ends) at the node $i$. Then, $W_{k}\left(i\right)<0$
if the node $i$ is in an unbalanced cycle. Otherwise, $W_{k}\left(i\right)>0$.
Therefore, $W_{k}\left(G\right)=\sum_{i=1}W_{k}^{+}\left(i\right)-\left|\sum_{i=1}W_{k}^{-}\left(i\right)\right|$,
where $W_{k}^{+}\left(i\right)$and $W_{k}^{-}\left(i\right)$ are
positive and negative CWs of length $k$ starting at node $i$. Let
us now designate $W^{+}\left(G\right)=\sum_{k=0}^{\infty}\sum_{i=1}\dfrac{\beta_{rel}^{k}}{k!}W_{k}^{+}\left(i\right)$
and $W^{-}\left(G\right)=\sum_{k=0}^{\infty}\sum_{i=1}\dfrac{\beta_{rel}^{k}}{k!}W_{k}^{-}\left(i\right)$.
It is straighforward to realize that in $G'$ $tr\left(\exp\left(\beta_{rel}A\left(G'\right)\right)\right)=W^{+}\left(G\right)+\left|W^{-}\left(G\right)\right|$,
which implies that

\begin{equation*}
K=\dfrac{W^{+}\left(G\right)-\left|W^{-}\left(G\right)\right|}{W^{+}\left(G\right)+\left|W^{-}\left(G\right)\right|}.
\end{equation*}

We can see now that $K=1$ only when the WSSN does not have any unbalanced
cycle, i.e., $W^{-}\left(G\right)=0.$ Also, $K$ departs from one
as the number of unbalanced walks growth, $K=1-\dfrac{2\left|W^{-}\left(G\right)\right|}{W^{+}\left(G\right)+\left|W^{-}\left(G\right)\right|}$,
which approaches asymptotically to zero as $2\left|W^{-}\left(G\right)\right|\rightarrow\left(W^{+}\left(G\right)+\left|W^{-}\left(G\right)\right|\right)$.\\ \\
\Large{\textbf{Materials}}\\ \\
\large \textbf{D. Data}\normalsize

We construct stock networks for some European countries, the US and Japan. The seven selected European countries are Greece, Italy, Ireland, Portugal, Spain (those five known as GIIPS) plus 
Germany and France as core countries. The US and Japan stock markets are considered as they belong to the main drivers of the world economy. 
The data set contains equities' daily closing price and volume data from 01-03-2005 to 09-15-2020, obtained from the Morningstar database. This data set is available from the authors on reasonable request. 

The criteria used to consider a company as a candidate to form the database for each country is the trading volume at the end of April 2020. 
Initially a set of around 250 companies with the highest trading volume at the 28th of April 2020 is considered, except for the US and Japan for which the initial sets are extended to 794 and 427 companies respectively because their
stock markets are very diversified in terms of volume. Then, companies with a big amount of missing values (more than three market years of consecutive missing values or more than 30 \% of non-consecutive missing values) according to the daily closing prices series have dropped out. There are many possible reasons for those missing values, such as trading cancellation for a period of time, stock market exit, late entry into the stock market, merger of companies, bankruptcy, etc.  We find that allowing for the 30\% of missing values in a stock price series is a reasonable threshold. 
Nevertheless there are some exceptions, companies with more than 30\% of missings but with a leading trading volume such as Abengoa S.A Class B (ABG.P) from the Spanish stock market, which constitutes the $39\%$ of the total trading volume on the reference date, are excluded from this screening. 

After the cleaning process, the database is composed by a set of representative companies according to the sector to which they belong (under Russell Global Sector classification) and their trading volume.

\begin{table}[htb]
\renewcommand{\arraystretch}{0.5}
\caption{Number of assets considered and the percentage of the total trading volume.}\label{assets and vol} {\scriptsize
\begin{center}
\begin{tabular}{lccccccccccc}
& Germany & France & Greece & Italy & Ireland & Portugal & Spain & US & Japan\\
\hline \rule[0mm]{-4pt}{0mm}\\
\vspace{0.5em}
no. of assets & 81 & $78$ & $73$ & $83$ & 32 & 36 & 78  & 119 & 120\\
\vspace{0.5em}
 volume & $70.86\%$ & $76.9\%$ & $97.41\%$& $90.56\%$ & $85.79\%$ & $98.75\%$ & $86.62\%$ & $45.66\%$ & $54.58\%$\\
\hline
\end{tabular}\label{companies-no}
\end{center}
}
\end{table}

Table \ref{companies-no} shows the number of companies considered and the percentage of the total trading volume that constitute such assets for each country.

As the inverse temperature $\beta$ of the networks we use monthly EPU index data \cite{baker2016measuring} for each country from January 2005 to September 2020. Portugal is an exception since there is no specific EPU index available for it. Therefore, the European EPU index is considered instead. EPU index data is publicly available in the Economic Policy Uncertainty website (\url{https://www.policyuncertainty.com/index.html}).\\ \\
\large \textbf{E. Events that may have triggered BUTs}\normalsize

In the paper we have observed a lack of capacity of national EPU indexes to explain the balance transition in those countries that have experienced it. This might be due to the relationship between the US EPU index and foreign stock markets being stronger than that between these stock markets and their corresponding national EPU indexes. While there are some contradictory results depending on the chosen methodology \cite{wu2016causal}, a large amount of evidence shows an association between the EPU index and greater stock price volatility for the US \cite{antonakakis2013dynamic,pastor2013political, liu2015economic, brogaard2015asset, baker2016measuring}, with this association being particularly strong during very high uncertainty episodes \cite{bijsterbosch2013characterizing}. However, the evidence for the relationship between national or European EPU indexes and stock market volatility for other countries is generally weaker \cite{li2016causal}, state dependent \cite{ko2015international} and shows important levels of heterogeneity \cite{wu2016causal, vskrinjaric2020economic}. Indeed, Mei et al. \cite{mei2018does} find that models including the US EPU index can achieve better forecasting performance for European stock markets volatility, while the inclusion of the corresponding national own EPU index does not significantly increase forecasts accuracy. In this same direction, Ko and Lee \cite{ko2015international} show that the negative link between the national EPU and stock prices changes over time, being importantly influenced by the co-movement of the national EPU index and the US one.

In the period considered in our study, the US EPU index attains its highest level before the COVID-19 crisis in August 2011, coinciding with the Black Monday. The Black Monday 2011 refers to August 8, when the US and global stock market crashed, following the credit rating downgrade by Standard and Poor’s of the US sovereign debt from AAA for the first time in history. This same period coincides with extremely high values for the US, European and Asian stock market volatility indexes, VIX, VSTOXX and VKOSPI respectively. These indexes - with an important leading role of the VIX on the other two through strong spillover effects- are established measures of fear, risk and uncertainty in international stock markets, proved to be important in explaining stock returns \cite{shu2019spillovers}. The three of them attained their highest point between the 2008 financial crisis and the current COVID-19 crisis in September 2011, coinciding with the balance transition found in national stock markets. 

These findings point towards the events associated to the Black Monday 2011 as the most likely triggers for the balance transition we observe in five of the countries in our sample. The association between a period of increased -economic policy- uncertainty and a sudden loss of balance seems coherent with the reduction of balance observed for most of the countries also after the crisis produced by COVID-19. However, this line of reasoning conflicts with the lack of a generalized, strong-enough drop in balance during the financial crisis of 2007-2008 in most of the countries. The US stock market did suffer significant balance fluctuations in this period, even if these were different from a strict loss of balance since balance was recovering after falling. 
 A potential explanation might be the interruption of the negative time-varying correlation between policy uncertainty and stock market returns that took place during the financial crisis. This could be a consequence of the unprecedented bailout package for the US banking sector of 2008 and the stimulus package of 2009, which pushed the market into positive returns even if the policy uncertainty remained high \cite{antonakakis2013dynamic}. 

A detailed exploration of the causes of the heterogeneous behavior of the different countries is out of the scope of this paper. Italy, Japan and Germany, the three countries without a balance transition during the period had lower long-term GDP growth before the 2007-2008 financial crisis, but the obvious commonalities finish there. A promising avenue for the further exploration of this question might consider the differential impact of the European debt crisis in each country: the ranking of the countries according to the long-term interest rates of their national debt during this period matches the intensity of the observed balance transitions, with the exception of Italy and France. \\ \\
\large \textbf{F. Quasi-CSG-WSSN}\normalsize

Here we define quasi-CSG-WSSN which are used in the main text of the
manuscript. Let $n$, $m_{-}$ and $m_{+}$ be the number of nodes,
of negative and positive edges in a real WSSN, respectively. We create
a CSG with a clique of $s<n$ nodes and $\dfrac{1}{2}s\left(2n-s-1\right)<m_{-}$
edges. We complete the quasi-CSG-WSSN by adding randomly and independently
$m_{-}-\dfrac{1}{2}s\left(2n-s-1\right)$ negative and $m_{+}$ positive
edges among the nodes not in the clique. Additionally, we create the
random-WSSN by adding randomly and independently $m_{-}$ and $m_{+}$
edges among $n$ nodes a la Erd\H{o}s-R\'enyi \cite{erdHos1960evolution}. A Matlab code for building
such graphs is given in Algorithm \ref{Agave}.

\begin{algorithm}
\begin{lstlisting}
%%%Input data%%%
n=50; 
m_neg=500; 
m_pos=100;
s=10;

%%%Construction of the adjacency matrix A(G)%%%
B=-ones(n-s,s); 
n_rest=n-s; 
m_rest=m_neg-(s*(n-s)+s*(s-1)/2); 
C=full(erdrey(n_rest,m_rest));
A=[-ones(s,s) B' B -C]; 
A=A-diag(diag(A));
A=triu(A); 
K=tril(ones(n,n)); 
A=A+K+eye(n);
[row,col] = find(~A); 
data=[row col]; 
y = datasample(data,m_pos,'Replace',false);
z=length(y);

for i=1:z
    A(y(i,1),y(i,2))=1;
end;

A=triu(A); 
A=A+A'; 
A=A-diag(diag(A)); 
\end{lstlisting}

\caption{Matlab code for constructing a quasi-CSG-WSSN with given $n$, $m_{-}$
and $m_{+}$ and $s$.}

\label{Agave}
\end{algorithm}

\begin{figure}[htb]
\centering
\includegraphics[width=0.4\linewidth]{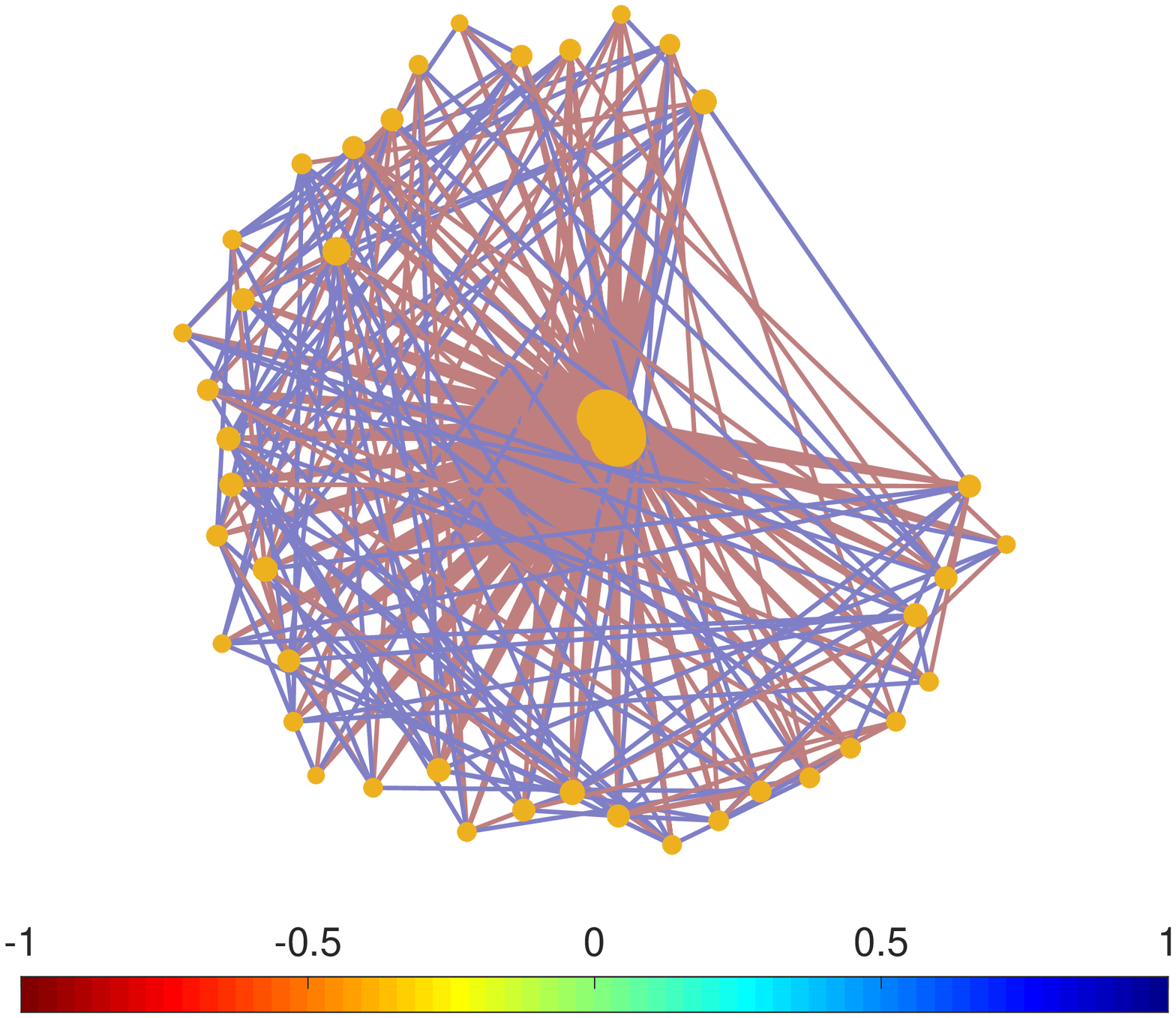}
\caption{Illustration of a quasi-CSG-WSSN with $n=50$, $m_{-}=500$, $m_{+}=100$,
and $s=10$.}
\label{quasi-CSG-WSSN}
\end{figure}

In Figure \ref{quasi-CSG-WSSN} we illustrate an example of quasi-CSG-WSSN
with $n=50$, $m_{-}=500$, $m_{+}=100$, and $s=10$.

We then tested the following two hypothesis:
\begin{itemize}
\item[1.] Unbalanced WSSNs depend only on $n$, $m_{-}$ and $m_{+}$ and not
on any specific structure.
\item[2.] Unbalanced WSSNs depend on the existence of specific quasi-CSG structure.
\end{itemize}
In the first case, the real-world WSSN would be more ``similar''
to its random version than to the quasi-CSG-WSSN and in the second
case it would be more similar to the quasi-CSG-WSSN. For the ``similarity''
between the different networks we focus here on the spectrum of their
adjacency matrices. The reason for that is that the eigenvalues of
this matrix determine the balance of the network as we have seen from
its definion before. For determining the value of $s$ we explore
all possible values and find the root mean square error (RMSE) between
the spectrum of the real WSSN and that of the quasi-CSG-WSSN. The
smallest RMSE, $s_{opt}$, determines the ``best'' value of $s$.\\ \\


\begin{table}[htpb]
\caption{Analysis of WSSNs with BUT. The number of companies $s$ forming a
central negative clique for each WSSN as well as the acronym for each
company in such clique. The last three columns correspond to the analysis
of the simulations of WSSNs with random and with quasi-CSG structures,
where $s_{opt}$ is the optimal value of $s$ to minimize the root
mean square error (RMSE) in the spectrum of the adjacency matrix of
the quasi-CSG network relative to the real one. Underlined stocks
correspond to the financial sector.}
\vspace{0.2cm}
\begin{centering}
\begin{tabular}{|>{\raggedright}m{1.5cm}|>{\centering}p{0.8cm}|>{\centering}p{7cm}|>{\centering}p{1cm}|>{\centering}p{1.3cm}|>{\centering}p{1.4cm}|}
\hline 
\multirow{3}{1.5cm}{\centering{}Market} & \multicolumn{2}{c|}{Real WSSN} & \multicolumn{3}{c|}{Simulated WSSN}\tabularnewline
\cline{2-6} \cline{3-6} \cline{4-6} \cline{5-6} \cline{6-6} 
 & \multirow{2}{0.8cm}{\centering{}$s$} & \multirow{2}{5cm}{\centering{}Companies in clique} & \multicolumn{2}{c|}{quasi-CSG} & Random\tabularnewline
\cline{4-6} \cline{5-6} \cline{6-6} 
 &  &  & $s_{opt}$ & RMSE & RMSE\tabularnewline
\hline 
Greece & 26 & BIOSK, BYTE, DROME, EKTER, ELTON, \uline{EUPIC}, EUROM, EVROF,
FIER, HAIDE, IATR, INTEK, INTET, KORDE, KRI, KTILA, KYRI, LAVI, MEDIC,
NAYP, OTOEL, SAR, SATOK, SELO, SPIR, XYLEK & 27 & 0.728 & 2.620\tabularnewline
\hline 
USA & 21 & ABCE, \uline{AHIX}, AMAZ, ARSN, ARTR, AYTU, \uline{BTSC}, CBDL,
DPLS, ETEK, FBCD, GFTX, GRSO, \uline{KPAY}, PLYZ, SGMD, SNRY, SNVP,
TPTW, VISM, WDLF & 21 & 1.810 & 4.516\tabularnewline
\hline 
Portugal & 12 & ESON, FCP, \uline{FEN}, GPA, LIG, LIT, MCP, ORE, RED, \uline{SANT},
SCP, VAF & 12 & 1.095 & 2.465\tabularnewline
\hline 
Ireland & 6 & DEL, DOY, DQ7A, GYQ, OVXA, P8ET & 6 & 1.131 & 2.255\tabularnewline
\hline 
Spain & 5 & \uline{ISUR}, NXT, \uline{NYE}, SPS, PVA & 5 & 1.684 & 3.037\tabularnewline
\hline 
France & 5 & ALEUP, ALSPW, ATA, CIB, GECP & 5 & 1.607 & 2.868\tabularnewline
\hline 
\end{tabular}
\par\end{centering}

\label{Core_companies}
\end{table}

\newpage{}

\large \textbf{G. WSSNs by financial and non-financial sectors}\normalsize\\

\begin{figure}[htpb]
\centering %
\begin{tabular}{@{}ccc@{}}
{\includegraphics[width=0.3\linewidth]{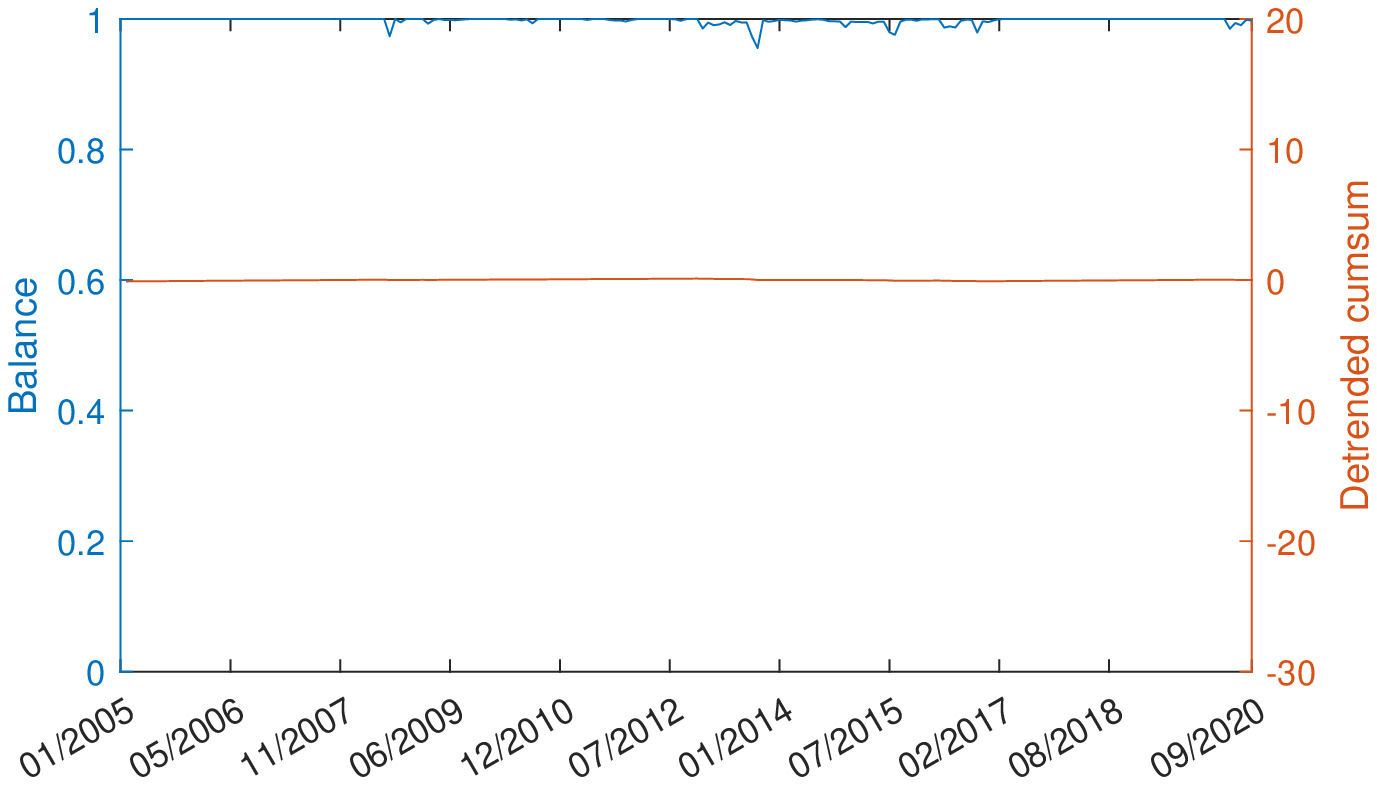}} & {\includegraphics[width=0.3\linewidth]{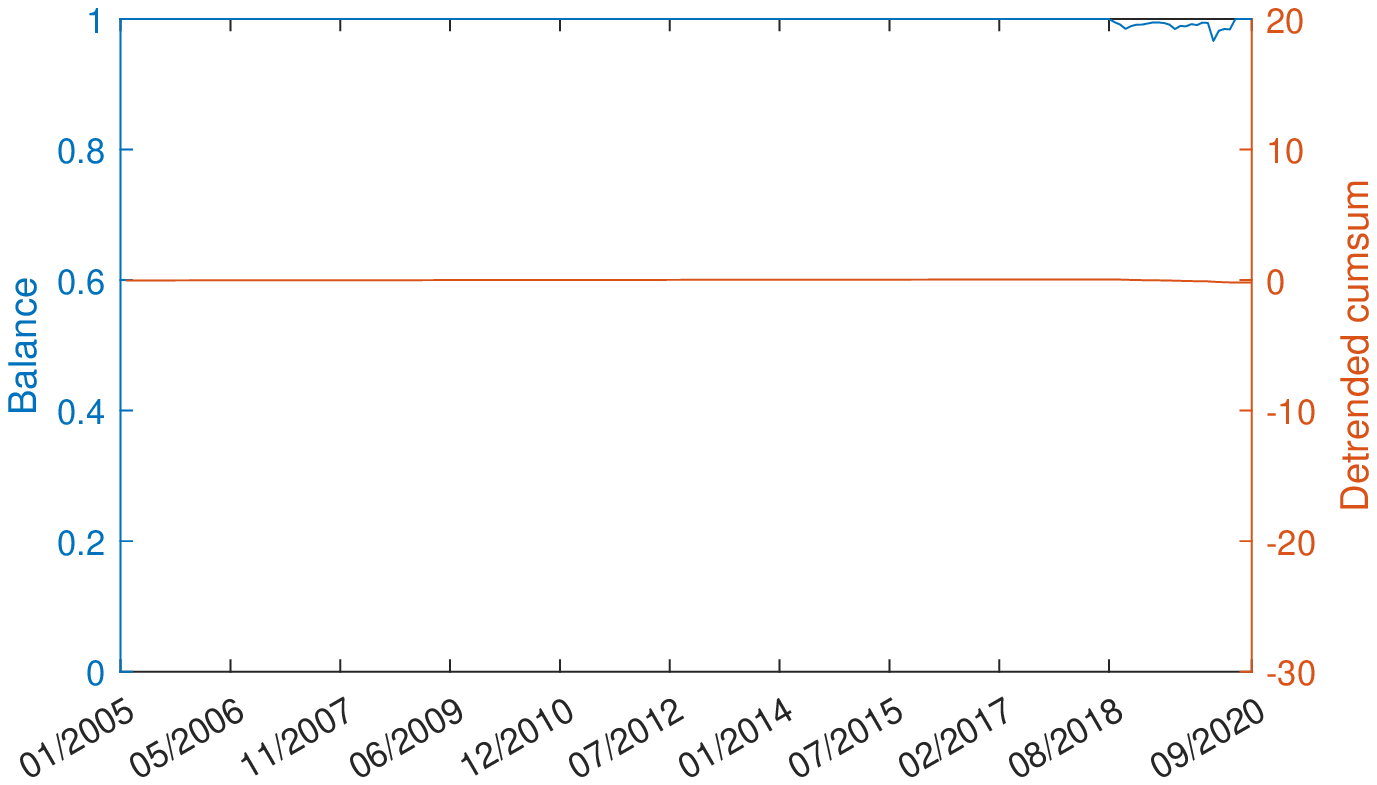}} & {\includegraphics[width=0.3\linewidth]{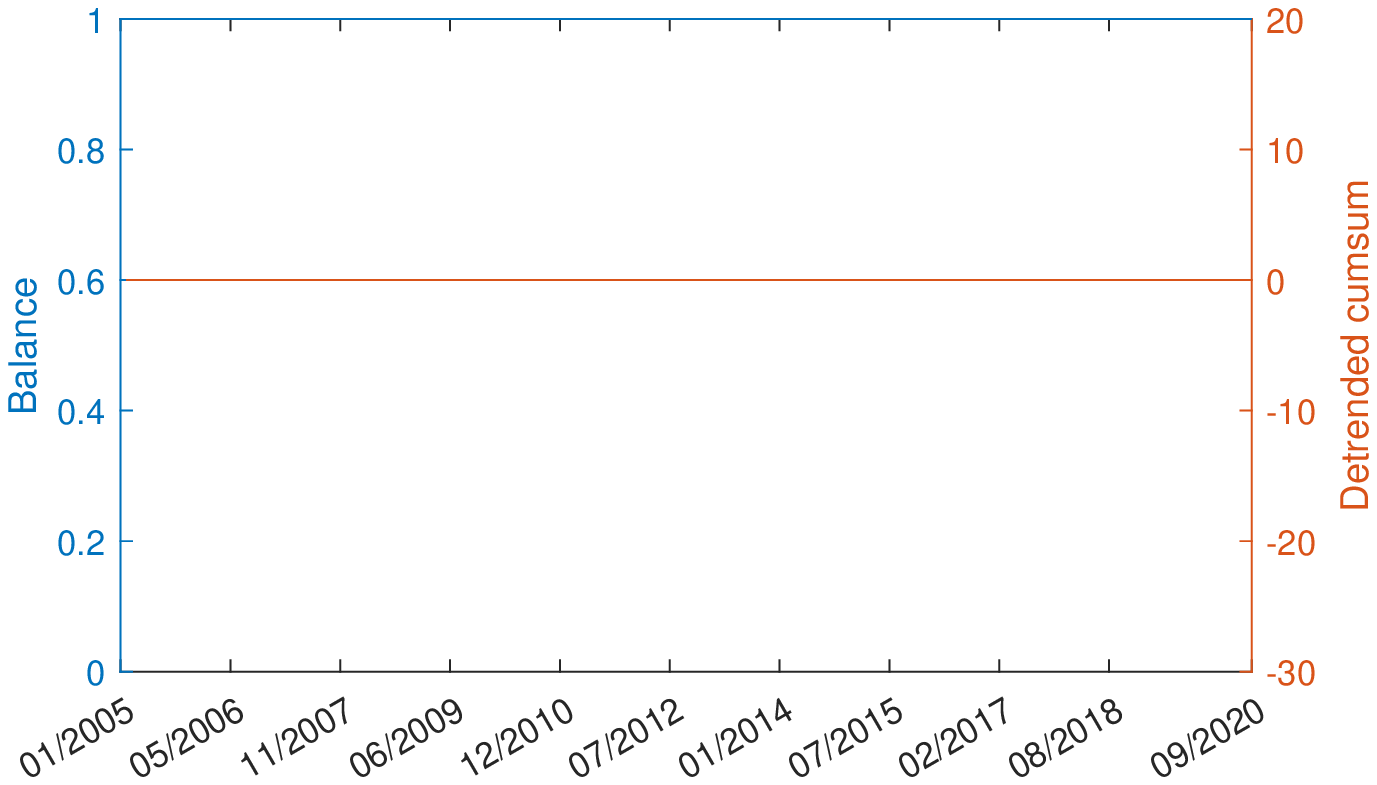}} \tabularnewline
{(a) the US} & {(b) Portugal} & {(c) Ireland}\tabularnewline \tabularnewline\tabularnewline {\includegraphics[width=0.3\linewidth]{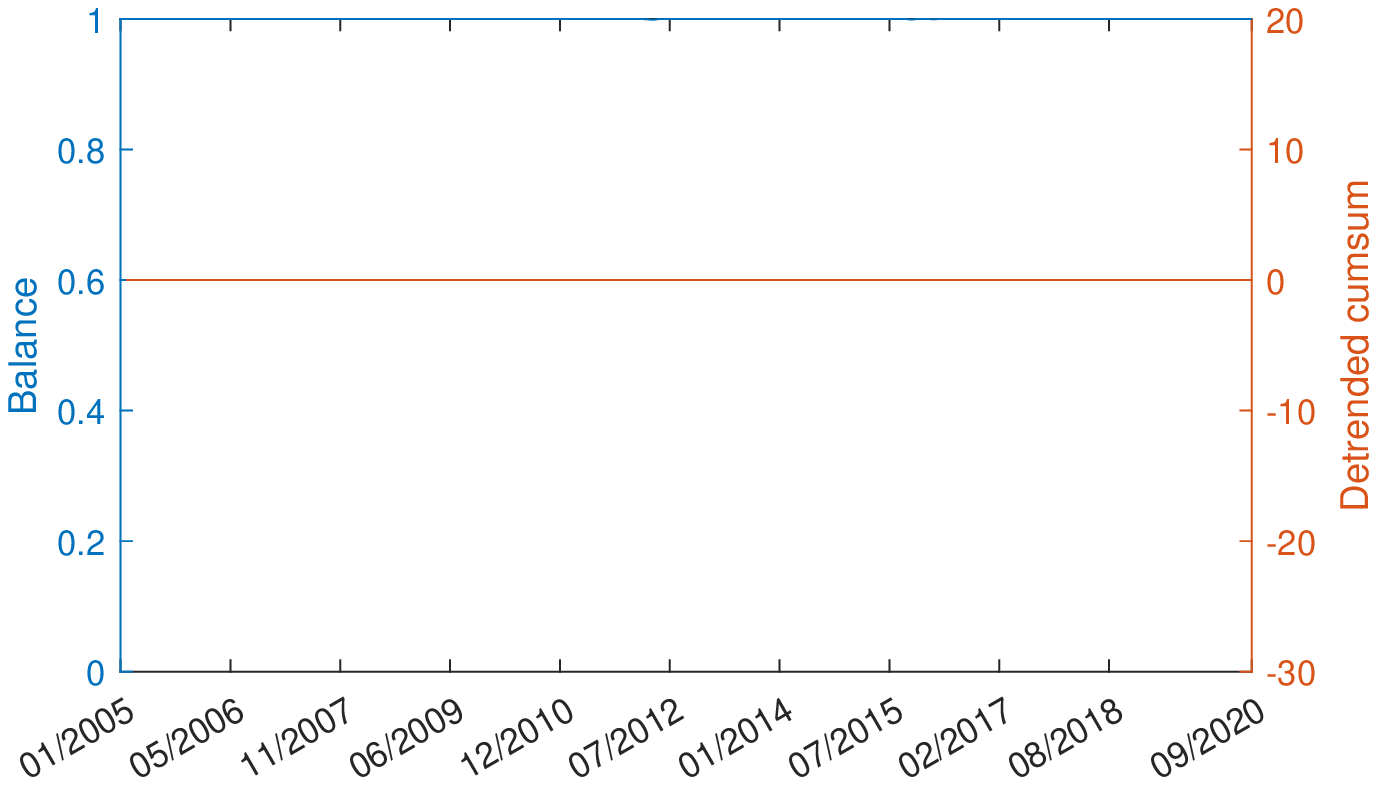}} & {\includegraphics[width=0.3\linewidth]{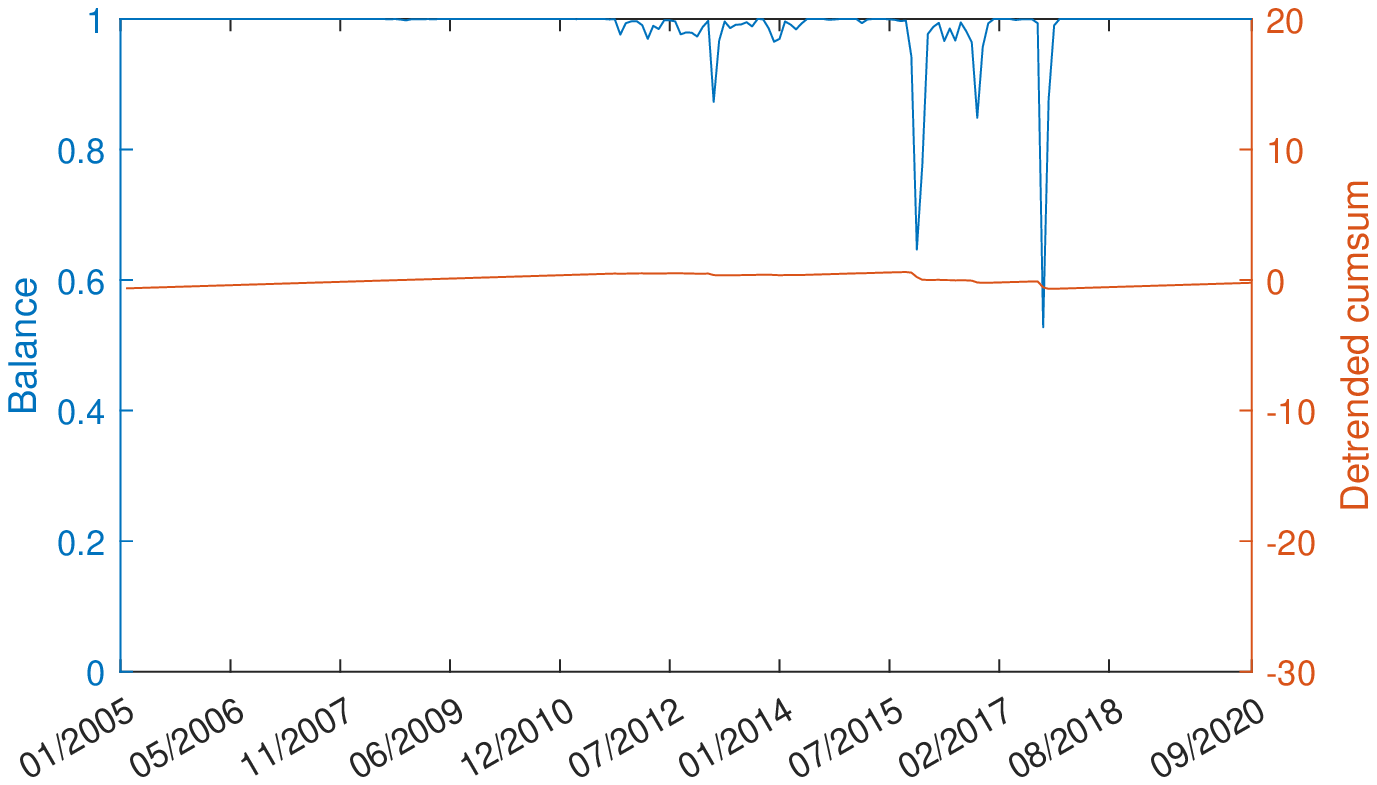}} & {\includegraphics[width=0.3\linewidth]{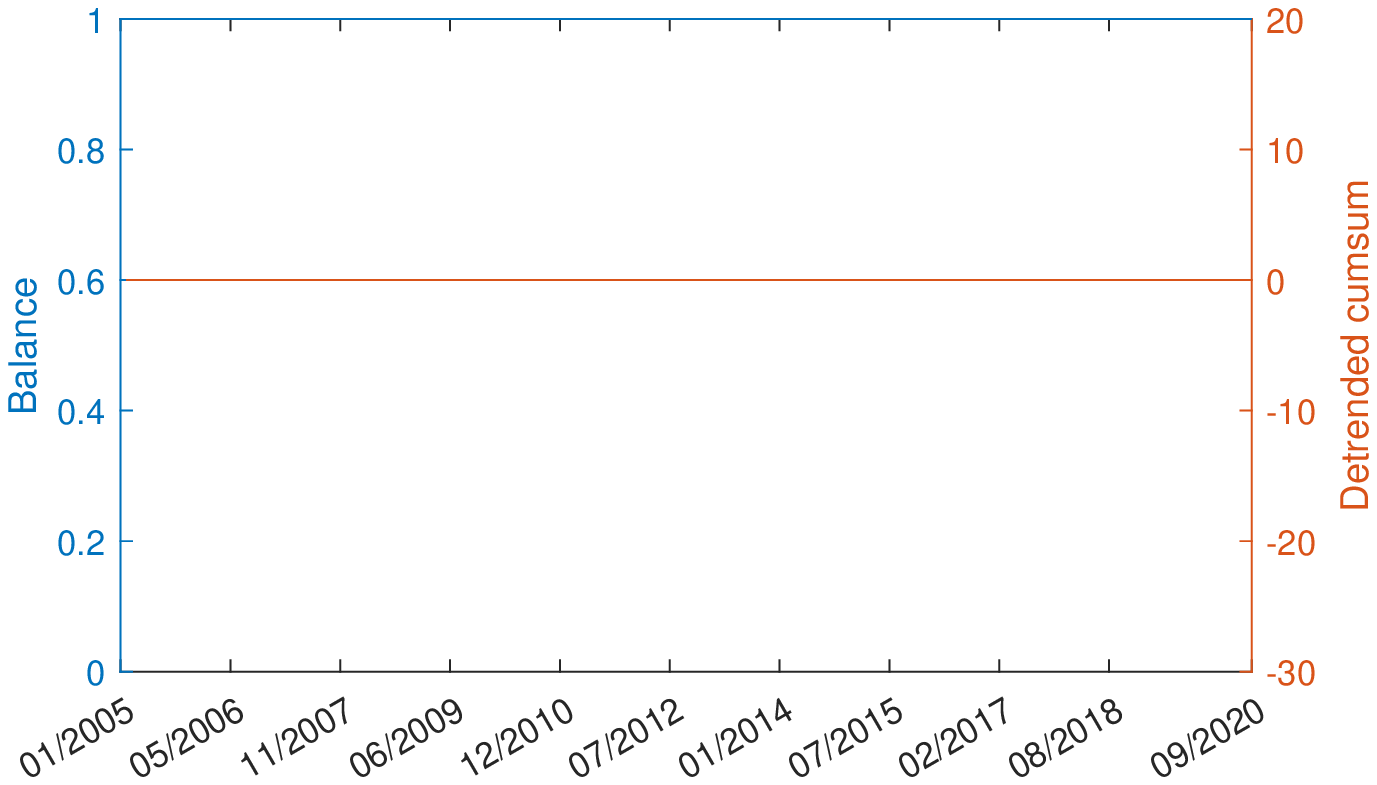}}\tabularnewline
 {(d) Greece} & {(e) Spain} & {(f) France}\tabularnewline
\end{tabular}\caption{Evolution of the balance for stocks in the financial sector between January 2005 and September 2020 in the WSSN (blue line) and of its detrended cumulative sum (red line).}

\label{Transition_countries_fin}
\end{figure}

\begin{figure}[htpb]
\centering %
\begin{tabular}{@{}ccc@{}}
{\includegraphics[width=0.3\linewidth]{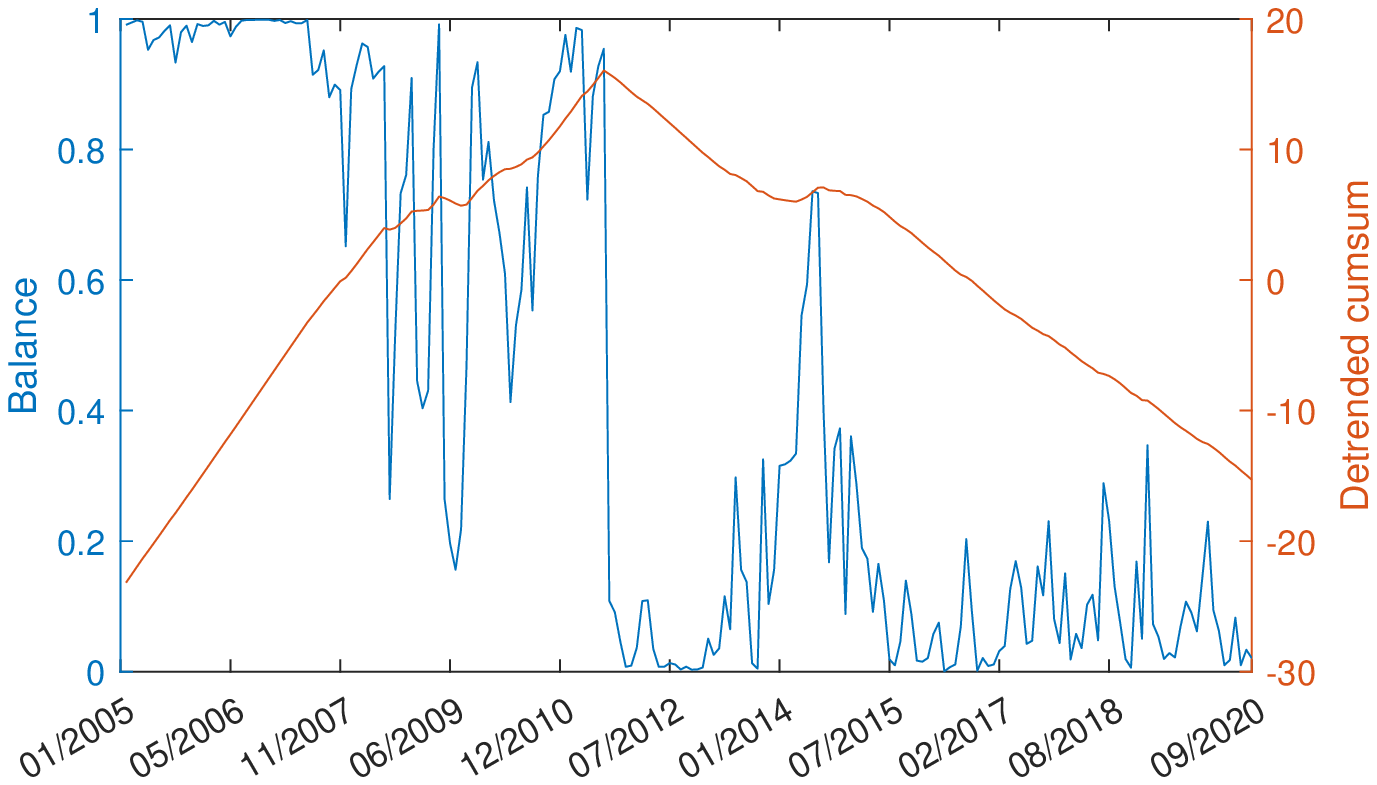}} & {\includegraphics[width=0.3\linewidth]{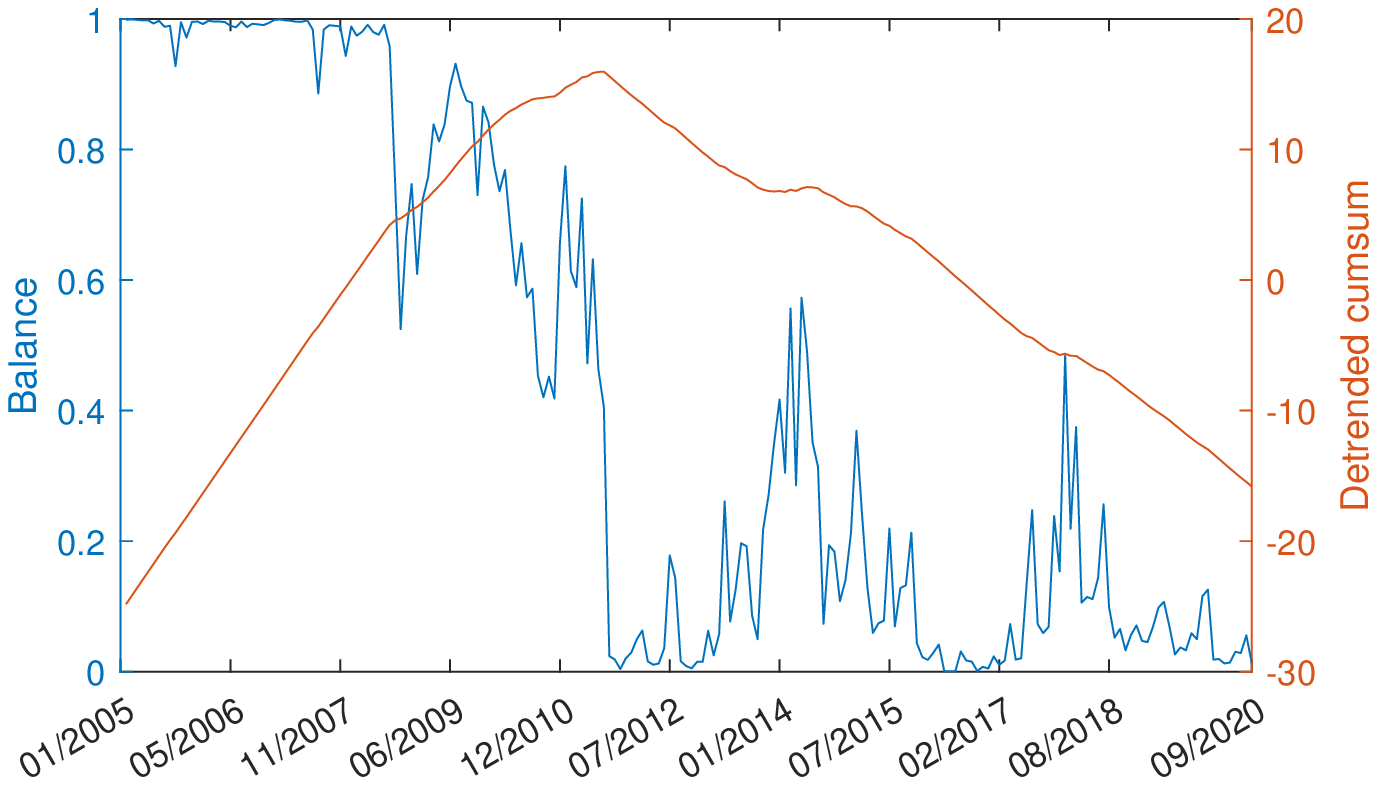}} & {\includegraphics[width=0.3\linewidth]{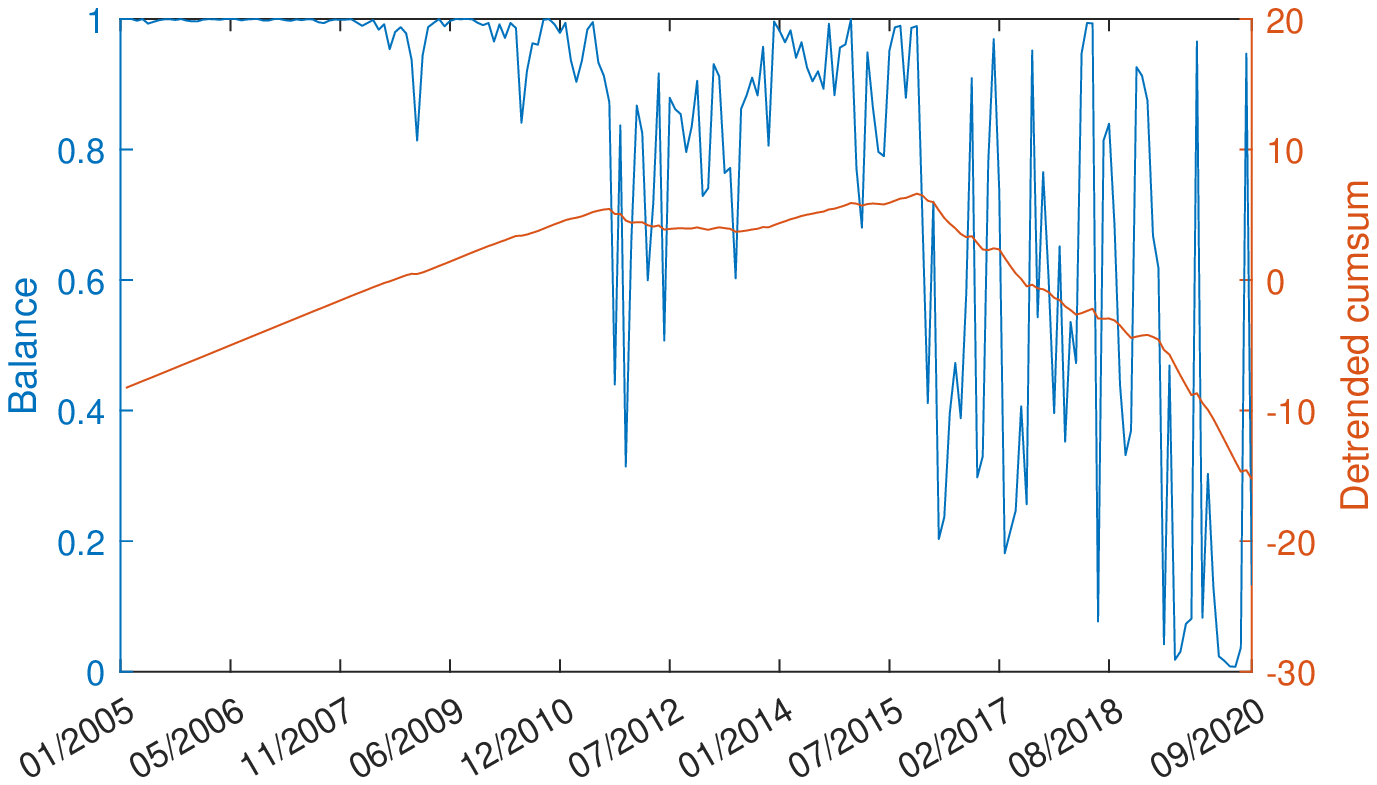}} \tabularnewline
{(a) the US} & {(b) Portugal} & {(c) Ireland} \tabularnewline\tabularnewline\tabularnewline
 {\includegraphics[width=0.3\linewidth]{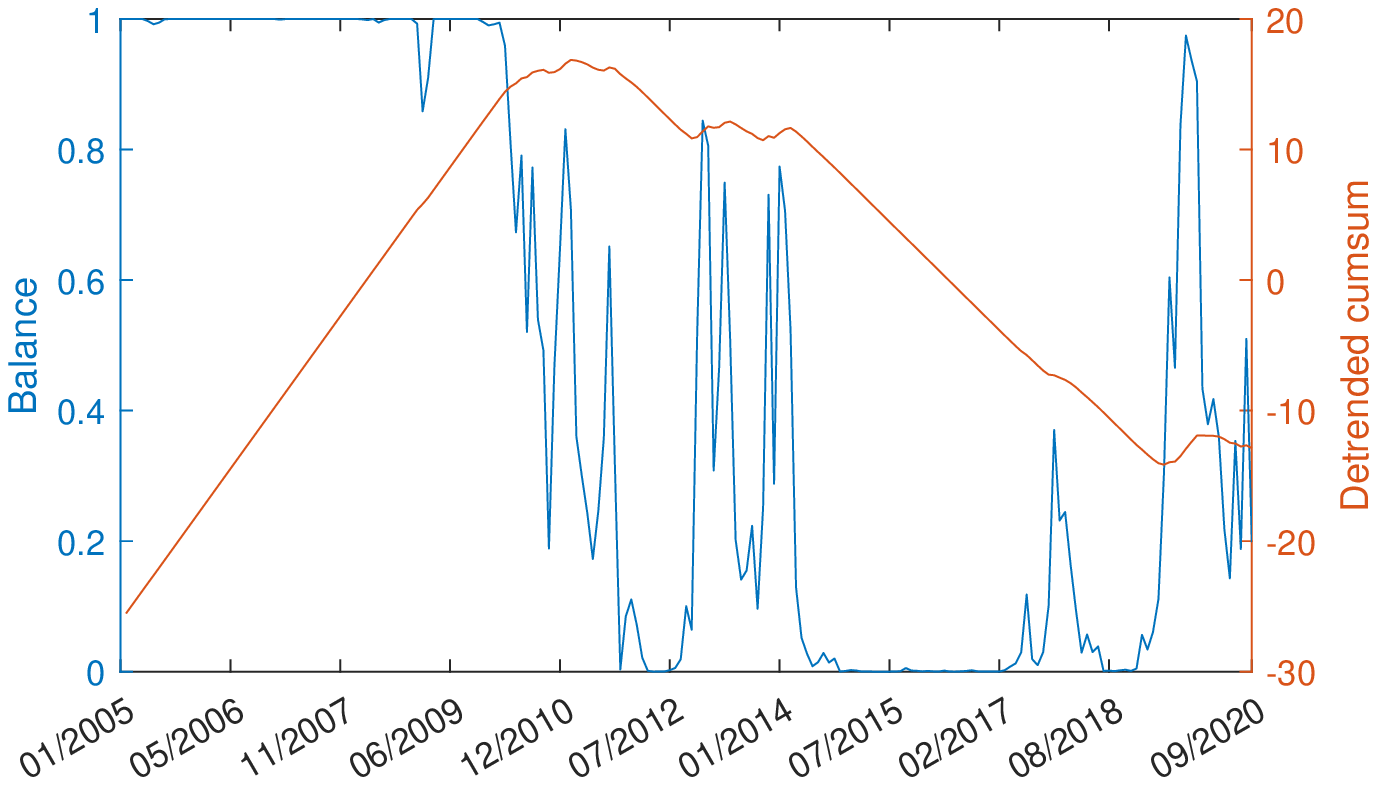}} &  {\includegraphics[width=0.3\linewidth]{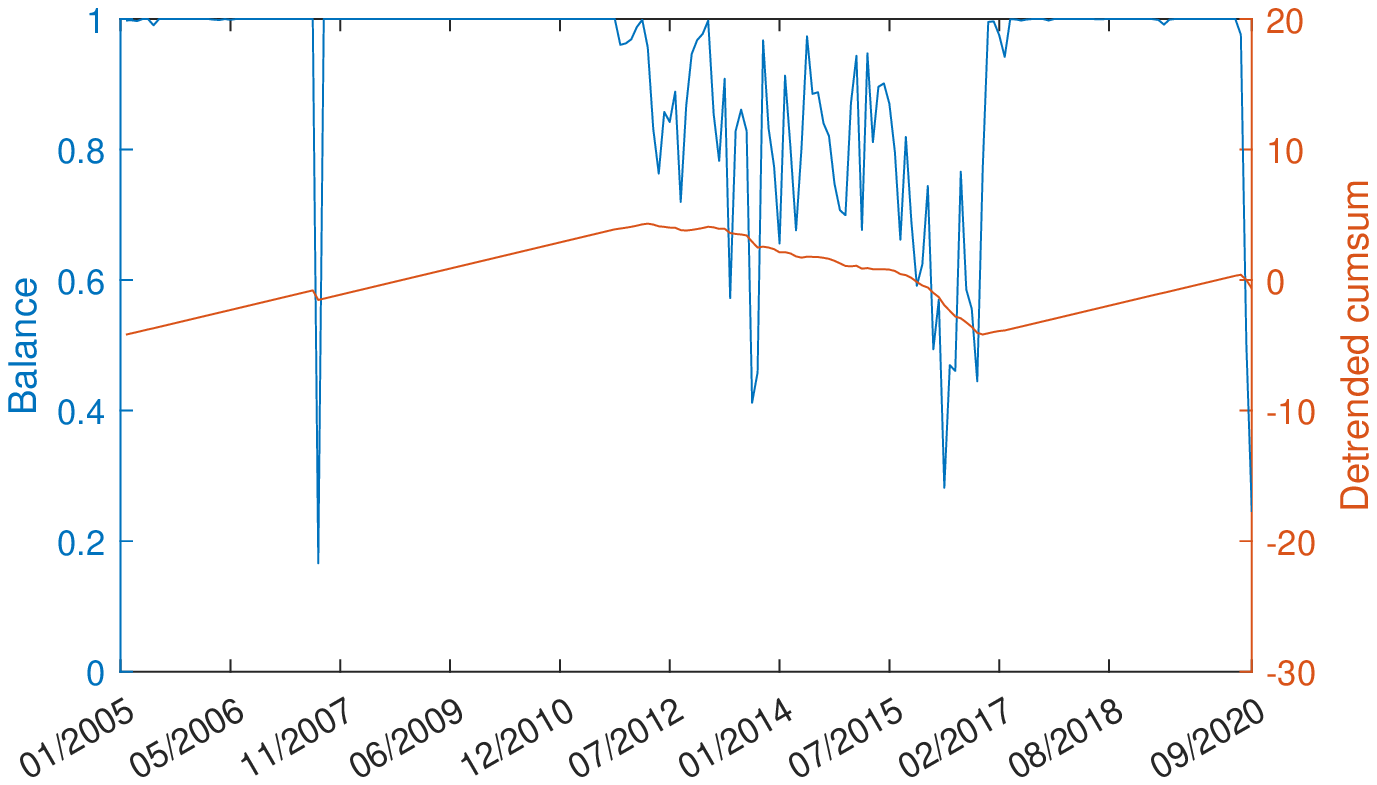}} & {\includegraphics[width=0.3\linewidth]{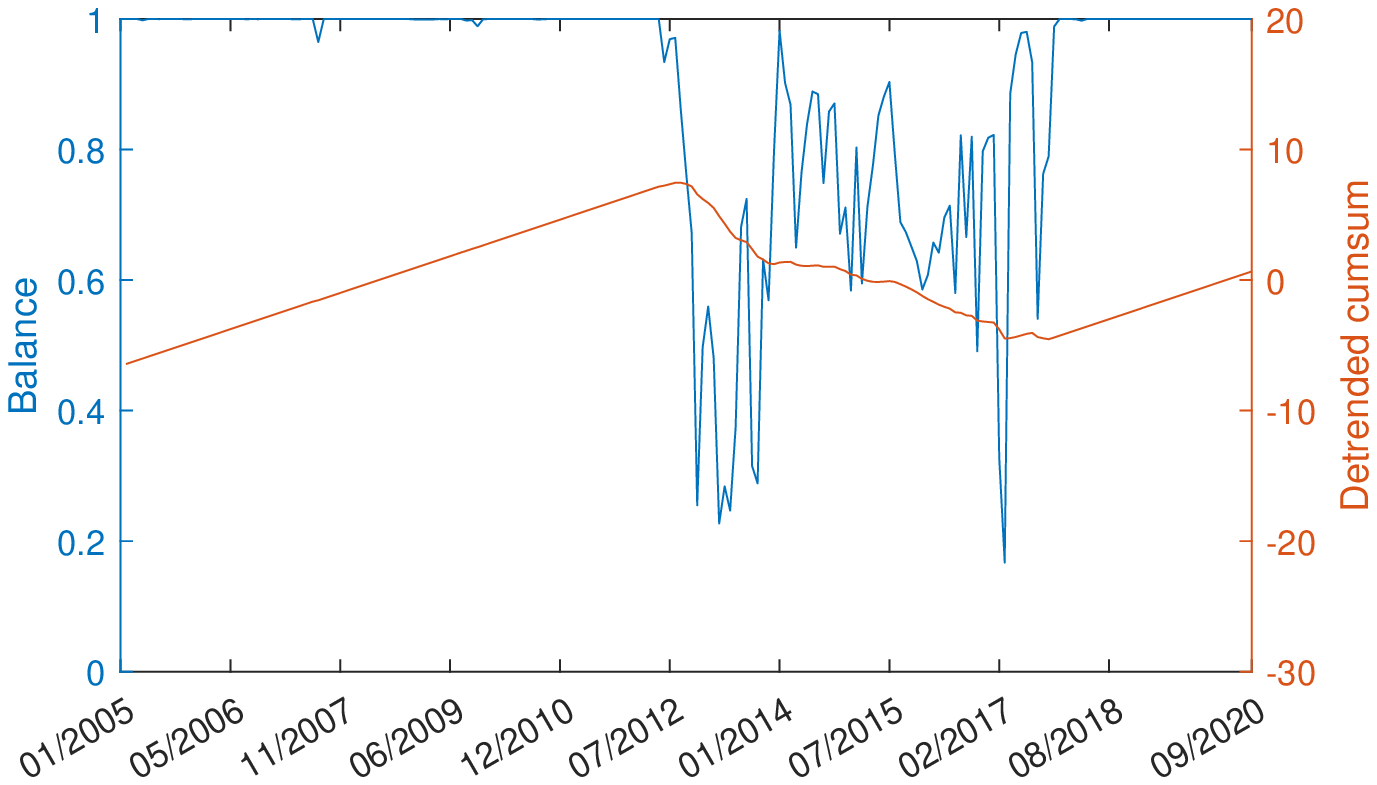}}\tabularnewline
 {(d) Greece} & {(e) Spain} & {(f) France}\tabularnewline
\end{tabular}\caption{Evolution of the balance for stocks in the non-financial sector between January 2005 and September 2020 in the WSSN (blue
line) and of its detrended cumulative sum (red line).}

\label{Transition_countries_nofin}
\end{figure}

\begin{figure}[htpb]
\centering %
\begin{tabular}{@{}ccc@{}}
{\includegraphics[width=0.3\linewidth]{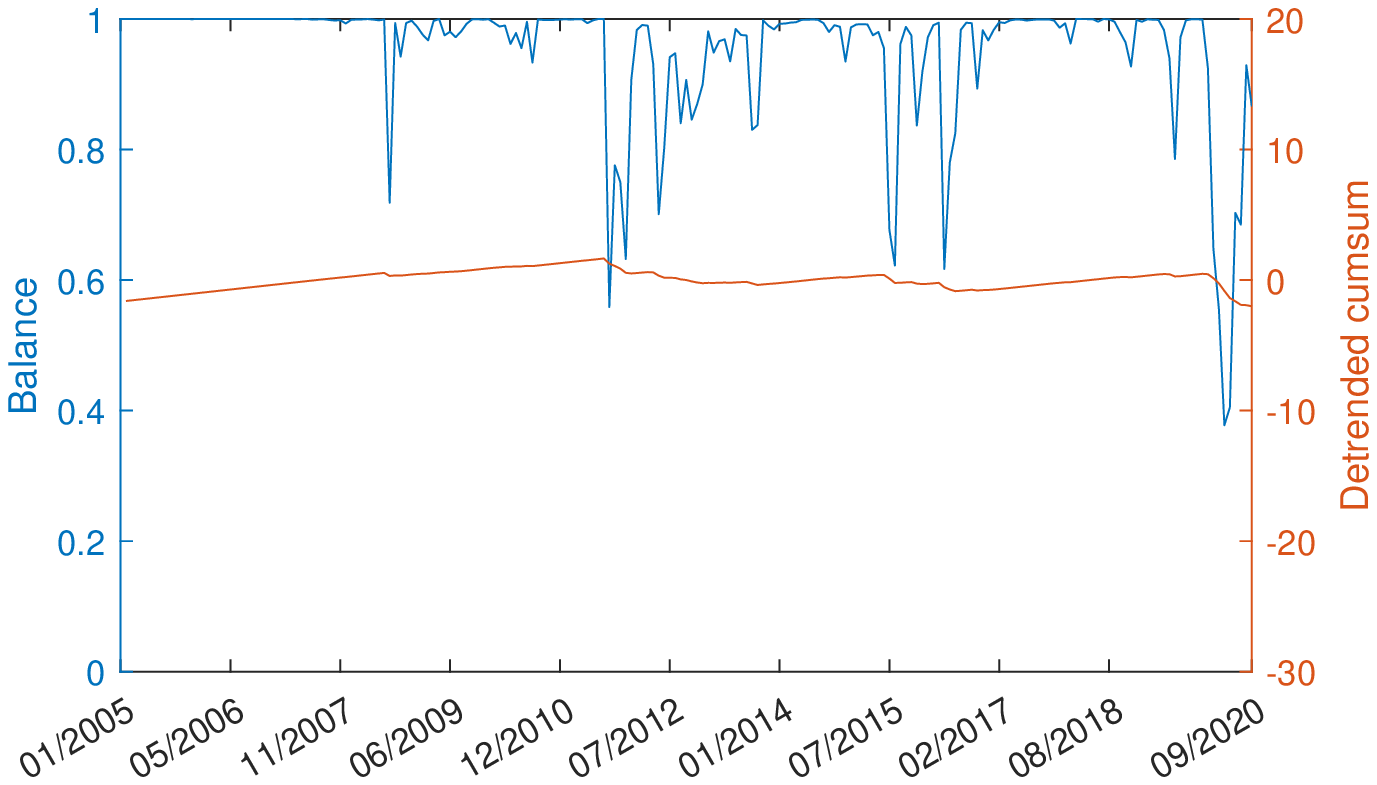}} & {\includegraphics[width=0.3\linewidth]{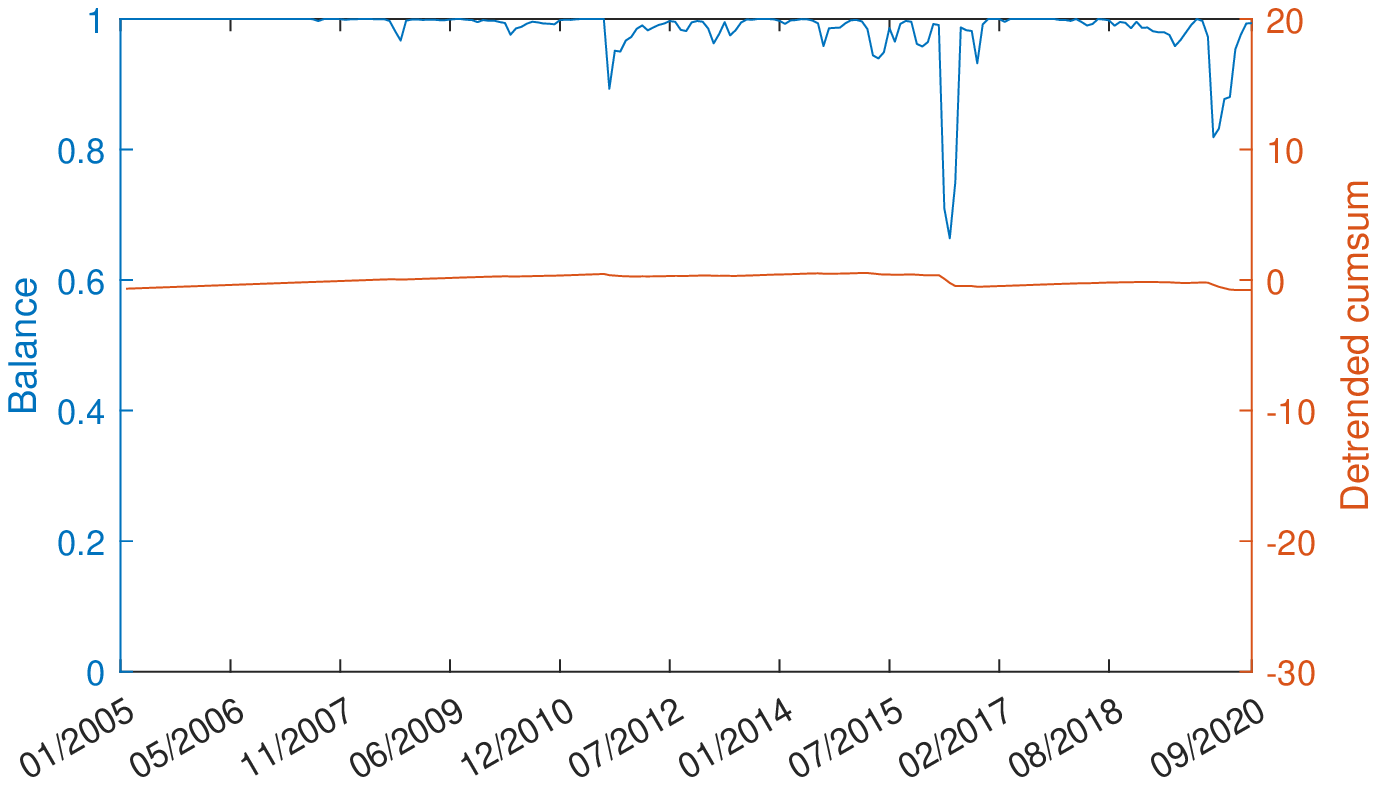}} & {\includegraphics[width=0.3\linewidth]{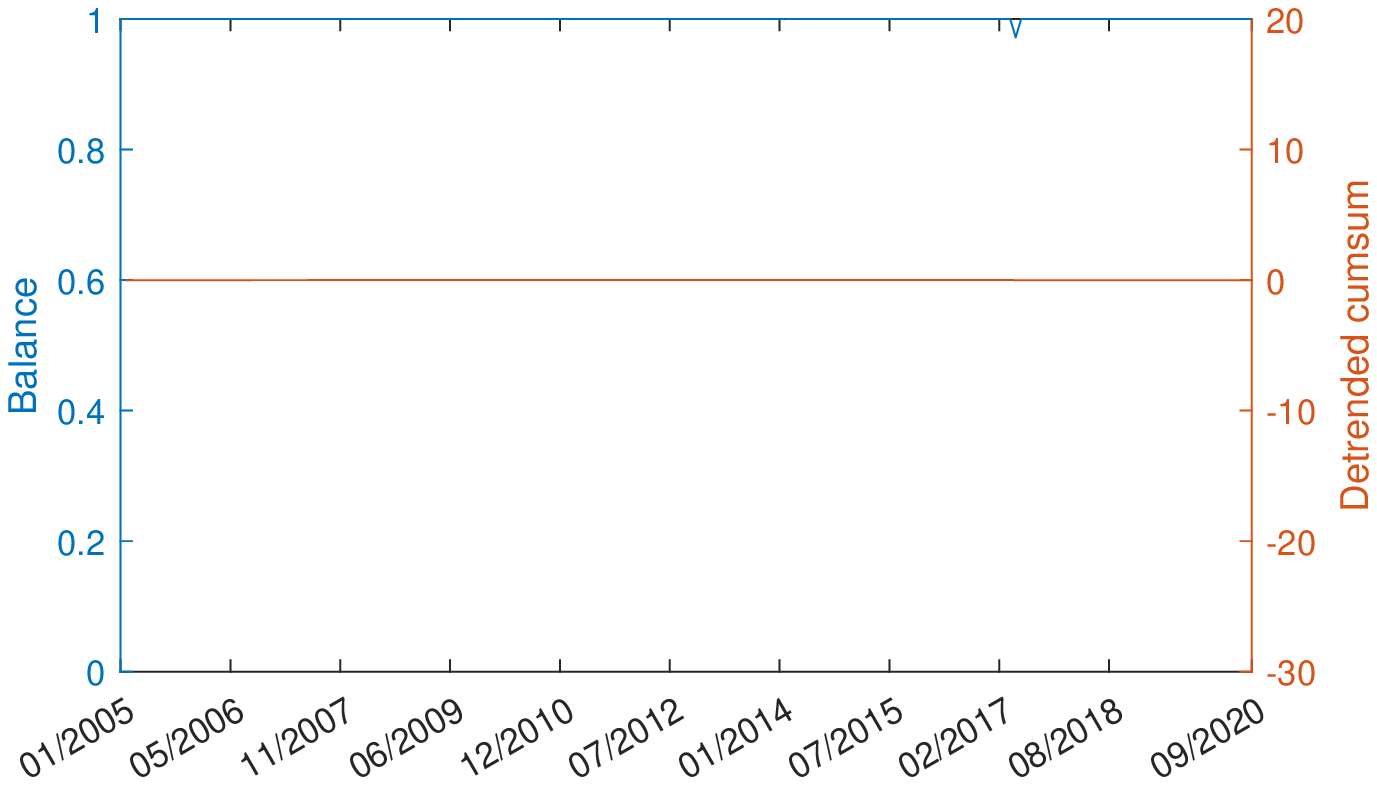}} \tabularnewline
{(a) the US} & {(b) Portugal} & {(c) Ireland} \tabularnewline\tabularnewline\tabularnewline {\includegraphics[width=0.3\linewidth]{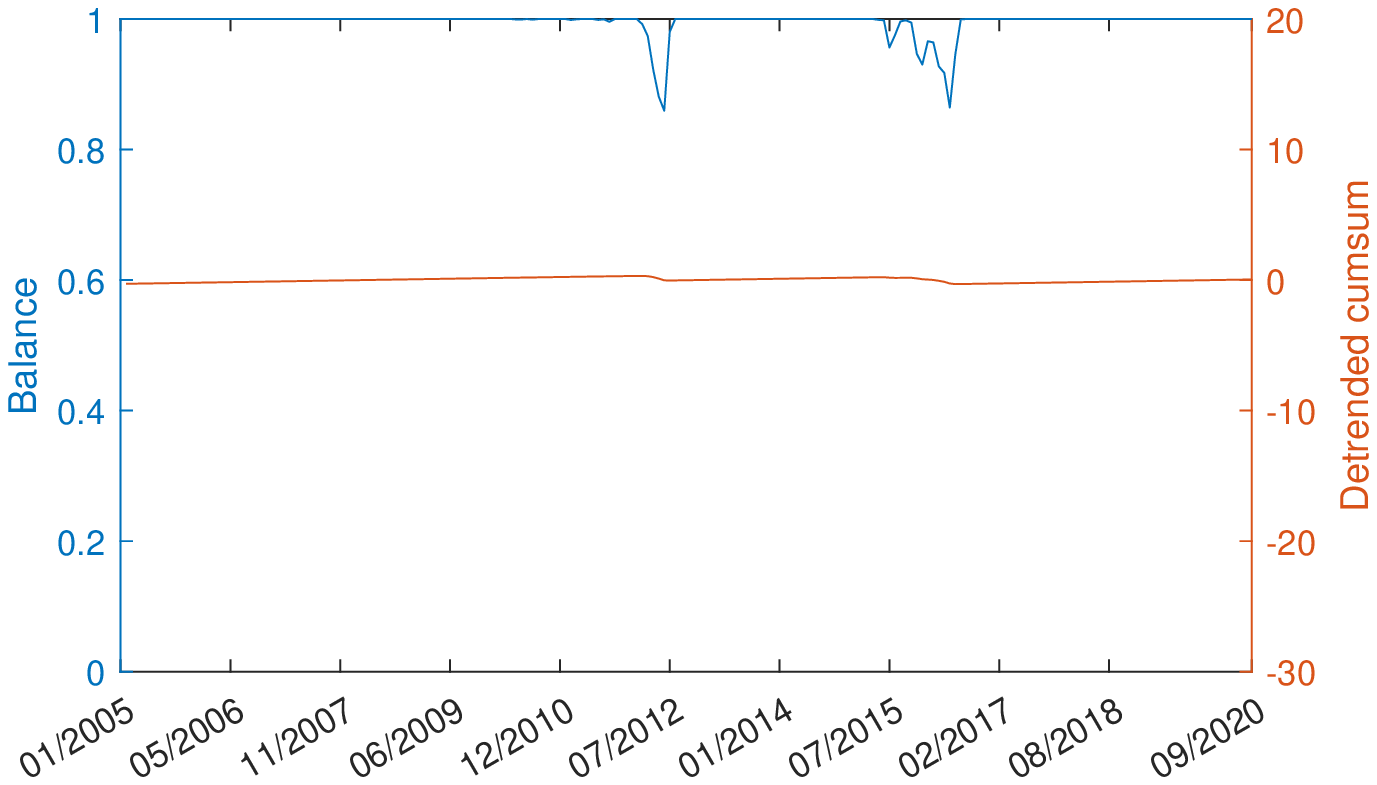}} & {\includegraphics[width=0.3\linewidth]{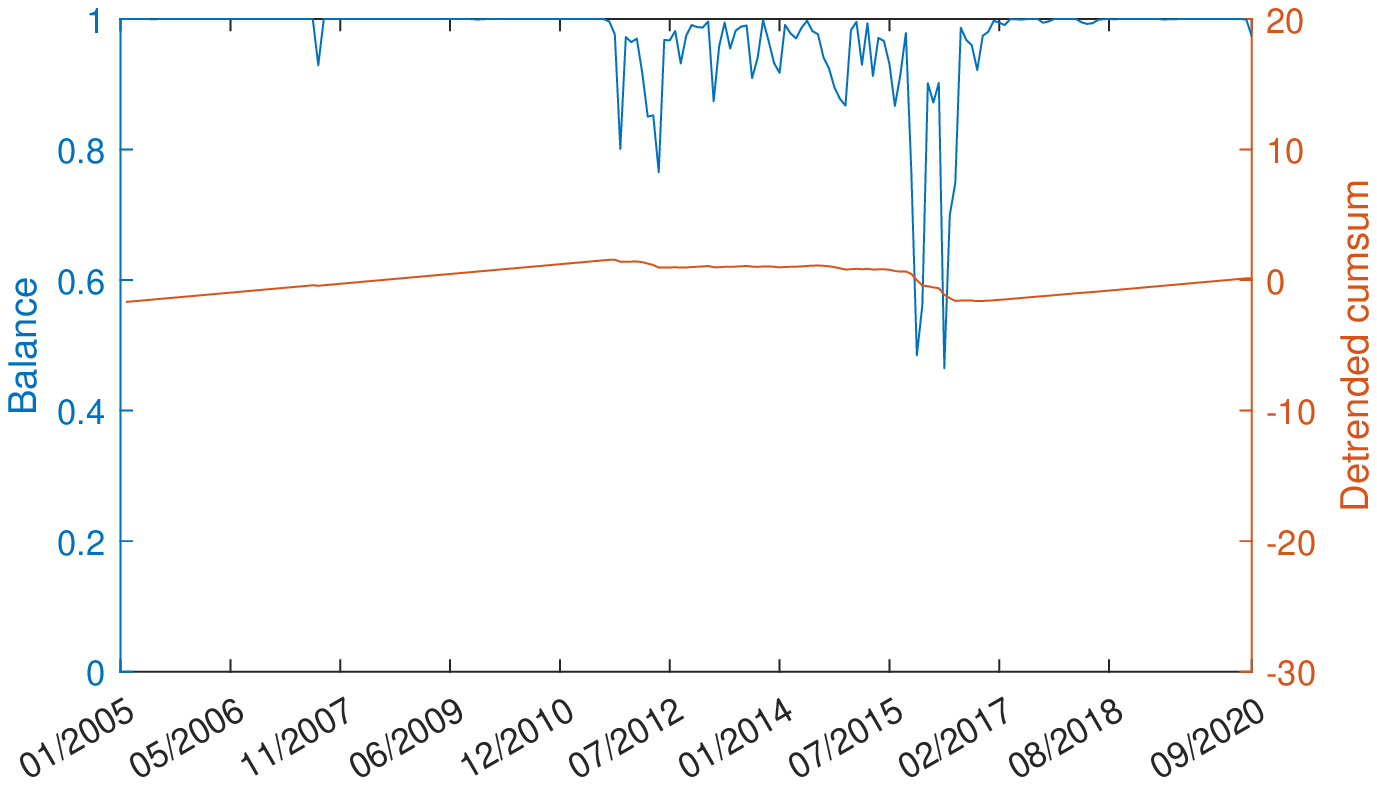}} & {\includegraphics[width=0.3\linewidth]{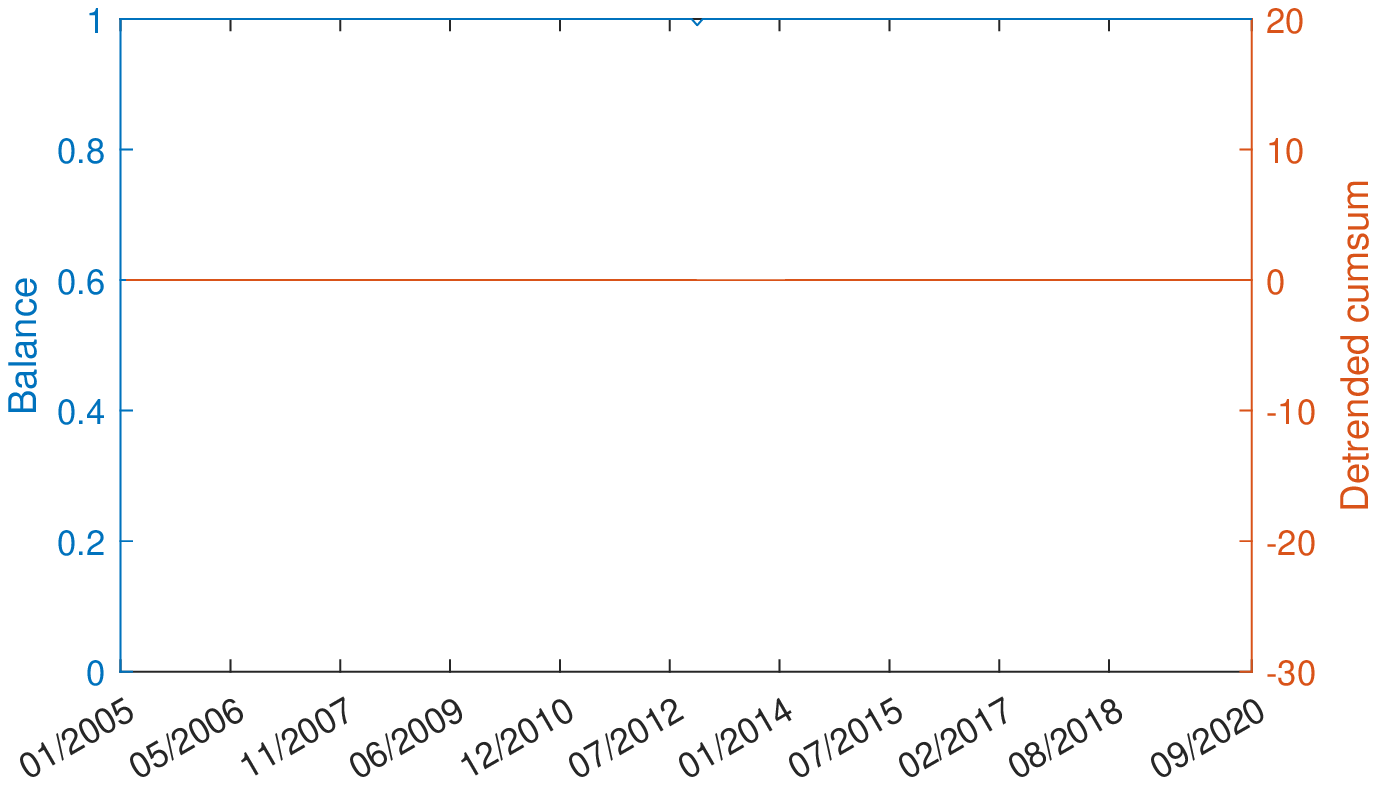}}\tabularnewline
 {(d) Greece} & {(e) Spain} & {(f) France}\tabularnewline
\end{tabular}\caption{Evolution of the balance in the WSSN for interactions between financial and non-financial sectors' stocks from January 2005 until September 2020 (blue
line), and of its detrended cumulative sum (red line).}

\label{Transition_countries_fin_nofin}
\end{figure}

Figures \ref{Transition_countries_fin} and \ref{Transition_countries_nofin} show the evolution of the balance when we split the WSSNs by financial and non-financial sectors. Figure \ref{Transition_countries_fin_nofin} presents the balance evolution for the WSSNs with cross interactions between financial and non-financial sectors' stocks.


\end{document}